\documentclass[sigconf]{acmart}

\usepackage[toc,page]{appendix}
 
\usepackage{pgfplotstable}
\usepackage{paralist}
\usepackage{amsmath}
\usepackage{listings}
\usepackage{multirow}

\usepackage{epsfig,endnotes}
\usepackage{tikz}
\usepackage{pgfplots}
\usepackage[ruled,vlined,linesnumbered] {algorithm2e}
\usepackage{amsfonts}
\usepackage{url}
\usepackage{tikz}
\usepackage{array}
\usepackage{booktabs,adjustbox}
\usepackage{etex}
\usepackage{color}
\usepackage{pstricks}
\usetikzlibrary{calc,positioning,shapes.geometric,shapes.symbols,shadows,arrows}
\usepackage{wasysym}
\usepackage{shortcuts}
\usepackage{paralist}
\renewcommand{\paragraph}[1]{\vspace{0.05in}\noindent{\bf{#1}.}}
\usepackage[justification=justified]{caption}
\usepackage{pgfplots}
\usetikzlibrary{arrows}
\usetikzlibrary{patterns}
 
\usepackage{listings}
\usepackage{multicol}
\usepackage{lipsum}
\usepackage{adjustbox}
\usepackage[font=bf,skip=\baselineskip]{caption}
\usepackage{tabularx}
\usepackage{multirow}

\usepackage[english]{babel}
\usepackage[utf8x]{inputenc}
\usepackage{listings}
\usepackage{color}
\usepackage{xcolor}
\usepackage{subcaption}

  \usepackage[frozencache=true,cachedir=minted-cache]{minted}

\usepackage{hyperref}
\hypersetup{
    colorlinks = true,
    linkcolor = blue,
    anchorcolor = blue,
    citecolor = blue,
    filecolor = blue,
    urlcolor = blue
}

\newenvironment{packeditemize}{
\begin{list}{$\bullet$}{
\setlength{\itemsep}{1.5pt}
\setlength{\labelwidth}{8pt}
\setlength{\leftmargin}{10pt}
\setlength{\labelsep}{3pt}
\setlength{\listparindent}{\parindent}
\setlength{\parsep}{1.5pt}
\setlength{\parskip}{1.5pt}
\setlength{\topsep}{1.5pt}}}{\end{list}}

\newcommand{\wechat}{{\textsf WeChat}\xspace}
 
\usepackage{enumitem}
\setcopyright{none} 
\acmConference[Submission to ACM CCS 2023]{ACM Conference on Computer and Communications Security} 
\acmYear{2023}

 \usepackage{lipsum}
\usepackage[htt]{hyphenat}
\usepackage[all]{nowidow}

\usepackage{caption}
\usepackage{color}
 \usepackage{fancyhdr}
\renewcommand{\paragraph}[1]{\vspace{0.05in}\noindent{\bf{#1}.}}

\usepackage{textcomp}
\usepackage[utf8x]{inputenc}
\usepackage{lipsum}
\usepackage[htt]{hyphenat}
\begin{document}
 

\title{Uncovering and Exploiting Hidden APIs in Mobile Super Apps} 
\author{Chao Wang}
\affiliation{%
  \institution{The Ohio State University}
}
\email{wang.15147@osu.edu}

\author{Yue Zhang}
\affiliation{%
  \institution{The Ohio State University}
}
\email{zhang.12047@osu.edu}

\author{Zhiqiang Lin}
\affiliation{%
  \institution{The Ohio State University}
}
\email{zlin@cse.ohio-state.edu}




  \settopmatter{printacmref=false} 
 \renewcommand\footnotetextcopyrightpermission[1]{} 
 \begin{abstract}
Mobile applications, particularly those from social media platforms such as \textsf{WeChat} and \textsf{TikTok}, are evolving into ``super apps'' that offer a wide range of services such as instant messaging and media sharing, e-commerce, e-learning, and e-government. These super apps often provide APIs for developers to create ``miniapps'' that run within the super app. 
%
%
These APIs should have been thoroughly scrutinized for security. Unfortunately, we find that many of them are undocumented and unsecured, potentially allowing miniapps to bypass restrictions and gain higher privileged access. 
   %
To systematically identify these hidden APIs before they are exploited by attackers, we developed a tool \sysname with both static analysis and dynamic analysis, where static analysis is used to recognize hidden undocumented APIs, and dynamic analysis is used to confirm whether the identified APIs can be invoked by an unprivileged 3rd-party miniapps. 
 We have applied \sysname to five popular super apps (i.e.,  \wechat, \textsf{WeCom}, \textsf{Baidu}, \textsf{QQ}, and \textsf{Tiktok}) and found that all of them contain hidden APIs, many of which can be exploited due to missing security checks. We have also quantified the hidden APIs that may have security implications by verifying if they have access to resources protected by Android permissions. Furthermore, we demonstrate the potential security hazards by presenting various attack scenarios, including unauthorized access to any web pages, downloading and installing malicious software, and stealing sensitive information. We have reported our findings to the relevant vendors, some of whom have patched the vulnerabilities and rewarded us with bug bounties. \looseness=-1
 

\end{abstract}

\maketitle
\section{Introduction}
\label{sec:intro}

%





Over the past a few years, we have witnessed a rapid growth of the miniapp paradigm~\cite{whitepaper20}, in which a mobile super app (e.g., \wechat~\cite{totalrevenue} and \textsf{TikTok}~\cite{TikTokMa29:online}) provides a seamless runtime environment for a miniapp, a web-app alike small application, for enhanced user experience (e.g., install-less) and stickiness with the super app (e.g., a user can access almost all the daily services without leaving it).
Today, more than 4.3 million miniapps~\cite{Wechat:miniapp:number} have been developed in \wechat (a super app with 1.2 billion monthly active users~\cite{wechat:stat}), surpassing the total number of Android apps in Google Play (which has about 2.7 million as of November 2022~\cite{GoogleP54:online}). 
 These miniapps offer a variety of daily services from transportation (e.g., ride hailing), e-commerce (e.g., online shopping), e-learning, e-government (e.g., pandemic control and contact tracing), mobile gaming, to entertainment (e.g., short-form user videos), and so on. They are developed by both the 1st-party (i.e., the one who makes the super app platform), as well as 3rd-party (i.e., developers who create additional software based on the platform provided by the 1st-party).
\looseness=-1 %
%


Obviously, since both the 1st-part and the 3rd-party miniapps are all built on top of the APIs provided by the super app platform, they would have used the same set of the APIs. However, by performing a manual analysis, we discovered discrepancies in the APIs used by these miniapps. For instance, privileged APIs like \texttt{openUrl} are present in 1st-party miniapps like \textsf{Tencent Doc}~\cite{tencentapp}, which has more than 200 million online consumers. \texttt{openUrl} can open arbitrary URLs, but the 3rd-party miniapps cannot use \texttt{openUrl} and must use the \texttt{wx.request} API to ensure that the URLs are checked by WeChat to prevent the loading of malicious content. Moreover, not all APIs are equally mentioned in the official documentation. The Chinese version of the development documentation comprises 975 APIs~\cite{wechatcngdoc}, while the English version has only 570 APIs~\cite{wechatengdoc}. Additionally, none of the privileged APIs, such as \texttt{openUrl} are ever referenced in the official documentation, regardless of the language. Thus, there may be undocumented APIs in the super app platforms (at least in WeChat). Such undocumented APIs may pose security risks. For example, they may have a higher level of privilege, as they are designed exclusively for use by 1st-party apps. In order to ensure security, super apps should implement proper access controls for these privileged APIs, such as allowing access solely through an approved list for 1st-party miniapps. Otherwise, they may be a weak spot for unauthorized access by 3rd-party miniapps.\looseness=-1 

Although our manual analysis with the host app and its 1st-party miniapp implementation has yielded surprising findings, it is certainly not scalable nor complete. Meanwhile, given the fact that so many super apps are 
available today, it will be extremely helpful if we can have a tool to identify all of the hidden APIs if that is possible from their implementations. Also, since privileged APIs without any checks can be easily exploited by malicious miniapps, we must inform the super app vendors to patch the missing or misplaced checks. Motivated by these pressing needs, in this paper, we present \sysname, a binary analysis tool combined with both static and dynamic analysis to systematically scrutinize hidden {API}s, which are undocumented, from super app implementations. 
 \looseness=-1

 

Multiple challenges must be addressed while developing \sysname. Particularly, several programming languages have been used to implement a super app at various layers (e.g., JavaScript at the miniapp layer, C/C++ at the JavaScript runtime layer, and Java at the service abstraction layer provided by the host app), and consequently it is challenging to recognize how APIs across these different languages and interfaces are invoked. Second, after identifying an undocumented API, it is also challenging to classify whether it is an API that can be invoked by third-party miniapps. Fortunately, we have addressed these challenges and successfully implemented \sysname.
There are two key components inside \sysname: \textit{Static API Recognition} and \textit{Dynamic API Classification}. At a high level, it takes a super app binary as well as its list of public APIs as input, and identifies the hidden APIs based on the invariants of the functions and interfaces from the public APIs in the super apps using \textit{Static API Recognition}. Next, it dynamically executes the identified APIs to confirm whether they are true APIs, and further classifies them into checked and unchecked ones based on whether it can only be invoked by the  
1st-party miniapps using \textit{Dynamic API Classification}.

We have tested \sysname with five popular super apps: \wechat, \textsf{WeCom}, \textsf{Baidu}, \textsf{QQ},  and \textsf{TikTok}.  
Our evaluation results show that all the tested super apps contained hidden APIs.  Interestingly, our study found hidden APIs in different categories, with some super apps having more hidden APIs than documented ones. For example, the API category of \textsf{Payment} of \wechat contains 28 hidden APIs, which is significantly more than its documented ones (i.e., only one). We also measure the usage of hidden APIs in both 1st party miniapps and 3rd party miniapps. We found that the use of undocumented APIs is common among both 1st-party miniapps and 3rd-party miniapps regardless of their category.

It is evident that not all hidden APIs may pose security risks when misused. Therefore, our objective was to dive into the security implications of hidden APIs. Specifically, we focused on the hidden APIs that lack security checks but can access sensitive Android OS resources. To achieve this, we proposed the use of dynamic analysis techniques. Our dynamic analysis approach involves identifying APIs that call native APIs, which can access sensitive resources. We achieved this by hooking APIs that access sensitive resources and monitoring their use by unchecked and undocumented APIs. After conducting our investigation, we found that WeChat has 39 hidden unchecked APIs (7.77\%) that invoke Android APIs protected by permissions. Similarly, WeCom has 40  (6.75\%), 
Baidu has 8 
(7.61\%), Tiktok has 32 
(26.23\%), and QQ has 38 
(12.88\%) such APIs, which can have security risks. 

To further validate our findings, we conducted several attack case studies by developing a number of malicious miniapp using these hidden APIs. Specifically, in \wechat, we developed a malicious mini-app to exploit the hidden \texttt{private\_openUrl} API to access arbitrary malicious content without detection by the super apps. Additionally, by using the \texttt{installDownloadTask} hidden API, we developed a mini-app that can download and install harmful Android apps surreptitiously. Malicious apps have the capability to pilfer a user's sensitive information. Our demonstration reveals the utilization of hidden APIs such as \texttt{captureScreen}, which enables malicious miniapps to steal screenshots, \texttt{getLocalPhoneNumber}, which permits theft of the user's phone number, and \texttt{searchContacts}, which facilitates the theft of the user's contact information.

\paragraph{Contributions} We make the following contributions:
\begin{packeditemize}

\item 
We are the first to discover that super apps may provide hidden, i.e., undocumented, APIs (for the 1st-party miniapps), and those hidden APIs that do not have permission checks can be exploited by the 3rd-party miniapps for privileged accesses. 
\looseness=-1

\item 
We propose \sysname  to systematically identify and classify the hidden  APIs in super apps, with two novel techniques to statically recognize the APIs  and  dynamically execute and classify them. \looseness=-1

\item 
We implement \sysname,  and evaluate it with 5 
super apps  
and {find} all of them containing hidden APIs, some of which can be exploited by malicious 3rd-party miniapps. 
We have made the responsible disclosure to their vendors, and received bug bounties from some of them. \looseness=-1 

\end{packeditemize}

\ignore{
This rapid development has won over a great amount of users,
    however,
    the security evaluation of Super Apps has never been conducted.
Is MiniApp  running upon a concrete runtime implemented by Super Apps?
Is there any ways that MiniApp  may be leveraged to attack user?
Is vendors of Super Apps hiding privileged API from developers for other purpose?

\noindent
\textbf{The Ambition of Super Apps}
The more frequent user uses  MiniApps,
    the more stick user rely on Super Apps.
This is why Super Apps sell their products and make costumers use their  MiniApps .
With the developing market sharing of the  MiniApps,
    native apps have been squeeze out of the user endpoint,
    which makes the security of MiniApp  becomes more and more essential.

To make MiniApp  has the native-like user experience,
    Super App is playing a role of the operating system that handles all of the resources and permissions requests from MiniApp .
Hence,
    a critical security pitfall comes:
    Super Apps do not have a solid gate to efficiently isolate the execution of MiniApp .
To address this issue,
    we compare the operating system with Super Apps.
For the operating system,
    kernels utilizes the hardware features,
    like \code{Ring} in x86\_64 platform and \code{Exception Level} in ARM platform.
These hardware features draw a solid boundary to isolate the kernel of operating system with normal applications.
However,
    for Super Apps and MiniApps,
    despite  MiniApps  running in parallel process with Super Apps,
    they are still in the same layer,
    even the same user with the same privilege to their host execution platform,
    which is impossible to have a hardware-enhanced isolation protection.
Thus,
    the only way that Super Apps can do is to guard the MiniApp  by software gates,
    which is vulnerable and insufficient to isolate MiniApp  with themselves.
Whenever the gate is compromised,
    MiniApp  could be able to cross the boundary,
    bypass the limitation and permission,
    abuse the hidden power which is giving by Super Apps themselves,
    entirely compromise the Super Apps,
    and finally exploit the operating system.

\noindent
\textbf{The Implementation Flaw inside Super Apps}
As we introduced above,
    Super Apps need to implement software gates to draw the boundary of MiniApp  with themselves.
However,
    after our investigation,
    almost all of the Super Apps have implementation flaws.
Despite the boundary they have built,
    there is still insufficient isolation applying on  MiniApps .

To form the implementation flaw,
    they failed to restrict  MiniApps  accessing specific APIs which is provided by Super Apps.
For example of Android Super Apps,
    in \wechat,
     MiniApps  of unprivileged developers are prohibited to access \code{wx.openUrl} API,
    which can access arbitrary \code{WebView} of Android system
    \footnote{The arbitrary access of \code{WebView} is powerful but dangerous,
        this permission can letting developers access any scheme of URL,
        and initiate local even remote attack on the host device.}.
However,
    a implementation flaw exist in \wechat MiniApp  Runtime has provided the chances letting MiniApp  access arbitrary APIs,
    including accessing \code{wx.openUrl} with arbitrary parameters.
Thus,
    the root cause of this API access bypassing is the pitfalls inside their MiniApp  isolation implementation,
    which is inevitable as the permission of Super Apps is insufficient to draw a solid boundary.

\noindent
\textbf{The Impact of Vulnerabilities}
To address the impact,
    we first evaluated the capability a malicious attacker could obtain.
In our thread model,
    a malicious attacker can leverage implementation flaws of Runtime of  MiniApps,
    and obtain permission to access privileged interfaces or APIs.
Within the privileged interfaces,
    there are multiple categories of APIs can be exploit to gain a higher capability,
    such as writing file to a arbitrary path,
    accessing \code{WebView} with arbitrary URI,
    even accessing private data of Super Apps via SQL injection.
Though this procedure is transparent,
    however,
    malicious attackers can leverage dynamic features of MiniApp  to escape vetting of Super Apps.

Not only how critical this vulnerability is,
    the fast redistribution of MiniApp  is also pushing impact more serious.
Mini Program is designed as lite weighted,
    fast and instant at the very beginning of its birth.
Normally,
    users can simply scan a QR code or click a sharing link to open a MiniApp .
This simple way of redistribution could help attackers a lot to spread their malicious  MiniApps ,
    increasing the impact of these vulnerabilities.

\noindent
\textbf{Our Finds}
We first manually uncovered a vulnerability in \wechat MiniApp  Runtime,
    after further investigation,
    we developed our analysis tool set,
    and discovered multiple vulnerabilities existed in various implementation of Super Apps.
These vulnerabilities also has various security impact,
    from user private information leakage to exploit operating system.

\CW{here should have a brief vulnerabilities evaluation,
but the data is still under review.}

\noindent
\textbf{Our Contributions}
In this paper,
    we made the following contribution to the community.

\begin{itemize}
    \setlength\itemsep{0.2em}
    \item We conducted the first security analysis against Runtime of MiniApp  implemented by Super Apps.
    \item We developed a set of analysis tools to systematically uncover the vulnerabilities of Super Apps.
    \item We disclosed the vulnerabilities to Super App vendors,
            and received positives responses.
\end{itemize}

}

\section{Background} 
\label{sec:back}

Miniapps are programs that run on top of host apps instead of directly on the operating system. Host apps have to function like an operating system and provide resources (e.g., location, phone numbers, addresses, and social network information) to miniapps through APIs. Mobile super apps are organized in a layered architecture, with each layer focusing on different aspects like portability, security, and convenience, but working together to support miniapp execution within host apps, as shown in \autoref{fig:miniprogram-arch}:

\begin{packeditemize}
\item \textbf{Mini-Application Layer}, which is the top layer of a super-app runtime. All miniapps, including 1st-party and 3rd-party miniapps, are located in this layer. To prevent one miniapp from accessing resources of other miniapps, the host app creates an isolated process for each miniapp. If privileged access is given to 1st-party miniapps, it must be controlled and checked to prevent 3rd-party miniapps from using them. Typically, miniapps are implemented using JavaScript~\cite{whitepaper20}. \looseness=-1     
\begin{figure}[t]
  
        \centering
        \includegraphics[width=0.495\textwidth]{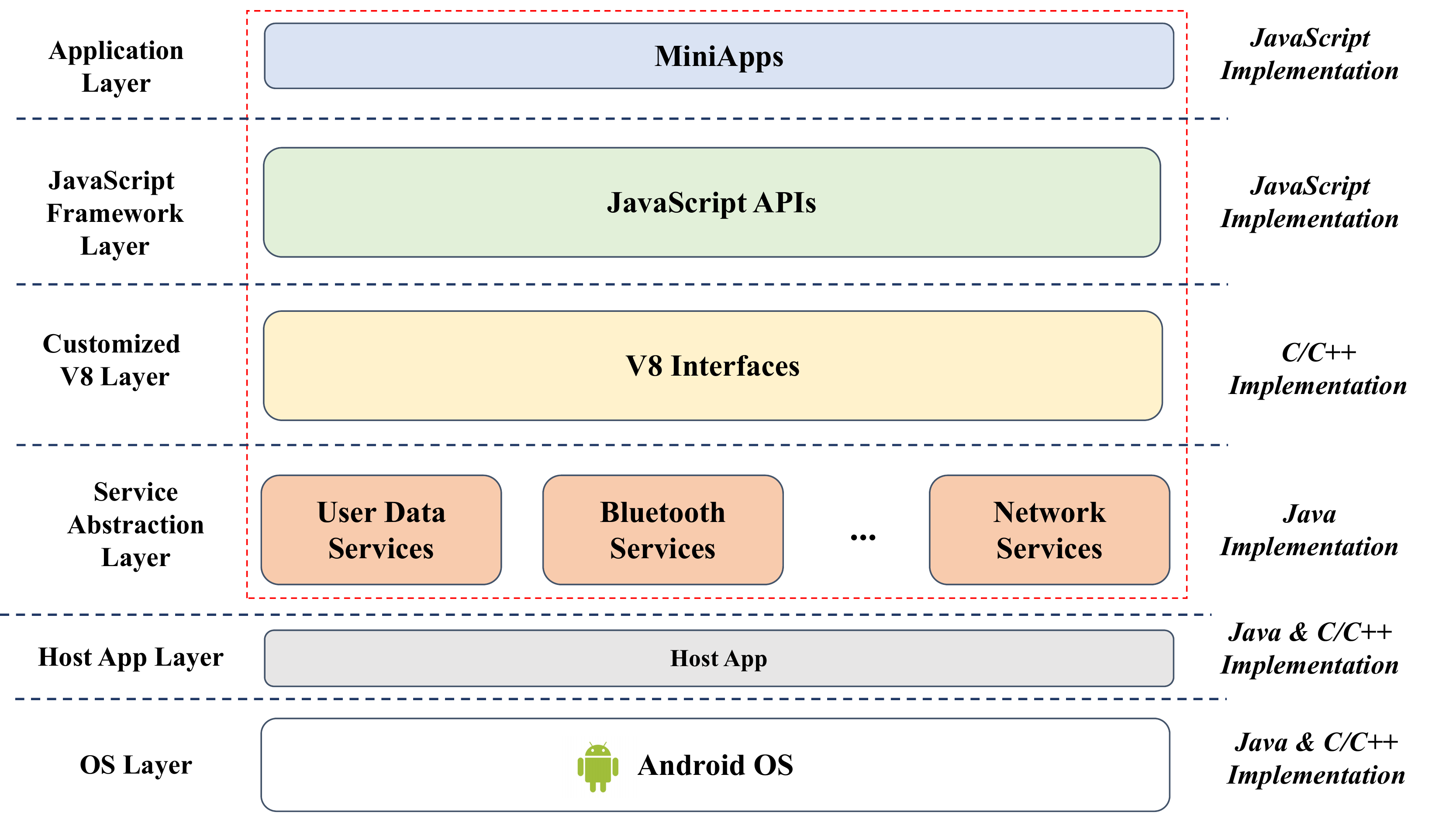} 
   \vspace{-0.2in}
        \caption{Architecture of Super App Runtime in Android}
        \label{fig:miniprogram-arch}
           \vspace{-0.2in}
\end{figure}

\item \textbf{JavaScript Framework Layer} provides APIs for resource accesses and management, which are consumed by miniapps in the Application Layer. These APIs allow miniapps to access resources (such as location-based services) and manage UI elements (such as opening a new UI window). The JavaScript Framework Layer is also implemented using JavaScript.\looseness=-1 
\item \textbf{Customized V8 Layer}, which provides support for native C/C++ libraries such as \texttt{WebGL} to power the execution of miniapps. It also acts as a bridge between the JavaScript Framework layer and lower-layers. When miniapps call APIs such as \texttt{wx.getLocation}, the Framework layer sends the API name and parameters to the Customized V8 layer, which then passes the request to the underlying layers. This layer is usually implemented using C/C++.\looseness=-1

\item \textbf{Service Abstraction Layer}, which provides an interface to access services from either the super apps (e.g., user account information) or the underlying OS (e.g., Bluetooth, location-based services). In the case of the \texttt{wx.getLocation} API, this layer communicates with the host app using IPC to invoke the Java API \texttt{getSystemService(LOCATION\_SERVICE)} to retrieve the current location. This layer is implemented using a combination of Java and C/C++ code for the Android platform.



\end{packeditemize}


\ignore{

\begin{figure}[t]
        {
        \centering
        \includegraphics[width=0.5\textwidth]{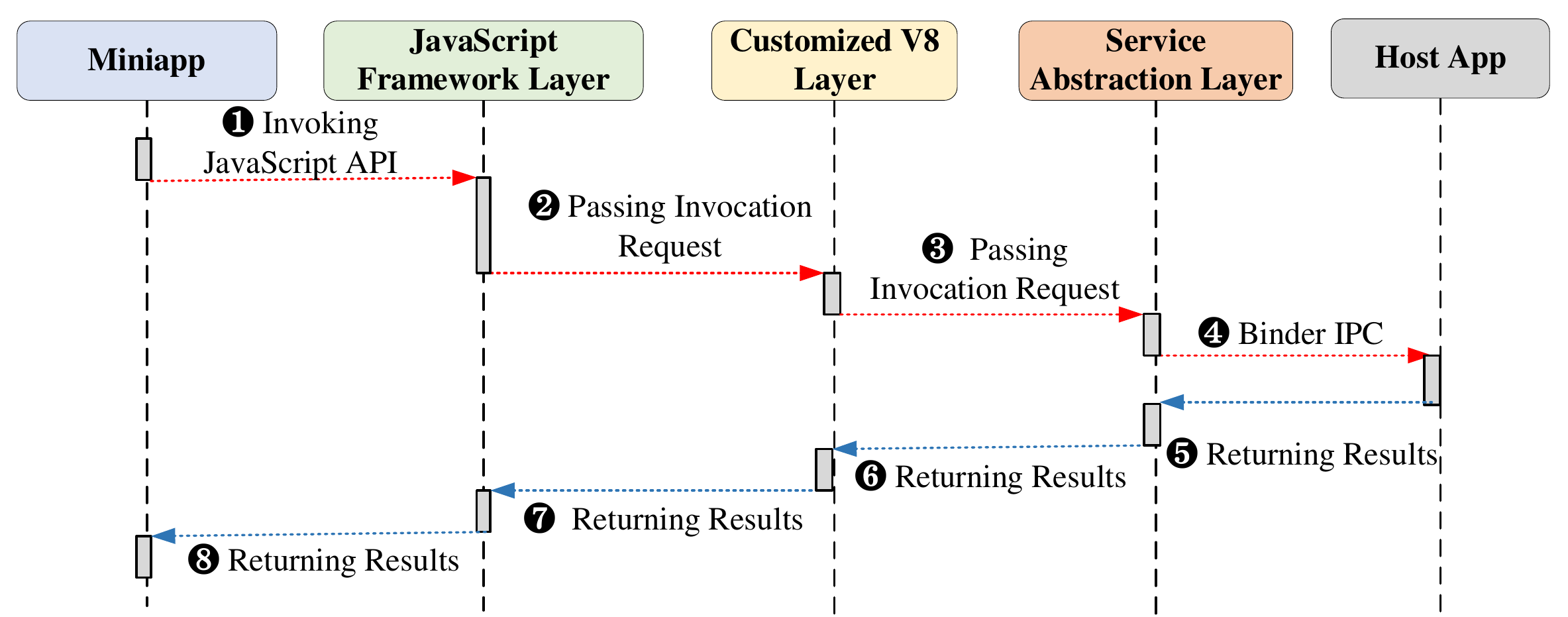}}

        \caption{ Typical Workflow of Miniapp API Execution}
        \label{fig:miniapp-api-invoke}
\end{figure}

\subsection{The Workflow of Miniapp API Execution}
\label{subsec:APIworkflow}

 Having explained the architecture of miniapp, we can now clearly notice that except the application layer, which only invokes the APIs provided by JavaScript Framework layer, all other layers   expose interfaces to their upper layers, allowing the miniapps running in the application layer to communicate with the super app and accomplish sophisticated functionalities. As shown in \autoref{fig:miniapp-api-invoke}, a miniapp needs to go through the following two stages with nine steps to complete the invocation of the interfaces. 
 
\begin{itemize}
\item \textbf{Invocation Request.} Assume that a mini-app attempts to get the current location of the device. To that end, the miniapp first invokes JavaScript API, e.g.,  \texttt{getLocation(object)} (through the parameter \texttt{object}, the miniapp can specify the format of coordinates and accuracy of the location data), which is exposed by JavaScript Framework Layer for location services (\textbf{Step} \ding{182}). JavaScript Framework Layer obtains the request,  and extracts the API name and the parameters, then passes the API name and the parameter through Interfaces  offered by the Customized V8 Layer (\textbf{Step} \ding{183}). Customized V8 Layer fetches the underlying API (i.e., \texttt{JsApigetLocation}) based on the passed API name, and feeds the fetched API the corresponding parameters extracted from the passed \texttt{object} (\textbf{Step} \ding{184}).  Next,  Service Abstraction Layer is notified and uses Binder IPC to communicate \wechat. Finally, \wechat handles the request, and obtains the location data through invocation of the Android system API (\textbf{Step} \ding{185}).  \looseness=-1
\item \textbf{Invocation Response.} When \wechat obtains the location data from Android OS, it needs to return the results to the miniapps. To that end, it first uses Binder IPC to communicate with the Service Abstraction Layer (\textbf{Step} \ding{186}), and the Service Abstraction Layer will pass the result t Customized V8 Layer (\textbf{Step} \ding{187}).Customized V8 Layer further encapsulates the results, and pass the results to JavaScript Framework Layer (\textbf{Step} \ding{188}), which allows the upper layer miniapps to call the corresponding callbacks to consume the results (\textbf{Step} \ding{189}). \looseness=-1
\end{itemize}
}

\noindent

\section{Motivation and Problem Statement} 
\label{sec:motivation}


%

This section describes the motivation of this work by providing some key observations  in \S\ref{sub:observation},  then define the problem, the scope and the threat model in \S\ref{sub:problem:statement}. 

\subsection{Key Observations}
\label{sub:observation}

\begin{figure} [t]
\centering






\includegraphics[width=0.495\textwidth]{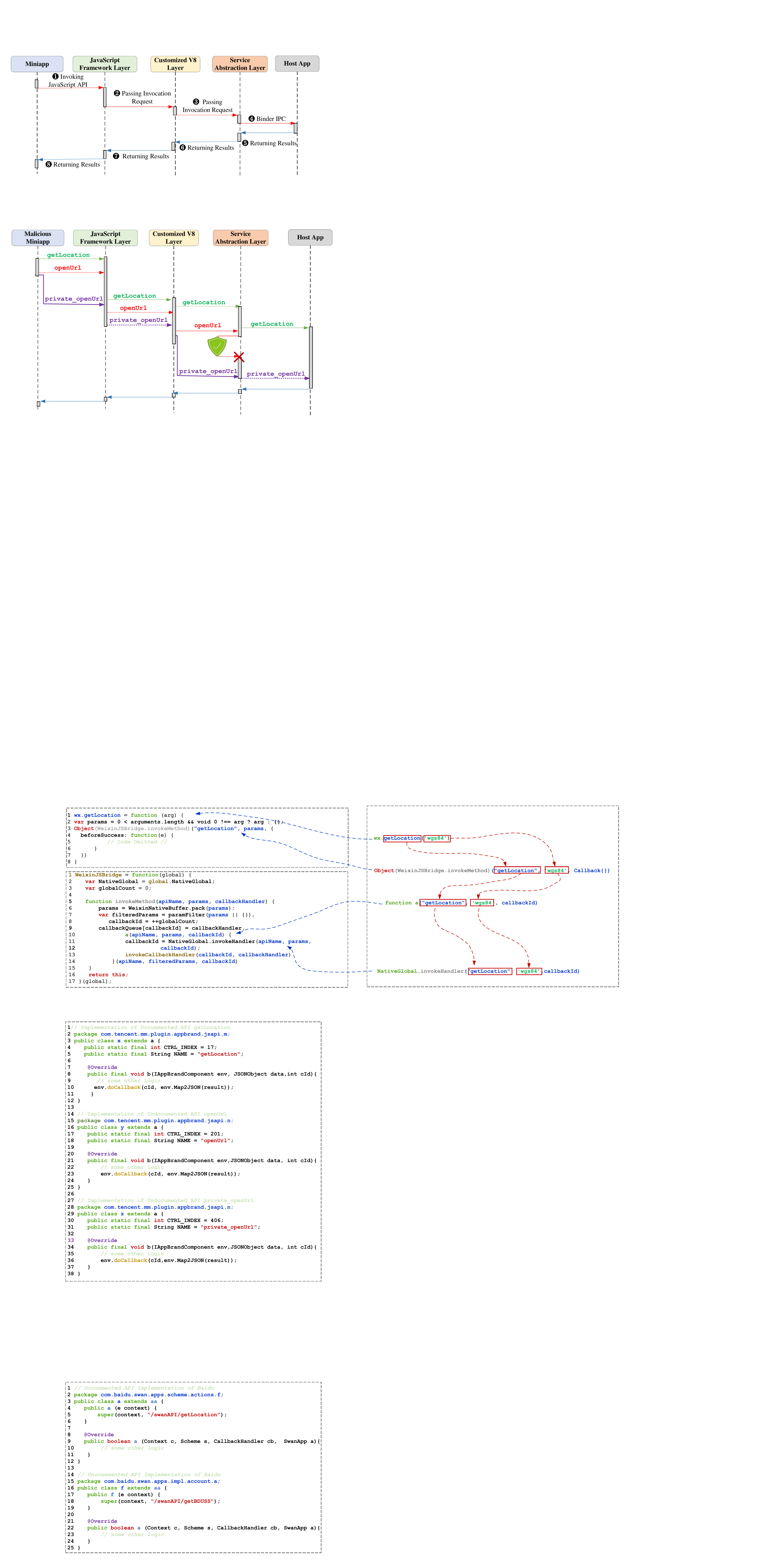}
 \vspace{-0.25in}
 \caption{APIs implementations of \wechat. }
 \label{lst:jsapiclass}
\vspace{-0.1in}
\end{figure}

\begin{figure*} 
\centering
    
 
\includegraphics[width=0.985\textwidth]{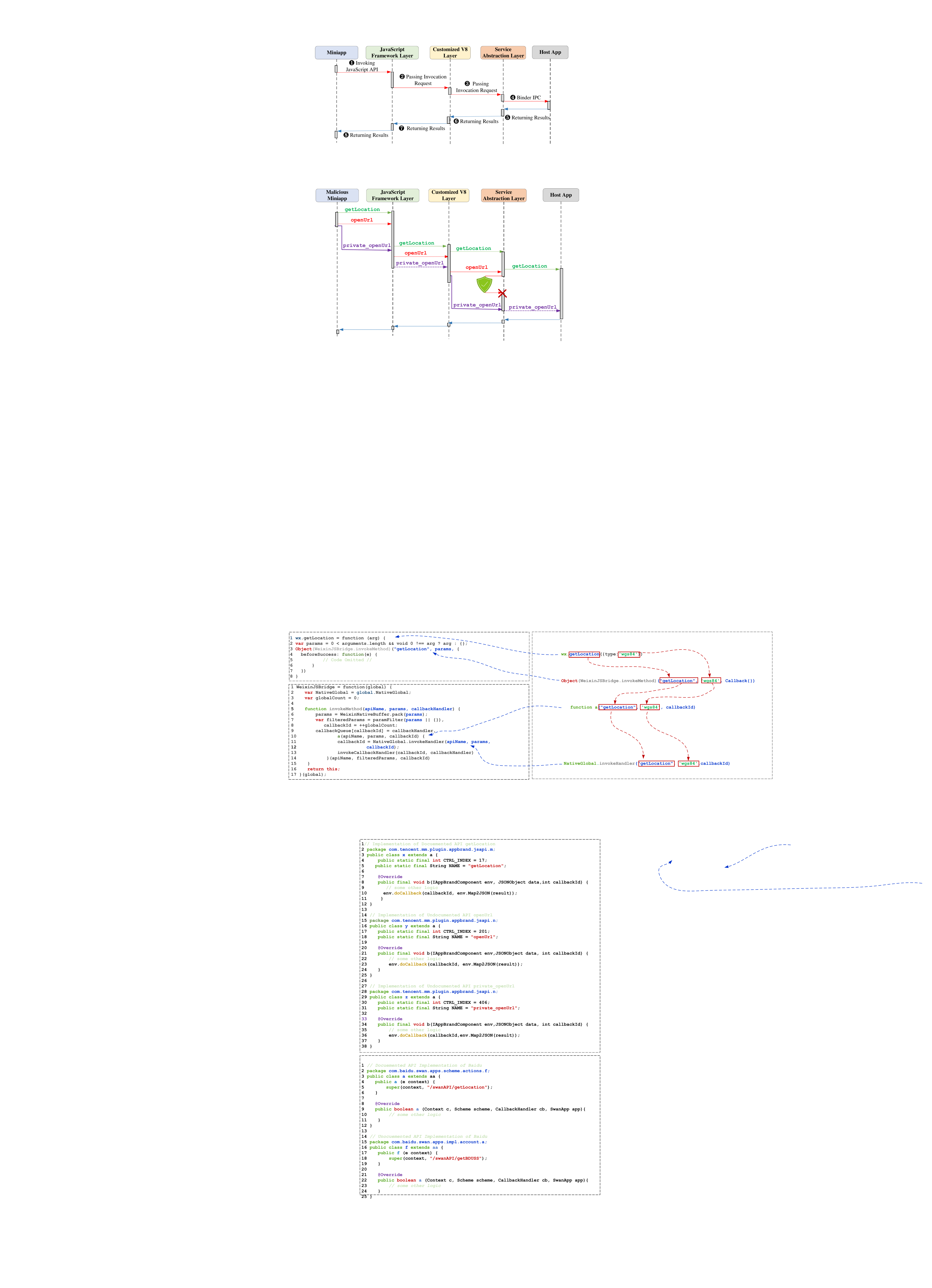}
\vspace{-0.15in}
\caption{An Example of \wechat API Invocation At JavaScript Framework Layer.  }
 \label{lst:jsapibridge}
 
\end{figure*} 

As alluded earlier, when manually inspecting the implementation of some of the 1st-party miniapps offered by \wechat, we found that other than the public APIs that all the miniapps can access without restrictions, the 1st-party miniapp \textsf{Tencent Doc} actually uses some  undocumented APIs (e.g., \texttt{openUrl} for opening arbitrary URLs). Moreover, the designers of \wechat do not make the APIs available to be public (their documentation does not even mention \texttt{openUrl}),  and have placed security checks to prevent \texttt{openUrl} from being accessed by arbitrary miniapps. For example, whenever a 3rd-party miniapp attempts to invoke \texttt{openUrl}, \wechat will throw an insufficient permission exception (i.e., ``{\tt fail: no permission}'')  and terminate its execution. The use of \texttt{openUrl} in the 1st-party \textsf{Tencent Doc} miniapp prompted us to investigate the possibility of other hidden APIs offered by \wechat without proper security checks. This inspired us to explore the feasibility of identifying and exploiting these APIs, but we faced two challenges: (i) identifying the hidden APIs and (ii) properly invoking them to test for potential vulnerabilities. Through further exploration, we made two key observations to address these challenges.\looseness=-1


 
\paragraph{Observation-I: Undocumented API Recognition} By manually inspecting the implementation of \wechat, we found that multiple suspicious undocumented functions are co-located with their documented APIs. That is, those functions and the public APIs are located in the same super app packages, and their implementations look similar to that of the documented APIs (e.g., they have similar function signature, similar parameter type and return value type). We start by inferring whether those functions are indeed undocumented APIs, since intuitively the public APIs and undocumented APIs are APIs, and the developers would have followed the same practice to implement them. 
Without surprise, we found the implementation of \textsf{openUrl}, which confirms our observation. In \autoref{lst:jsapiclass}, {{we show 3 API implementations of \wechat. Although the code is highly obfuscated (where the names of the classes and methods are replaced with meaningless letters, such as ``{\tt a}'',``{\tt b}''), we still can observe some invariants:}} \wechat's public API \texttt{getLocation} (line 1--13) and its undocumented  API \texttt{openUrl} (line 14--25) both have the same parameter types and return types, as well as the same superclass  (i.e., class \texttt{b}). 
As such, we can use these invariants (e.g., the superclass of the API, the parameters of the API) collected from the public APIs to search for possible undocumented APIs. For instance, as shown in \autoref{lst:jsapiclass}, we identified another function \texttt{private\_openUrl} (lines 28--38) that has the same function signature, which is very likely an undocumented API. \looseness=-1

\paragraph{Observation-II: Undocumented API Invocation} Although there may be undocumented APIs (e.g., \texttt{private\_openUrl}) provided by \wechat, we have to find a way to invoke them (if they are indeed APIs). Interestingly, when we directly invoke undocumented APIs such as \texttt{private\_openUrl} in a miniapp, we obtain an error, ``\texttt{fail: not supported}'', which is different from the error we observed when invoking \texttt{openUrl} with ``\texttt{fail: no permission}''. As such, we infer that the accessibility of the API \texttt{private\_openUrl} is not the same as that of \texttt{openUrl} (since the observed error messages are different), and there may be a way to invoke it. As such, we further inspected the normal invocation of the documented APIs,  and seek to obtain insights from the process. 
\looseness=-1

\begin{figure}[t]
        \centering
        \includegraphics[width=0.45\textwidth]{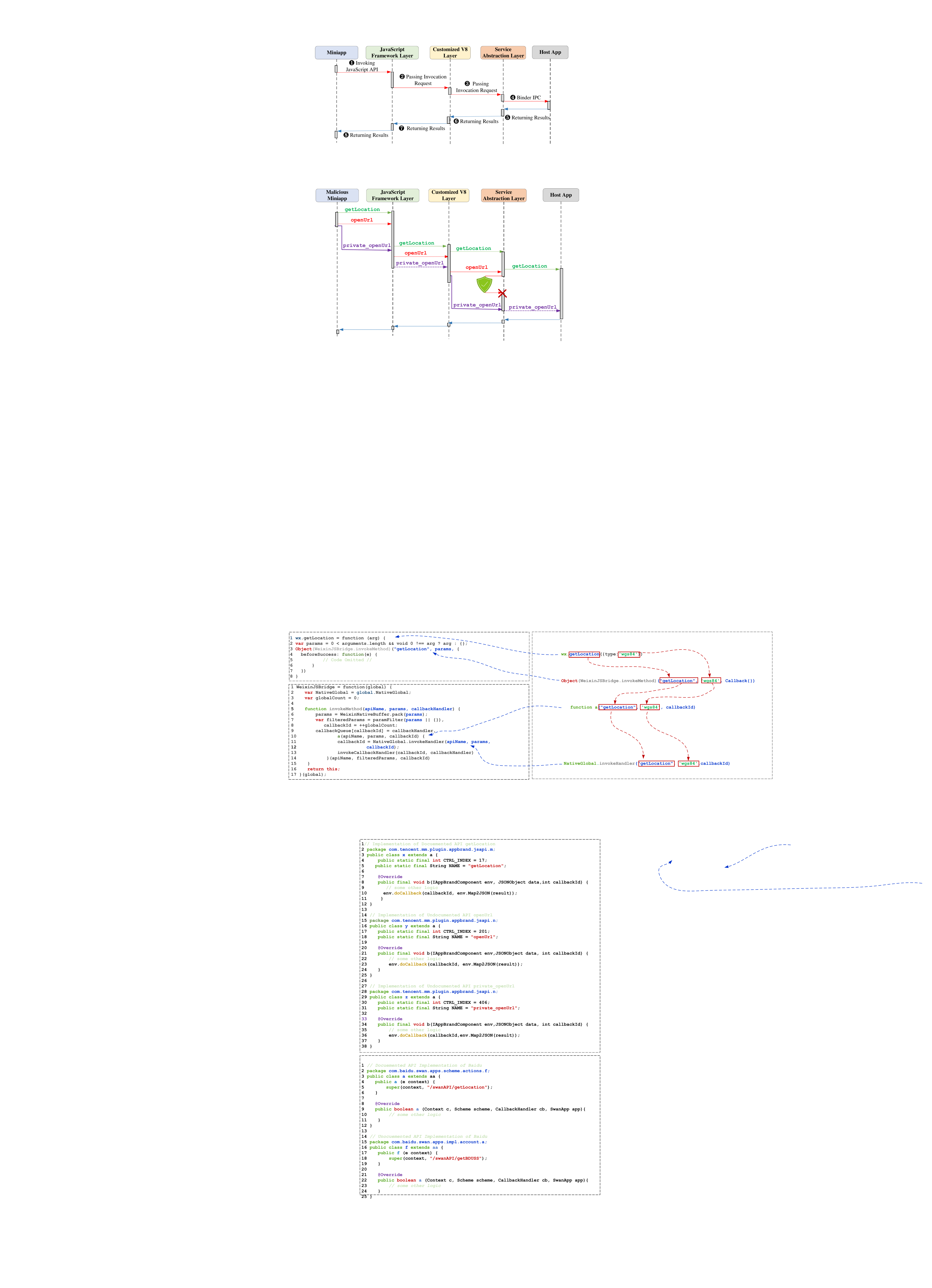} 
          \vspace{-0.1in}
        \caption{The Workflow of API invocations. Public API invocation \texttt{getLocation} (\textcolor{green}{green} line);  Checked Undocumented API \texttt{openUrl} (\textcolor{red}{red} line);  Unchecked Undocumented API \texttt{private\_openUrl} (\textcolor{violet}{purple} line).   }
        \label{fig:vul}
        \vspace{-0.2in}
\end{figure}

To be more precise, as described in \S\ref{sec:back}, the JavaScript Framework Layer acquires the invocation request during a regular API call and transfers it to the lower layers via the interfaces exposed by the Customized V8 Layer. In  \autoref{lst:jsapibridge}, we provide a code snippet illustrating the API invocation chain of \wechat, where the invocation request for the \texttt{getLocation} API (line 3 in the top-left frame) is eventually passed to the \code{NativeGlobal.invokeHandler} function (line 11 in the bottom-left frame), which in turn conveys the API invocation request to the underlying layers. Notably, the \code{NativeGlobal.invokeHandler} function receives three inputs: the API name (e.g., \texttt{getLocation}), the API parameters, and a callback function ID (which enables the API to manage the asynchronous call).\looseness=-1


Given that \code{NativeGlobal.invokeHandler} can deliver the normal invocation request to the underlying layers, we conclude that it also has the capabilities to deliver undocumented API invocation requests. Therefore, we feed the API name \texttt{private\_openUrl} and its parameter (which is a URL) to the interface and let it pass the API name and the URL to the underlying layers. Interestingly, we find that the underlying layers handle the passed API name and the parameter as normal API invocations and further pass the invocation requests to the host apps. As shown in \autoref{fig:vul}, while \wechat restricts the undocumented APIs to be accessed by mini-apps, unfortunately we find that not all undocumented APIs are protected through security checks. In particular, \wechat has enforced the security check for the undocumented API \texttt{openUrl}, but it does not add the security checks for the undocumented API \texttt{private\_openUrl}, which has the exact same functionalities as \texttt{openUrl}. Also, the API name and parameters are not obfuscated since they have to be passed to lower layers.  \looseness=-1

\subsection{Problem Statement and Scope}
\label{sub:problem:statement}

Since our manual investigation has revealed that there are indeed hidden APIs in the super app platform and some of them can be exploited, the goal of this work is to develop techniques to uncover them. More specifically, we need to recognize the hidden APIs based on how documented APIs are implemented and executed, and meanwhile test them to determine whether they can be invoked by 3rd-party miniapps to bypass security restrictions (or those APIs themselves may have vulnerabilities). Please note that we do not consider all those 3rd-party invocable APIs as exploitable, since whether an API is exploitable depends on the functionalities of the APIs (e.g., the API implements privileged operations).  

Also, since there are multiple super apps available today, ideally, we would like to develop generic techniques to cover them all. However, our observation is heavily based on the miniapp run-time architecture presented in~\autoref{fig:miniprogram-arch}. Therefore, the super apps that do not follow this architecture, e.g., do not use V8 engine to execute their miniapp code, will be out of our scope.
Finally, because of the convenience and also our expertise, we focus on the super apps running on Android platform, though in theory our approach should also work for the iOS platform.

\subsection{Threat Model}

As previously discussed, our objective is to develop techniques for detecting hidden APIs that lack security checks before a malicious app exploits them. In this context, the attacker is a malware that has been installed on the user's mobile device. We will not delve into the details of how this malware can be installed, as we believe it is practical to assume that super apps are not aware of such types of malware until we report our findings to them. It is worth noting that previous research on super apps has also made similar assumptions~\cite{lu2020demystifying}. 
Undocumented APIs refer to functions or APIs that are not included in the official documentation, regardless of whether it is in English or Chinese. An attacker could acquire knowledge about the existence of these hidden APIs by reverse engineering the super app client or by reading technical blogs on the internet. Specifically, undocumented APIs may have access to sensitive resources that are safeguarded by Android OS. If an attacker exploits these APIs, they can launch attacks against the victim users. \looseness=-1

\ignore{
In this work, we seek to understand the undocumented API exploitation, its consequences, automatic detection approaches, and  possible countermeasures. 
While the problem is a general problem regardless of the OS platforms (since the official miniapps that invoke the undocumented APIs can run on multiple OSes), we particularly focus on Android, and use the super apps running on Android to demonstrate the principles. Moreover, we require that the super apps use V8 engine to execute their JavaScript code, and for the super apps that do not, our approach may need extra efforts to work.      \looseness=-1  
}

\section{Challenges and Insights} 
\label{subsec:challenges}

\paragraph{(I) Challenges in API Recognition} The first step of our \sysname is to identify undocumented APIs when given a host app.  Intuitively, it sounds trivial, since when given an API, we could compare it with the APIs released on the official documentation to decide whether it is documented or not.  However, it is challenging to determine whether an internal function or an interface is an API.  For instance,  there are 3,702 functions and interfaces implemented in JavaScript, not to mention those implemented in 92 native C/C++ libraries, and 56,492 Java classes in \wechat's latest version. Note that we do not have to consider the functions at lower-layer's implementations (i.e., any layer below the JavaScript framework), since the hidden APIs are not exposed at these layers. Obviously, we cannot directly treat all these functions as APIs.\looseness=-1 

\begin{figure*}[t]
        \centering
       \includegraphics[width=1.02\textwidth]{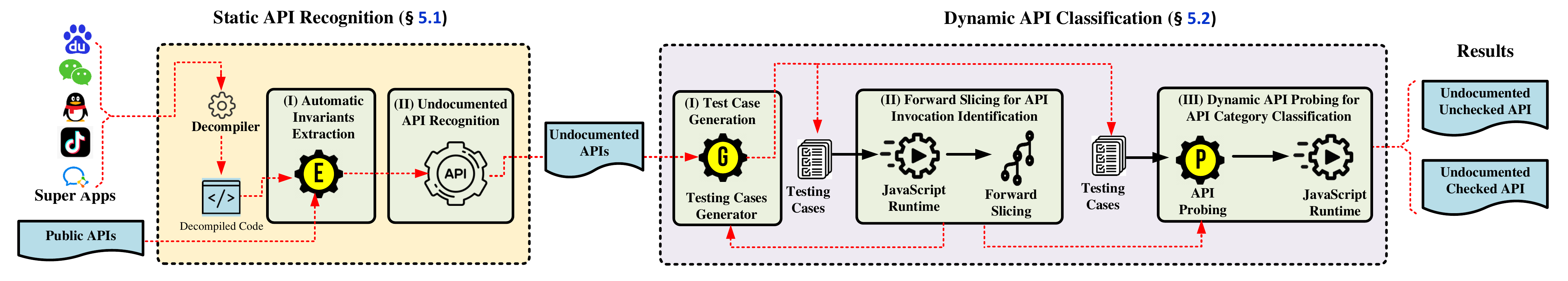}
 \vspace{-0.25in}
        \caption{\sysname Architecture  }
        \label{fig:tool}
         
\end{figure*}

Also, although for a specific implementation of host apps (e.g., \wechat), simple pattern matching approaches can be applied to recognize APIs.  For example, when implementing the callbacks of the APIs, \wechat uses \texttt{android.webkit.ValueCallback} at the Service Abstraction layer to handle all the callback results. From the callbacks, we can locate the corresponding APIs 
and extract patterns to pinpoint the rest APIs.
However, there are multiple super apps, each of which could have different implementations. 
For example, unlike the implementation of \wechat, \textsf{TikTok} uses \texttt{com.he.jsbinding.JsContext.ScopeCallback} at the Service Abstraction layer to handle the callback results of their APIs, and the pattern for \wechat will fail when dealing with \textsf{TikTok}. 
Moreover, such a pattern-matching approach requires recognizing callbacks first, which may be challenging due to the code obfuscation. As discussed in \S\ref{sub:observation}, the miniapp is executed on top of the super apps (e.g., Android apps), which is often heavily obfuscated. It is hard to recognize callbacks statically unless we fully understand the obfuscated code, and as such, we need a more obfuscation-resilient approach instead of simple pattern matching.   \looseness=-1

\paragraph{Insights} We notice that there exist some invariants such as the method signatures of public APIs and their superclasses in the API implementations, as illustrated in \S\ref{sub:observation} based on super app \wechat (e.g., every API has the same superclass {\tt a}, though this name is obfuscated; every public API must contain the name of the API for the references by the miniapps, and this cannot be obfuscated {  but} can be easily recognized). As such, we can first extract these API invariants based on these public API implementations, 
from which to recognize the rest of the APIs. This process can be automated since it is easy to identify these API invariants when the implementation of public APIs is provided. \looseness=-1 

\paragraph{(II) Challenges in API Classification} 
Once we have identified all these hidden APIs, we still need to further classify them into different categories and determine whether they are invocable (when there is no security check). It will be very challenging if we only use static analysis to decide this, and thus we need to rely on dynamic analysis to dynamically invoke them.  However, to invoke a hidden API, we still need to recognize the interface that can communicate with the underlying layers. 
Although we have already known that the interface communicates with the underlying layers takes the API name as its inputs (as described in \S\ref{sub:observation}), it is still challenging to know whether this interface accepts the API name as its input before we actually execute it (due to the obfuscated JavaScript code).  
Meanwhile, although multiple dynamic tools are available for JavaScript, they cannot be applied to our case directly due to the highly customized JavaScript framework implementations. For example,  most JavaScript analysis tools (e.g., Jalangi2~\cite{sen2013jalangi}) are designed for traditional web browsers. They cannot run with the super apps since the offered APIs are different. Moreover, most of these tools need to instrument the testing instances, which involves the modification of the testing instances. In our case, the testing instances are the miniapps (not web applications), which usually have integrity checks and cannot be modified easily. \looseness=-1

\paragraph{Insights} To invoke the API for its behavior classification, we need to find the interface, e.g., \code{NativeGlobal.invokeHandler} as shown in~\autoref{lst:jsapibridge}. Interestingly, to identify this interface, we can monitor how a public API is executed, e.g., how it is invoked (its name, parameters), and when it is passed between the boundary of the layers. More specifically, we notice that we can use function trace analysis to identify interfaces such as \code{NativeGlobal.invokeHandler}, since the API execution starts from the invocation, and ends at the interface boundary. By tracing all of the function executions with their parameters and then identifying them based on the use of the API name, which is passed as parameters, we can automatically identify the interface, which is typically the last invocation point in the JavaScript layer. 
With the identified invocation point, we can then feed it with different API names and invoke them to classify further (e.g., whether they can be invoked by the 3rd-party miniapps). \looseness=-1

\section{\sysname}
\label{subsec:tooloverview}

As shown in \autoref{fig:tool}, our developed \sysname consists of two phases of analysis---static analysis first and then dynamic analysis, with the following two key components: 
\ignore{
Based on the challenges above, we can notice that to identify and exploit the undocumented API vulnerability, we must first identify all the APIs, and then classify the APIs into different categories using the interfaces that are used to invoke APIs.  As such,  as shown in \autoref{fig:tool}, we propose the architecture of our \sysname,
which also contains two major components to serve the two purposes:
}
\looseness=-1
\begin{packeditemize}
\item \textbf{Static API Recognition (\S\ref{subsce:apifinder}).} This component takes the binary code of super apps (i.e., APKs) and the list of the official APIs in the documentation as input, and produces the undocumented APIs as output. At a high level, it first decompiles the APKs by {\sf Soot}~\cite{SootSoot43:online}, automatically extracts the invariants based on the public APIs, and then uses the invariants to recognize the hidden APIs from the implementations of  super apps. \looseness=-1 

\item \textbf{Dynamic API Classification (\S\ref{subsce:apiexploiter}). } This component takes the hidden APIs as input, and classifies them into three different categories: unchecked hidden APIs (exploitable by 3rd miniapps), 
checked APIs (available to only 1st-party miniapps), and non-APIs, as the final output. At a high level, it first uses the Test Case Generator to produce two types of test cases: one is for API invocation identification executed by a lightweight tracing engine for the monitored execution, and the other is for API classification. With these test cases, \sysname eventually identifies the interfaces as well as the categories of the APIs.


\end{packeditemize}




\subsection{Static API Recognition}
\label{subsce:apifinder}

To recognize APIs, \sysname first needs to extract the invariants based on the decompiled code of public APIs. With the invariants, it then recognizes the hidden APIs. Therefore, it is a two-step process. In the following, we describe these two steps in greater details.

\begin{figure} 
\centering
    

    
 
\includegraphics[width=0.495\textwidth]{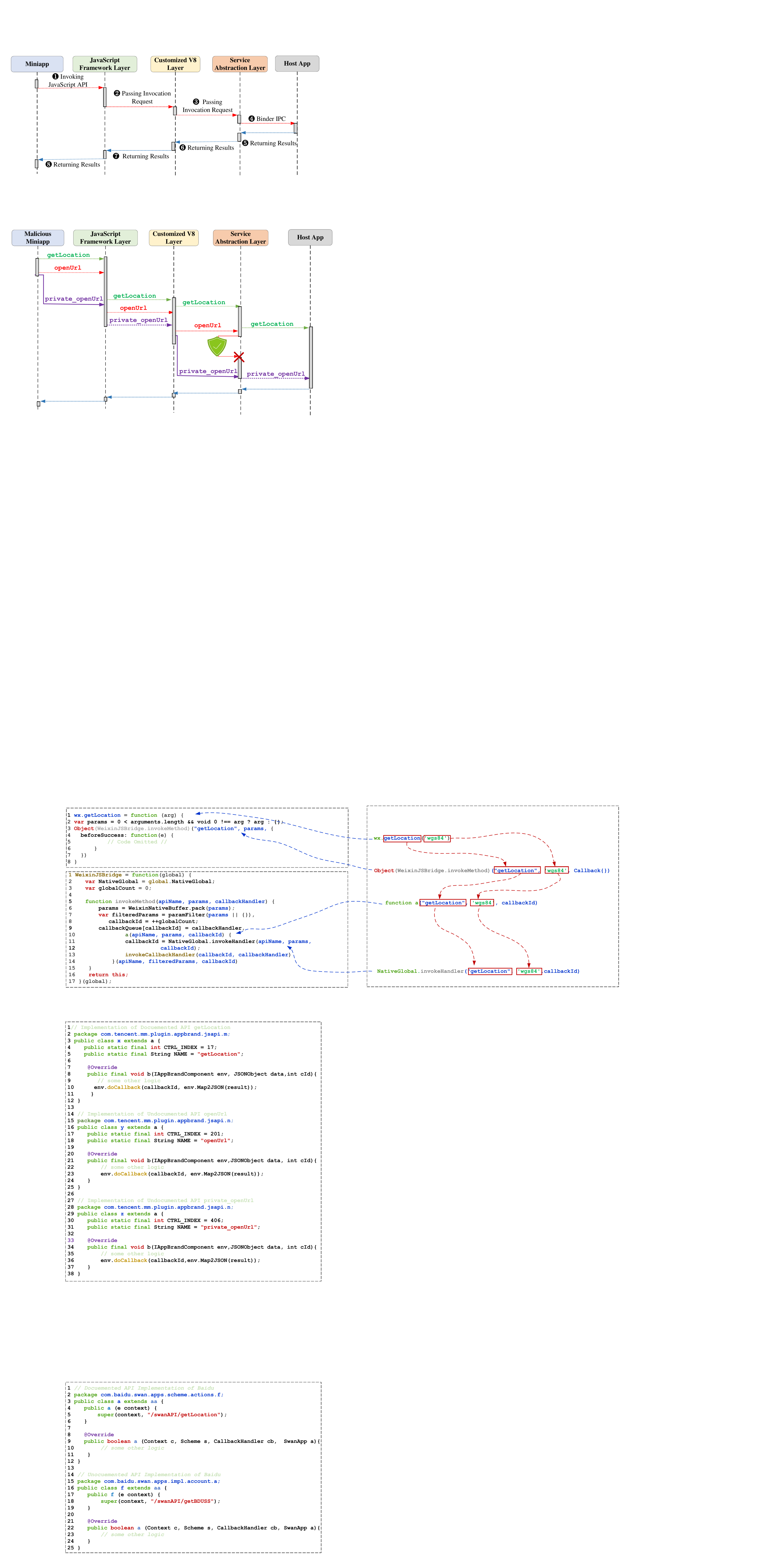}
   \vspace{-0.3in}
 \caption{APIs implementations of \textsf{Baidu}. Note that lines 1 -- 12 contain a documented API, and lines 14 -- 25 contain an undocumented API.}
 \label{lst:bdapiclass}
   \vspace{-0.2in}
\end{figure}

\paragraph{Step-I: Automatic Invariants Extraction}
\sysname first needs to extract the invariants based on the decompiled code of the public APIs. 
from the implementations of the super apps.  In particular, when given an API, \sysname will aggressively identify as many invariants as possible from the implementation, and these invariants include: (i) the method signatures (e.g., the return type, the number of the parameters, and parameter types); (ii) the superclass; (iii) the super packages (e.g., in super app \textsf{Baidu} \texttt{com.baidu.swan.apps} is the super package of 
\texttt{com.baidu.swan.apps.scheme.actions.f} as shown in  \autoref{lst:bdapiclass}), and (iv) their callers.  
Again, they are invariants because they will not be changed in the API implementation (both public and undocumented) for a specific super app, though the specific content for the invariant may be changed across super apps.  For instance, 
in the superclass invariant of APIs, in \wechat, when comparing any two implementations of the provided APIs (e.g., \texttt{getLocation} and \texttt{private\_openUrl}), we can easily recognize that they are both extended from the superclass \texttt{a}, as shown in \autoref{lst:jsapiclass}; similarly, the superclass of APIs provided by \textsf{Baidu} is extended from the same superclass \texttt{aa}, as shown in \autoref{lst:bdapiclass}.

\paragraph{Step-II: Undocumented API Recognition}
With the invariants, \sysname  then recognizes the undocumented APIs.   In particular, it iterates each of the function implementations again, by matching the invariants extracted; if it matches with all the invariants as in the public APIs and it has not been added in the undocumented set yet, this function's implementation is an undocumented API. That is, we have used quite restrictive patterns that need to exist in all public API implementations for a particular super app, and a function must contain all of these invariants in order to be considered an undocumented API. Regarding how exactly \sysname identifies them, we present a detailed algorithm in Appendix \S\ref{appendixa} for the readers of interest.

\ignore{
If it fails  out by our for a specific public API, there could be multiple implementations that can be identified, and  only one implementation is the actual implementation. As such, when we attempt to extract the API invariant, the tool may be subject to a false positive, since we may extract the API invariant from a non-APIs. Consequently, if we use those API invariants to search the undocumented API, we may collect some functions that are not the APIs. To reduce the false positive, we only consider a pattern is the API invariant, when the pattern appears in all the implementation of its public APIs. For example, as shown in \autoref{lst:bdapiclass}, 
all implementations of public APIs provided by \textsf{Baidu}  are co-located in the super package \texttt{com.baidu.swan.apps}, and the super package can be considered as the API invariant, but \texttt{com.baidu.swan.apps.impl} cannot be considered as the API invariant, since not all implementations are co-located in this super package. 
}

\ignore{
In theory, the invariant we used may allow us to identify some internal functions that are not APIs. That is, it might have false positives. However, this false positive API can actually be identified by our {\it Dynamic API classification} component, which will generate a test case to invoke the identified API. 

If it fails  out by our for a specific public API, there could be multiple implementations that can be identified, and  only one implementation is the actual implementation. As such, when we attempt to extract the API invariant, the tool may be subject to a false positive, since we may extract the API invariant from a non-APIs. Consequently, if we use those API invariants to search the undocumented API, we may collect some functions that are not the APIs. To reduce the false positive, we only consider a pattern is the API invariant, when the pattern appears in all the implementation of its public APIs. For example, as shown in \autoref{lst:bdapiclass}, 
all implementations of public APIs provided by \textsf{Baidu}  are co-located in the super package \texttt{com.baidu.swan.apps}, and the super package can be considered as the API invariant, but \texttt{com.baidu.swan.apps.impl} cannot be considered as the API invariant, since not all implementations are co-located in this super package.  
}
 



\ignore{
As shown in \autoref{alg:a1}, we now explain how we extract all the four API invariants when given a super app. We first use the names of the public APIs as the keywords and search the reverse-engineered code of the super apps to find out the possible implementations of these public APIs, which form the candidate API set $PAPI$ (line 1 -- line 6). For each candidate API, we get its function signature (e.g., the parameter types, return values), and compare the function signature with function signatures of other API candidates. If we find the all function signatures are the same in different implementations, we consider the function signature as one of the API invariants that can be used to match the APIs (line 11 -- 18).   We  use similar approaches to collect other API invariants, including their superclasses (line  19-- line 26), their super packages (line  28 -- line 35), and their callers (line 37 -- line 44). As shown in \autoref{tab:platfroms}, we summarize the extracted API invariant of super apps (by using our algorithm). It can be observed that not all invariants will appear in the implementations of a given super app.  \looseness=-1

The above approach requires us first to identify the implementation of public APIs.  To that end, we extract the names of the public APIs and use the names to search the reverse-engineered code to find methods that contain the API names (in their method bodies). The identified code may be the implementation of public APIs. This is because implementations of the API must contain the names of the APIs, for the miniapps running at the application layer to fetch. For example, as shown in the 5th line of \autoref{lst:jsapiclass}, in the implementation of API \texttt{wx.getLocation}, we can identify the string  ``getLocation''.  As such, we can use the methods to find out the possible implementations of the APIs. 
}

 \vspace{-0.1in}
\subsection{Dynamic API Classification}
\label{subsce:apiexploiter}

With the identified undocumented APIs, next we need to invoke each of them to decide whether they can be exploited by attackers based on the error messages obtained while executing the corresponding test cases for each of the API. This is a three-step process, starting from test case generation, followed by API invocation identification using function trace analysis, and finally the API classification through dynamic API probing. 


\paragraph{Step-I: Test Case Generation} In this step, we use our test case generator to produce test cases. The test cases are the JavaScript code snippets that contain the APIs to be invoked (with their parameters configured). For example, \texttt{wx.getLocation(\{type: "wgs84"\})} is a test case for testing API \texttt{wx.getLocation} (how to invoke such test cases will be described in API invocation identification).   
There are two types of test cases: one for API invocation identification and the other for API classification. The goal of API invocation identification is to execute the documented API, and use the function trace analysis to identify the invocation point. Therefore, we only need to generate a few test cases (which are the test cases of documented APIs). However, in API classification, which invokes the undocumented APIs and categorizes them based on their outputs, we need to produce at least one valid test case for each undocumented API (to obtain the outputs). In particular, since each API may accept one or multiple parameters, to produce a valid test case,  we have to identify all the types (e.g., \texttt{Integer}, \texttt{Boolean}) of the parameters, through which we can further feed each API a list of parameter instances in the right order (e.g., \texttt{testAPI(true, 1234)}): \looseness=-1 
%

\begin{packeditemize}
 
\item \textbf{Parameter Type Extraction.} While \sysname could identify the types of parameters through documentation analysis, such an approach cannot identify the types of parameters for undocumented APIs. Therefore, we need a more reliable approach to ensure that we can extract parameter types for both documented and undocumented APIs. Our idea is to analyze the implementations of the APIs, since we have already identified the implementations for both documented and undocumented APIs as described in \S\ref{subsce:apifinder}.
For instance, in \wechat's implementation, we notice that the types of the parameters of an API can be recognized by inspecting the methods invoked by JSON instances, e.g., in the implementation of \texttt{getLocation}, we can notice that a JSON object invokes method \texttt{optString("paramname", paramvalue)}, which indicates that \texttt{getLocation} has a   ``\texttt{paramname}'' parameter with type \texttt{String}. Similarly, if the API accepts a Boolean value as its parameter, there will be a method \texttt{optBoolean("paramname", paramvalue)} in its implementation.

\item \textbf{Parameter Instance Generation.} 
The parameters must be instantiated before being fed into the APIs. We used a pre-defined template-based approach to instantiate the parameters. 
At a high level, the template specified the appropriate values with different types that can be used to produce the parameters (e.g., ``1'' and ``0'' are used when the ``type'' of the parameter is of type ``number'', and ``test'' was used when the ``type''  of the parameter is of type ``string''). For instance, \wechat API \texttt{showToast} (which shows a message to the user) has two parameters \texttt{title} and  \texttt{duration}, with types string and number, respectively. As such, 
we produced an instance with the predefined template, where \texttt{title} is set to ``test'' and \texttt{duration} is set to ``1''. Using such a template method, we successfully instantiated all the parameters.

\item \textbf{Parameter Order Permutation.} 
Although we have instantiated the parameters, we still do not know the orders of those parameters for the undocumented APIs, as the parameters in the Service Abstraction layer are all encapsulated in JSON objects. Therefore, we have to properly order the parameters, and we use a brute-force approach. For example, \texttt{true} and \texttt{1234} are two parameters of  \texttt{testAPI}, which could have two possible combinations: \texttt{testAPI(true, 1234)}) and \texttt{testAPI(1234, true)}. We just assume that all those combinations are valid and invoke them one-by-one (the invalid ones will be filtered out during the API classification, which will be described later). Given that one API can accept no more than 4 parameters (which results in 24 combinations), according to our static analysis with the code, 
we believe such a brute-force approach is acceptable.  




\end{packeditemize}

Specifically, we would like to clarify certain technical details. First, during our dynamic analysis, we only explore a limited range of inputs. This is because dynamic tracing does not require a broad range of input to expose hidden APIs. Additionally, the test case generation is sufficient for testing security checks, such as whether the hidden API is protected by security checks. In other words, as long as valid inputs are provided to the API, our tool can trigger the API if there are no security checks. If there are security checks, we can observe errors. Our objective is not to enumerate all possible inputs, as we are not fuzzing the actual hidden API. 
Second, hidden APIs may require complex parameter types, such as JSON-objects. These complex parameter types are combinations of other basic parameter types (e.g., integer, string), and can be recursively derived until they become primitive types. For instance, an object may contain a string, an integer, and a boolean. We can simply inflate each parameter based on its respective parameter type.
As APIs implemented in the Service Abstraction Layer lack states or context, it is unnecessary to determine their execution state within this layer. Our testing process involves providing our tool with a code snippet containing the API to be tested, which is sufficient for our purposes. The JavaScript Framework Layer handles most of the checks, so the API invocation is checked before its order or dependency state is resolved.

\paragraph{Step-II: API Invocation Identification} Next, \sysname needs to execute the generated test cases on top of our customized V8 engine to identify how the documented API is invoked, so that it can later similarly invoke the undocumented ones. Intuitively, when we test a specific API, we need to compile and produce a testing miniapp that contains the API for our test. However, this approach is not scaled and can slow down our testing performance. Interestingly, we notice that we can let the V8 engine directly inject the JavaScript code into the JavaScript Framework Layer (the V8 engine has a function named \texttt{script}, which accepts JavaScript code as input, and injects the code for the JavaScript Framework Layer to execute). Since the JavaScript code is injected into the JavaScript Framework layer, the super apps will handle the code as they handle the code in a regular miniapp. \looseness=-1

\ignore{
The test case generator can also produce test cases for dynamic API probing, which involves the invocation of undocumented API, and the use of the interface that communicate with the underlying layers (the approach to identify such interface will be introduced next). If this is the case, in the generated JavaScript code, the undocumented API is fed as an parameter of the identified interface (e.g., \texttt{invokeHandler("private\_openUrl","test.com",\\0)}), and the rest procedure is the same as that of invoking a documented API. We omit the details for brevity.    
}

Also, in most cases, V8 Engine has a built-in Profiler, 
but the super apps do not directly expose any interfaces for developers to use. Meanwhile, although it is true that different platforms may customize the V8 Engine to enable their desired functionalities, they will not intentionally remove the built-in Profiler since it is also helpful for their own debugging purposes. Therefore, as long as we can find a way to invoke Profiler, we will be able to collect the traces. Fortunately, 
we can use Frida~\cite{druffel2020davinci}, an Android hooking tool, to dynamically instrument the V8 Engine to invoke \texttt{startProfiling} of Profiler and let it start profiling, and collect the function traces of documented API execution. 
\looseness=-1


With the collected function traces, we then present how to find the desired interface using function trace analysis, a standard technique widely used in program analysis. As discussed in \S\ref{sub:observation}, API invocation is a complicated process involving multiple layers. Fortunately, the Profiler only runs inside the JavaScript Framework layer, and we can just monitor the function traces produced at this layer since we aim to identify how to invoke an API from the JavaScript layer. In particular, our analysis starts from the API of our interests (e.g., {\tt wx.getLocation}), identifies all the functions involved based on the dependencies of parameter and API names, and eventually identifies the last invocation function, e.g., \code{NativeGlobal.invokeHandler} (see \autoref{lst:jsapibridge}), which is the desired interface we aim to discover. Specifically, the dependencies are indeed the chained relationship, and we actually build such dependencies based on the parameters that are fed into the functions (we can monitor the changes of parameters of the functions). For example, when we execute \texttt{wx.getLocation}, we will observe a function named \texttt{NativeGlobal.invokeHandler} that takes a parameter named \texttt{getLocation} as its inputs. Therefore, we know that \texttt{wx.getLocation} and \texttt{NativeGlobal.invokeHandler} have dependencies. \looseness=-1

\ignore{
Therefore, we directly invoke (taint) one of the public APIs (i.e., test cases) dynamically and let our profiler collect the tracers of the API invocation, which allows us to build an API call stack (where the bottom of the stack is the invoked API, and the
top of the stack is the interface exposed by Customized V8 Layer for passing API names and parameters). Given that there could be other function invocations that may occur together with the API invocation, and consequently, our profiler may produce multiple stacks instead of just one. As such, we inspect all the functions located on the top of the function call stacks to see whether the functions' inputs are the fed API names (i.e., tainted API). If so, we then know that the interface is the desired interface, which can be used to invoke APIs. For example, we identify the interface of \wechat is \code{NativeGlobal.invokeHandler}, which is the same as results achieved through the manual efforts.   
}

 
\begin{figure*}

\includegraphics[width=0.98\textwidth]{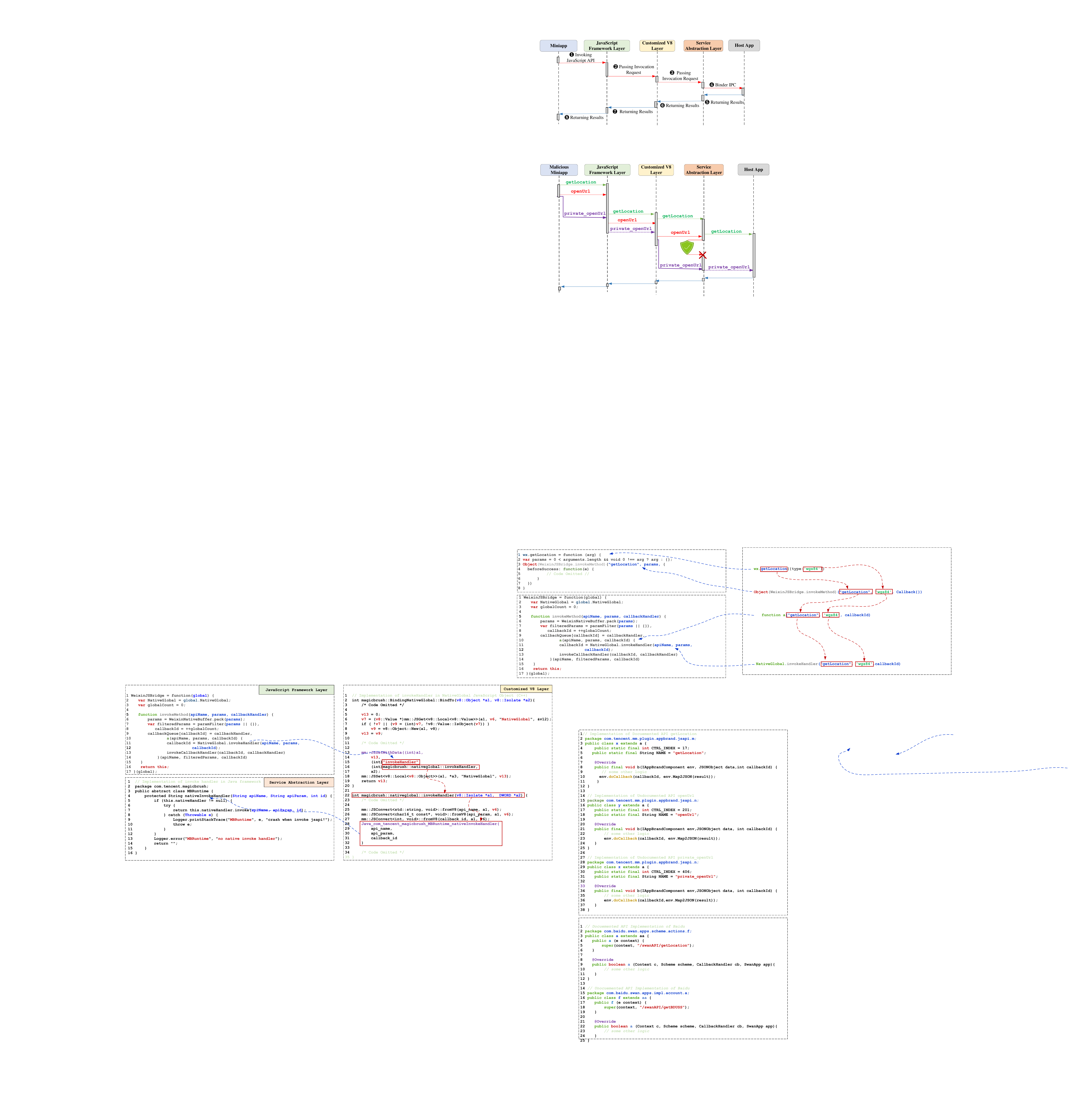}
 \vspace{-0.1in}
\caption{The implementations of API invocations across three layers (\wechat)}
 \label{fig:invokehandlerflow}
 
\end{figure*} 


To provide a detailed explanation of how our trace analysis works, we will utilize an example that features the implementations of API invocations across three layers, namely the JavaScript Framework layer, the Customized V8 layer, and the Service Abstraction layer. The process begins with the JavaScript Framework layer, which initiates the API invocation by calling \texttt{NativeGlobal.invokeHandler}. This invocation is then handed over to the Customized V8 layer, which is responsible for handling it. As shown in \autoref{fig:invokehandlerflow}, this step is represented line 10 of the JavaScript Framework layer's implementation. Next, the Customized V8 layer extracts critical information from the API invocation, including the API name, its parameters, and any corresponding callbacks. This information is obtained from lines 28--32 of the Customized V8 layer's implementation. The Customized V8 layer then proceeds to invoke the relevant APIs at the Service Abstraction Layer through the use of the Java Native Interface (JNI)~\cite{liang1999java}. Finally, during the API invocations at the Service Abstraction layer (line 4), this layer may need to communicate with the Customized V8 layer for additional operations, such as performing permission checks if the API requires them. We have omitted this code for the sake of brevity. In summary, our trace analysis provides insight into the entire process of API invocations across the three layers of the system. We track the flow of control and collect data on API names, parameters, and callbacks to enable a more comprehensive analysis of the system's behavior. \looseness=-1


 
 \paragraph{Step-III: Dynamic Probing for API Category Classification} With the identified interfaces of how to invoke a public API, we then use it to similarly invoke undocumented APIs, by first generating the corresponding test cases, and then injecting the JavaScript code using the \text{script} function into the V8 engine, as described earlier.
 %
When executing a particular test case, there could be three types of outcomes:  the tested ``API'' is a checked API (when invoked, a permission denial will be observed based on the standard error messages),  the tested ``API'' is an unchecked API (which can be invoked successfully), the tested ``API'' is not an API. As such, we can use the following strategies to identify them. 
\begin{packeditemize}
\item \textbf{Unchecked APIs.}  Similar to the public APIs, the unchecked undocumented APIs can be invoked without requiring additional permissions. As such, we first deliver a public API invocation request, such as \texttt{getLocation}, and record the feedback of the host app. For example, \wechat and \textsf{Baidu} will not print any errors when the invocation request gets approved, and we then use this as a signature to see whether an invocation request is successfully executed.

\item \textbf{Checked APIs.} The checked APIs are the APIs that are protected by security checks, which can only be invoked by their 1st-party miniapps. In the event of a security check failure, the super apps will generate error messages notifying the user of insufficient permissions. This exception applies to all APIs within various super apps, albeit with minor variations in the error messages displayed. For example, when 3rd-party mini-apps attempt to invoke a checked API of \wechat, the host app will throw an error message ``{\tt fail: no permission}''. For \textsf{WeCom}, the error message becomes ``{\tt fail: access denied}''. Therefore, we use keywords such as ``{\tt fail}'', ``{\tt no permission}'' and ``{\tt access denied}'' to match and decide whether the invocation request gets denied. If so, it is a checked API. 

\item \textbf{Non-APIs.} Theoretically, \sysname may have false positives, and as such, our tool may mistakenly recognize some non-APIs.
Therefore, we need to filter them out. To that end, we first create an invalid request and then send it to the host app to see the feedback. For example, if we initiate an invalid request and send it to \wechat, \wechat will reject the invocation request and throw an error message ``{\tt fail: not supported}''.  Then, such an error message is used as a signature to match the non-APIs. 
\end{packeditemize}

As an example, in the case of WeChat, if we attempt to use the API \texttt{openUrl}, the super app will generate an error message stating ``\texttt{fail: no permission}''. This error message implies that the API is a checked hidden API. On the other hand, if we use the API \texttt{private\_openUrl}, the super app will handle the invocation request as a regular request without displaying any error message. As a result, we can conclude that this API is an unchecked hidden API.\looseness=-1

\section{Evaluation}
\label{sec:eval}







We have developed a prototype of \sysname with 5K lines of code on top of open source tools such as \textsf{Soot}~\cite{bartel2012dexpler} for decompilation and \textsf{Frida}~\cite{druffel2020davinci} for tracing. In this section, we present the evaluation results. We first describe our experimental setup in \S\ref{subsec:setup}, and then \sysname's effectiveness in \S\ref{sub:effectiveness}. The efficiency of \sysname is presented in Appendix-\S\ref{sub:efficiency} for readers of interests. 
\looseness=-1

\subsection{Experiment Setup}
\label{subsec:setup}

\paragraph{The Tested Host Apps} Today, there are quite a number of super apps that support the execution of miniapps. Although we wish to test all of them, eventually we selected five of them,  as shown in \autoref{tab:runtime}, and these include 
\wechat,  \textsf{WeCom} and \textsf{QQ} from Tencent Holdings Ltd.,  \textsf{Baidu} from Baidu Inc., and \textsf{TikTok} from ByteDance Ltd. 
We excluded other super apps such as \textsf{Alipay} and \textsf{Snapchat} particularly because they do not build on the V8 engine (making our tool unsuitable for them at this moment). 
Also, to study the security issues of the tested super apps correspondingly,  we registered an account in each platform, downloaded their development tools and SDKs, built miniapps by following their official documents, and inspected their code. Among them, \textsf{Baidu} has a relatively closed ecosystem, where only the enterprise developers are allowed to register as their developers. However,  they allow individuals to apply for trial accounts to use their development tools to develop miniapps, and therefore, we tested \textsf{Baidu} using their trial accounts.

\begin{table}[t]
	\centering
	\begin{threeparttable}
	\belowrulesep=0pt
\aboverulesep=0pt
\setlength\tabcolsep{3pt}
\scriptsize
	\begin{tabular}{llrlcrc}
		\toprule[1.5pt]
			\bfseries{Name} &
			\bfseries{Vendor} &
			\bfseries{Version} &
			\bfseries{V8} & 
			\bfseries{Date} &
			\bfseries{Installs} & 
			\multicolumn{1}{c}{\textbf{\begin{tabular}[c]{@{}c@{}}1st-party miniapp \\ being tested?\end{tabular}}}
			\\
		\midrule[0.5pt]
	    Baidu  & Baidu     & 12.21  & 7.6 & 08/13/2021 & 5,000,000+  & \tickYes \\
		QQ     & Tencent   & 8.8    & 7.2 & 10/05/2021 & 10,000,000+   & \tickYes \\
		TikTok & ByteDance & 17.9   & 7.2 & 10/19/2021 & 1,000,000,000+ & \xmark \\
		WeChat & Tencent   & 8.0    & 8.0 & 07/21/2021 & 100,000,000+   & \tickYes \\
		WeCom  & Tencent   & 3.1    & 8.0 & 09/14/2021 & 100,000+     & \tickYes \\
		\bottomrule[1.5pt]
	\end{tabular}

	\end{threeparttable}
 
	\caption{Summary of the Tested Super Apps}
    \label{tab:runtime}
 \vspace{-0.35in}
\end{table}


\begin{table*}[]
\centering
\scriptsize
\setlength\tabcolsep{3pt}
\begin{tabular}{lccccccccccccccccc} 
\toprule[1.5pt]
\multirow{6}{*}{\textbf{Name}} & \multicolumn{3}{c}{\textbf{Input}}                                                                                                                     & \multicolumn{5}{c}{\textbf{Static Analysis}}                                                                                                                                                                                                            & \multicolumn{3}{c}{\textbf{Dynamic Analysis}}                                                                                                                                                                                                                                                  & \multicolumn{3}{c}{\textbf{Output}}                                                                                                                 \\ 
\cmidrule(lr){2-4} \cmidrule(lr){5-9} \cmidrule(lr){10-12} \cmidrule(lr){13-15}
                                    &\multirow{3}{*}{\begin{tabular}[c]{@{}c@{}}\#\textbf{Size}\\\textbf{(MBs)}\end{tabular}}   
                                    & \multirow{3}{*}{\begin{tabular}[c]{@{}c@{}}\#\textbf{ of  }\\\textbf{LoC}\end{tabular}} 
                                    & \multirow{3}{*}{\begin{tabular}[c]{@{}c@{}}\#\textbf{ of Public}\\\textbf{API}\end{tabular}} & \multicolumn{4}{c}{\textbf{API Invariants}}                                                                                                                                                                                                             &  \multirow{3}{*}{\begin{tabular}[c]{@{}c@{}}\#\textbf{ of Hidden API  }\\\textbf{Candidates}\end{tabular}} & \multirow{3}{*}{\begin{tabular}[c]{@{}c@{}}\textbf{\textbf{Invocation}}\\\textbf{\textbf{Identification}} \\ (\# \textbf{of Traced} \\ \textbf{Functions}) \end{tabular}} & \multicolumn{2}{c}{\begin{tabular}[c]{@{}c@{}}\textbf{\textbf{API Classification}}\\\textbf{\textbf{(\# of Test Cases)}}\end{tabular}} &  \multirow{3}{*}{\begin{tabular}[c]{@{}c@{}}\#\textbf{ of  }\\\textbf{Checked}\\ \textbf{API} \end{tabular}} &  \multirow{3}{*}{\begin{tabular}[c]{@{}c@{}}\#\textbf{ of  }\\\textbf{Unchecked} \\\textbf{API}\end{tabular}} &  \multirow{3}{*}{\begin{tabular}[c]{@{}c@{}}\#\textbf{ of  }\\\textbf{Non}\\ \textbf{API} \end{tabular}}  \\ 
\cmidrule(lr){5-8}    \cmidrule(lr){11-12}
                                    &                                      &                                     &                                                                            & \begin{tabular}[c]{@{}c@{}}\textbf{Method }\\\textbf{ Signature}\end{tabular} & \begin{tabular}[c]{@{}c@{}}\textbf{Super }\\\textbf{ Class}\end{tabular} & \begin{tabular}[c]{@{}c@{}}\textbf{Super}\\\textbf{ Package}\end{tabular} & \textbf{Callers} &                                             &                                                                                                                            & \begin{tabular}[c]{@{}c@{}}\textbf{\# of Auto}\\\textbf{Generated}\end{tabular} & \begin{tabular}[c]{@{}c@{}}\textbf{\# of Manually} \\\textbf{Created}\end{tabular} &                                                                  &                                                               &          \\ 
\cmidrule{1-15} 
\textsf{Baidu}  & \multicolumn{1}{r}{123.6} & \multicolumn{1}{r}{2,005,003} & 464 & \cmark & \cmark & \cmark & \xmark & 143 & 30 & \multicolumn{1}{r}{  423} & \multicolumn{1}{r}{56} & \multicolumn{1}{r}{25} & \multicolumn{1}{r}{113} & \multicolumn{1}{r}{5} \\
\textsf{QQ}     & \multicolumn{1}{r}{138.6} & \multicolumn{1}{r}{1,557,805} & 506 & \cmark & \cmark & \cmark & \xmark & 304 & 43 & \multicolumn{1}{r}{1,083} & \multicolumn{1}{r}{61} & \multicolumn{1}{r}{ 6} & \multicolumn{1}{r}{295} & \multicolumn{1}{r}{3} \\
\textsf{TikTok} & \multicolumn{1}{r}{  6.2} & \multicolumn{1}{r}{  718,395} & 383 & \cmark & \cmark & \cmark & \cmark & 124 & 37 & \multicolumn{1}{r}{  352} & \multicolumn{1}{r}{53} & \multicolumn{1}{r}{ 2} & \multicolumn{1}{r}{122} & \multicolumn{1}{r}{0} \\
\textsf{WeChat} & \multicolumn{1}{r}{199.2} & \multicolumn{1}{r}{1,609,650} & 590 & \cmark & \cmark & \cmark & \cmark & 575 & 28 & \multicolumn{1}{r}{2,184} & \multicolumn{1}{r}{66} & \multicolumn{1}{r}{65} & \multicolumn{1}{r}{502} & \multicolumn{1}{r}{8} \\
\textsf{WeCom}  & \multicolumn{1}{r}{224.8} & \multicolumn{1}{r}{1,067,273} & 606 & \cmark & \cmark & \cmark & \cmark & 683 & 31 & \multicolumn{1}{r}{2,315} & \multicolumn{1}{r}{70} & \multicolumn{1}{r}{82} & \multicolumn{1}{r}{593} & \multicolumn{1}{r}{8} \\

\bottomrule[1.5pt]
\end{tabular}
 
\caption{Effectiveness of \sysname with the tested super apps. The terms ``Signature'', ``Super Class'', ``Super Package'',  and ``Callers'' have consistent meanings with those defined in \S\ref{subsce:apifinder}.  
\vspace{-0.25in}
}
 
 \label{tab:effectiveness}

\end{table*}



		


\paragraph{The Tested Miniapps} We believe it is important to measure the usage of undocumented APIs in 1st-party and 3rd-party miniapps for two reasons. First, understanding how 1st-party miniapps use these APIs can help us comprehend the entire ecosystem. Second, if 3rd-party developers know about these APIs, they may use them, which can lead to security issues if these APIs have access to sensitive resources.
To analyze the usage of undocumented APIs in 1st-party miniapps, we searched for interfaces provided by host apps and collected 236 miniapps from WeChat and WeCom, 340 miniapps from Baidu, and 24 miniapps from QQ. We could not find information about the 1st-party miniapps of TikTok, so we did not report their API usage.  We could not scan all 3rd-party miniapps because there is no public dataset or crawlers available. Therefore, we can only measure the usage of hidden APIs among 3rd-party miniapps within the \wechat ecosystem.  We collected $267,359$ miniapps using Mini-Crawler~\cite{zhang2021measurement} within 3 weeks. 

\paragraph{The Testing Environment}  We performed our static analysis on one laptop, which has 6 cores, Intel Core i7-10850H (4.90 GHz) CPUs and 64 GB RAM, and our dynamic analysis on a \textsf{Google Pixel 4} running Android 11 and a \textsf{Google Pixel 2} running Android 9, since we particularly focused on the Android version of miniapps.




\subsection{Effectiveness} 
\label{sub:effectiveness}

 \begin{table*}[h]
\centering
\belowrulesep=0pt
\aboverulesep=0pt
\setlength\tabcolsep{1.6pt}
\renewcommand{\arraystretch}{1.2}
\scriptsize

\begin{tabular}{clcrcrcr|crcrcr|crcrcr|crcrcr|crcrcr}
\toprule[1.5pt]

    \multicolumn{2}{c}{
        \multirow{2}{*}{\textbf{Available APIs}}} &

    \multicolumn{6}{c}{
        \textbf{WeChat}
    } &
    \multicolumn{6}{c}{
        \textbf{WeCom}
    } &
    \multicolumn{6}{c}{
        \textbf{Baidu}
    } & 
    \multicolumn{6}{c}{
        \textbf{TikTok}
    } & 
    \multicolumn{6}{c}{
        \textbf{QQ}
    } \\
    \cline{3-32}

    \multicolumn{2}{c}{} &
        \begin{tabular}[c]{@{}c@{}}
            D
        \end{tabular} & 
        \begin{tabular}[c]{@{}c@{}}
            \%
        \end{tabular} & 
        \begin{tabular}[c]{@{}c@{}}
            UU
        \end{tabular} &
        \begin{tabular}[c]{@{}c@{}}
            \%
        \end{tabular} & 
        \begin{tabular}[c]{@{}c@{}}
            UC
        \end{tabular} &
        \begin{tabular}[c]{@{}c@{}}
            \%
        \end{tabular} & 
        \begin{tabular}[c]{@{}c@{}}
            D
        \end{tabular} & 
        \begin{tabular}[c]{@{}c@{}}
            \%
        \end{tabular} & 
        \begin{tabular}[c]{@{}c@{}}
            UU
        \end{tabular} & 
        \begin{tabular}[c]{@{}c@{}}
            \%
        \end{tabular} & 
        \begin{tabular}[c]{@{}c@{}}
            UC
        \end{tabular} &
        \begin{tabular}[c]{@{}c@{}}
            \%
        \end{tabular} & 
        \begin{tabular}[c]{@{}c@{}}
            D
        \end{tabular} &
        \begin{tabular}[c]{@{}c@{}}
            \%
        \end{tabular} & 
        \begin{tabular}[c]{@{}c@{}}
            UU
        \end{tabular} &
        \begin{tabular}[c]{@{}c@{}}
            \%
        \end{tabular} & 
        \begin{tabular}[c]{@{}c@{}}
            UC
        \end{tabular} &
        \begin{tabular}[c]{@{}c@{}}
            \%
        \end{tabular} & 
        \begin{tabular}[c]{@{}c@{}}
            D
        \end{tabular} &
        \begin{tabular}[c]{@{}c@{}}
            \%
        \end{tabular} & 
        \begin{tabular}[c]{@{}c@{}}
            UU
        \end{tabular} &
        \begin{tabular}[c]{@{}c@{}}
            \%
        \end{tabular} & 
        \begin{tabular}[c]{@{}c@{}}
            UC
        \end{tabular} &
        \begin{tabular}[c]{@{}c@{}}
            \%
        \end{tabular} & 
        \begin{tabular}[c]{@{}c@{}}
            D
        \end{tabular} &
        \begin{tabular}[c]{@{}c@{}}
            \%
        \end{tabular} & 
        \begin{tabular}[c]{@{}c@{}}
            UU
        \end{tabular} &
        \begin{tabular}[c]{@{}c@{}}
            \%
        \end{tabular} & 
        \begin{tabular}[c]{@{}c@{}}
            UC
        \end{tabular} &
        \begin{tabular}[c]{@{}c@{}}
            \%
        \end{tabular} \\ 

\midrule[0.5pt]

\multirow{4}{*}{Base} 
 & Basic      &  5 & \ccell{ 5}{  7}{71.4} &  2 & \ccell{ 2}{  7}{28.6} & - & \ccell{ 0}{  7}{ 0.0} &  6 & \ccell{ 6}{  9}{66.7} &  3 & \ccell{ 3}{  9}{33.3} & - & \ccell{ 0}{  9}{ 0.0} &  8 & \ccell{ 8}{ 11}{72.7} &  2 & \ccell{ 2}{ 11}{18.2} &  1 & \ccell{ 1}{ 11}{ 9.1} &  7 & \ccell{ 7}{ 11}{63.6} &  4 & \ccell{ 4}{ 11}{36.4} & - & \ccell{ 0}{ 11}{ 0.0} &  3 & \ccell{ 3}{  3}{100.0} & - & \ccell{ 0}{  3}{ 0.0} & - & \ccell{ 0}{  3}{ 0.0} \\
 & App        & 13 & \ccell{13}{ 33}{39.4} & 14 & \ccell{14}{ 33}{42.4} &  6 & \ccell{ 6}{ 33}{18.2} & 13 & \ccell{13}{ 35}{37.1} & 16 & \ccell{16}{ 35}{45.7} &  6 & \ccell{ 6}{ 35}{17.1} &  8 & \ccell{ 8}{ 19}{42.1} & 10 & \ccell{10}{ 19}{52.6} &  1 & \ccell{ 1}{ 19}{ 5.3} &  6 & \ccell{ 6}{ 12}{50.0} &  6 & \ccell{ 6}{ 12}{50.0} & - & \ccell{ 0}{ 12}{ 0.0} &  9 & \ccell{ 9}{ 26}{34.6} & 17 & \ccell{17}{ 26}{65.4} & - & \ccell{ 0}{ 26}{ 0.0} \\
 & Debug      & 15 & \ccell{15}{ 17}{88.2} &  2 & \ccell{ 2}{ 17}{11.8} & - & \ccell{ 0}{ 17}{ 0.0} & 15 & \ccell{15}{ 17}{88.2} &  2 & \ccell{ 2}{ 17}{11.8} & - & \ccell{ 0}{ 17}{ 0.0} &  1 & \ccell{ 1}{ 30}{ 3.3} & 28 & \ccell{28}{ 30}{93.3} &  1 & \ccell{ 1}{ 30}{ 3.3} & - & \ccell{ 0}{0.1}{  0.0} & - & \ccell{ 0}{0.1}{  0.0} & - & \ccell{ 0}{0.1}{  0.0} & 20 & \ccell{20}{ 20}{100.0} & - & \ccell{ 0}{ 20}{ 0.0} & - & \ccell{ 0}{ 20}{ 0.0} \\
 & Misc       & 10 & \ccell{10}{ 17}{58.8} &  7 & \ccell{ 7}{ 17}{41.2} & - & \ccell{ 0}{ 17}{ 0.0} & 10 & \ccell{10}{ 18}{55.6} &  8 & \ccell{ 8}{ 18}{44.4} & - & \ccell{ 0}{ 18}{ 0.0} &  9 & \ccell{ 9}{  9}{100.0} & - & \ccell{ 0}{  9}{ 0.0} & - & \ccell{ 0}{  9}{ 0.0} & 10 & \ccell{10}{ 19}{52.6} &  9 & \ccell{ 9}{ 19}{47.4} & - & \ccell{ 0}{ 19}{ 0.0} &  9 & \ccell{ 9}{  9}{100.0} & - & \ccell{ 0}{  9}{ 0.0} & - & \ccell{ 0}{  9}{ 0.0} \\ \hhline{*{32}{-}}

\multirow{5}{*}{UI}
 & Interaction &  6 & \ccell{ 6}{ 13}{46.2} &  7 & \ccell{ 7}{ 13}{53.8} & - & \ccell{ 0}{ 13}{ 0.0} &  6 & \ccell{ 6}{ 13}{46.2} &  7 & \ccell{ 7}{ 13}{53.8} & - & \ccell{ 0}{ 13}{ 0.0} &  7 & \ccell{ 7}{ 17}{41.2} & 10 & \ccell{10}{ 17}{58.8} & - & \ccell{ 0}{ 17}{ 0.0} &  9 & \ccell{ 9}{ 11}{81.8} &  2 & \ccell{ 2}{ 11}{18.2} & - & \ccell{ 0}{ 11}{ 0.0} &  6 & \ccell{ 6}{ 15}{40.0} &  9 & \ccell{ 9}{ 15}{60.0} & - & \ccell{ 0}{ 15}{ 0.0} \\
 & Navigation &  4 & \ccell{ 4}{  9}{44.4} &  5 & \ccell{ 5}{  9}{55.6} & - & \ccell{ 0}{  9}{ 0.0} &  4 & \ccell{ 4}{ 10}{40.0} &  6 & \ccell{ 6}{ 10}{60.0} & - & \ccell{ 0}{ 10}{ 0.0} &  4 & \ccell{ 4}{  4}{100.0} & - & \ccell{ 0}{  4}{ 0.0} & - & \ccell{ 0}{  4}{ 0.0} &  5 & \ccell{ 5}{  5}{100.0} & - & \ccell{ 0}{  5}{ 0.0} & - & \ccell{ 0}{  5}{ 0.0} &  4 & \ccell{ 4}{ 12}{33.3} &  8 & \ccell{ 8}{ 12}{66.7} & - & \ccell{ 0}{ 12}{ 0.0} \\
 & Animation  & 32 & \ccell{32}{ 32}{100.0} & - & \ccell{ 0}{ 32}{ 0.0} & - & \ccell{ 0}{ 32}{ 0.0} & 32 & \ccell{32}{ 32}{100.0} & - & \ccell{ 0}{ 32}{ 0.0} & - & \ccell{ 0}{ 32}{ 0.0} & 21 & \ccell{21}{ 22}{95.5} &  1 & \ccell{ 1}{ 22}{ 4.5} & - & \ccell{ 0}{ 22}{ 0.0} &  1 & \ccell{ 1}{  1}{100.0} & - & \ccell{ 0}{  1}{ 0.0} & - & \ccell{ 0}{  1}{ 0.0} & 31 & \ccell{31}{ 31}{100.0} & - & \ccell{ 0}{ 31}{ 0.0} & - & \ccell{ 0}{ 31}{ 0.0} \\
 & WebView    & - & \ccell{ 0}{ 23}{ 0.0} & 22 & \ccell{22}{ 23}{95.7} &  1 & \ccell{ 1}{ 23}{ 4.3} & - & \ccell{ 0}{ 25}{ 0.0} & 24 & \ccell{24}{ 25}{96.0} &  1 & \ccell{ 1}{ 25}{ 4.0} & - & \ccell{ 0}{  4}{ 0.0} &  3 & \ccell{ 3}{  4}{75.0} &  1 & \ccell{ 1}{  4}{25.0} & - & \ccell{ 0}{  3}{ 0.0} &  3 & \ccell{ 3}{  3}{100.0} & - & \ccell{ 0}{  3}{ 0.0} & - & \ccell{ 0}{ 16}{ 0.0} & 16 & \ccell{16}{ 16}{100.0} & - & \ccell{ 0}{ 16}{ 0.0} \\
 & Misc       & 20 & \ccell{20}{ 74}{27.0} & 54 & \ccell{54}{ 74}{73.0} & - & \ccell{ 0}{ 74}{ 0.0} & 20 & \ccell{20}{ 78}{25.6} & 58 & \ccell{58}{ 78}{74.4} & - & \ccell{ 0}{ 78}{ 0.0} & 37 & \ccell{37}{ 48}{77.1} & 11 & \ccell{11}{ 48}{22.9} & - & \ccell{ 0}{ 48}{ 0.0} & 14 & \ccell{14}{ 19}{73.7} &  5 & \ccell{ 5}{ 19}{26.3} & - & \ccell{ 0}{ 19}{ 0.0} & 18 & \ccell{18}{ 42}{42.9} & 24 & \ccell{24}{ 42}{57.1} & - & \ccell{ 0}{ 42}{ 0.0} \\ \hhline{*{32}{-}}

\multirow{5}{*}{Network}
 & Request    &  5 & \ccell{ 5}{  9}{55.6} &  4 & \ccell{ 4}{  9}{44.4} & - & \ccell{ 0}{  9}{ 0.0} &  5 & \ccell{ 5}{  9}{55.6} &  4 & \ccell{ 4}{  9}{44.4} & - & \ccell{ 0}{  9}{ 0.0} &  2 & \ccell{ 2}{  3}{66.7} &  1 & \ccell{ 1}{  3}{33.3} & - & \ccell{ 0}{  3}{ 0.0} &  6 & \ccell{ 6}{ 10}{60.0} &  4 & \ccell{ 4}{ 10}{40.0} & - & \ccell{ 0}{ 10}{ 0.0} &  4 & \ccell{ 4}{  6}{66.7} &  2 & \ccell{ 2}{  6}{33.3} & - & \ccell{ 0}{  6}{ 0.0} \\
 & Download   &  7 & \ccell{ 7}{ 29}{24.1} & 21 & \ccell{21}{ 29}{72.4} &  1 & \ccell{ 1}{ 29}{ 3.4} &  7 & \ccell{ 7}{ 30}{23.3} & 22 & \ccell{22}{ 30}{73.3} &  1 & \ccell{ 1}{ 30}{ 3.3} & 11 & \ccell{11}{ 11}{100.0} & - & \ccell{ 0}{ 11}{ 0.0} & - & \ccell{ 0}{ 11}{ 0.0} & - & \ccell{ 0}{  4}{ 0.0} &  4 & \ccell{ 4}{  4}{100.0} & - & \ccell{ 0}{  4}{ 0.0} &  6 & \ccell{ 6}{ 10}{60.0} &  4 & \ccell{ 4}{ 10}{40.0} & - & \ccell{ 0}{ 10}{ 0.0} \\
 & Upload     &  7 & \ccell{ 7}{ 14}{50.0} &  5 & \ccell{ 5}{ 14}{35.7} &  2 & \ccell{ 2}{ 14}{14.3} &  7 & \ccell{ 7}{ 15}{46.7} &  6 & \ccell{ 6}{ 15}{40.0} &  2 & \ccell{ 2}{ 15}{13.3} &  6 & \ccell{ 6}{  6}{100.0} & - & \ccell{ 0}{  6}{ 0.0} & - & \ccell{ 0}{  6}{ 0.0} & - & \ccell{ 0}{  4}{ 0.0} &  4 & \ccell{ 4}{  4}{100.0} & - & \ccell{ 0}{  4}{ 0.0} &  6 & \ccell{ 6}{  8}{75.0} &  2 & \ccell{ 2}{  8}{25.0} & - & \ccell{ 0}{  8}{ 0.0} \\
 & Websocket  & 14 & \ccell{14}{ 15}{93.3} &  1 & \ccell{ 1}{ 15}{ 6.7} & - & \ccell{ 0}{ 15}{ 0.0} & 14 & \ccell{14}{ 15}{93.3} &  1 & \ccell{ 1}{ 15}{ 6.7} & - & \ccell{ 0}{ 15}{ 0.0} & 13 & \ccell{13}{ 13}{100.0} & - & \ccell{ 0}{ 13}{ 0.0} & - & \ccell{ 0}{ 13}{ 0.0} &  7 & \ccell{ 7}{  9}{77.8} &  2 & \ccell{ 2}{  9}{22.2} & - & \ccell{ 0}{  9}{ 0.0} & 13 & \ccell{13}{ 15}{86.7} &  2 & \ccell{ 2}{ 15}{13.3} & - & \ccell{ 0}{ 15}{ 0.0} \\
 & Misc       & 23 & \ccell{23}{ 26}{88.5} &  3 & \ccell{ 3}{ 26}{11.5} & - & \ccell{ 0}{ 26}{ 0.0} & 23 & \ccell{23}{ 27}{85.2} &  4 & \ccell{ 4}{ 27}{14.8} & - & \ccell{ 0}{ 27}{ 0.0} & - & \ccell{ 0}{0.1}{  0.0} & - & \ccell{ 0}{0.1}{  0.0} & - & \ccell{ 0}{0.1}{  0.0} & - & \ccell{ 0}{0.1}{  0.0} & - & \ccell{ 0}{0.1}{  0.0} & - & \ccell{ 0}{0.1}{  0.0} & 10 & \ccell{10}{ 18}{55.6} &  8 & \ccell{ 8}{ 18}{44.4} & - & \ccell{ 0}{ 18}{ 0.0} \\ \hhline{*{32}{-}}

\multicolumn{2}{c}{Storage} & 10 & \ccell{10}{ 15}{66.7} &  5 & \ccell{ 5}{ 15}{33.3} & - & \ccell{ 0}{ 15}{ 0.0} & 10 & \ccell{10}{ 15}{66.7} &  5 & \ccell{ 5}{ 15}{33.3} & - & \ccell{ 0}{ 15}{ 0.0} & 10 & \ccell{10}{ 10}{100.0} & - & \ccell{ 0}{ 10}{ 0.0} & - & \ccell{ 0}{ 10}{ 0.0} & 10 & \ccell{10}{ 11}{90.9} &  1 & \ccell{ 1}{ 11}{ 9.1} & - & \ccell{ 0}{ 11}{ 0.0} & 10 & \ccell{10}{ 12}{83.3} &  2 & \ccell{ 2}{ 12}{16.7} & - & \ccell{ 0}{ 12}{ 0.0} \\ \hhline{*{32}{-}}

\multirow{8}{*}{Media}
 & Map        &  8 & \ccell{ 8}{ 56}{14.3} & 48 & \ccell{48}{ 56}{85.7} & - & \ccell{ 0}{ 56}{ 0.0} &  8 & \ccell{ 8}{ 56}{14.3} & 48 & \ccell{48}{ 56}{85.7} & - & \ccell{ 0}{ 56}{ 0.0} &  7 & \ccell{ 7}{  7}{100.0} & - & \ccell{ 0}{  7}{ 0.0} & - & \ccell{ 0}{  7}{ 0.0} &  6 & \ccell{ 6}{  6}{100.0} & - & \ccell{ 0}{  6}{ 0.0} & - & \ccell{ 0}{  6}{ 0.0} &  9 & \ccell{ 9}{ 25}{36.0} & 16 & \ccell{16}{ 25}{64.0} & - & \ccell{ 0}{ 25}{ 0.0} \\
 & Image      &  6 & \ccell{ 6}{ 10}{60.0} &  4 & \ccell{ 4}{ 10}{40.0} & - & \ccell{ 0}{ 10}{ 0.0} &  6 & \ccell{ 6}{ 10}{60.0} &  4 & \ccell{ 4}{ 10}{40.0} & - & \ccell{ 0}{ 10}{ 0.0} &  6 & \ccell{ 6}{  7}{85.7} &  1 & \ccell{ 1}{  7}{14.3} & - & \ccell{ 0}{  7}{ 0.0} &  5 & \ccell{ 5}{  6}{83.3} &  1 & \ccell{ 1}{  6}{16.7} & - & \ccell{ 0}{  6}{ 0.0} &  6 & \ccell{ 6}{ 10}{60.0} &  4 & \ccell{ 4}{ 10}{40.0} & - & \ccell{ 0}{ 10}{ 0.0} \\
 & Video      & 14 & \ccell{14}{ 40}{35.0} & 26 & \ccell{26}{ 40}{65.0} & - & \ccell{ 0}{ 40}{ 0.0} & 14 & \ccell{14}{ 44}{31.8} & 30 & \ccell{30}{ 44}{68.2} & - & \ccell{ 0}{ 44}{ 0.0} & 19 & \ccell{19}{ 20}{95.0} &  1 & \ccell{ 1}{ 20}{ 5.0} & - & \ccell{ 0}{ 20}{ 0.0} &  8 & \ccell{ 8}{ 10}{80.0} &  2 & \ccell{ 2}{ 10}{20.0} & - & \ccell{ 0}{ 10}{ 0.0} & 14 & \ccell{14}{ 22}{63.6} &  8 & \ccell{ 8}{ 22}{36.4} & - & \ccell{ 0}{ 22}{ 0.0} \\
 & Audio      & 64 & \ccell{64}{ 76}{84.2} &  9 & \ccell{ 9}{ 76}{11.8} &  3 & \ccell{ 3}{ 76}{ 3.9} & 64 & \ccell{64}{ 81}{79.0} & 14 & \ccell{14}{ 81}{17.3} &  3 & \ccell{ 3}{ 81}{ 3.7} & 44 & \ccell{44}{ 44}{100.0} & - & \ccell{ 0}{ 44}{ 0.0} & - & \ccell{ 0}{ 44}{ 0.0} & 44 & \ccell{44}{ 54}{81.5} & 10 & \ccell{10}{ 54}{18.5} & - & \ccell{ 0}{ 54}{ 0.0} & 61 & \ccell{61}{ 71}{85.9} & 10 & \ccell{10}{ 71}{14.1} & - & \ccell{ 0}{ 71}{ 0.0} \\
 & Live       & 26 & \ccell{26}{ 56}{46.4} & 30 & \ccell{30}{ 56}{53.6} & - & \ccell{ 0}{ 56}{ 0.0} & 26 & \ccell{26}{ 66}{39.4} & 40 & \ccell{40}{ 66}{60.6} & - & \ccell{ 0}{ 66}{ 0.0} &  8 & \ccell{ 8}{  8}{100.0} & - & \ccell{ 0}{  8}{ 0.0} & - & \ccell{ 0}{  8}{ 0.0} & 19 & \ccell{19}{ 19}{100.0} & - & \ccell{ 0}{ 19}{ 0.0} & - & \ccell{ 0}{ 19}{ 0.0} & 23 & \ccell{23}{ 40}{57.5} & 17 & \ccell{17}{ 40}{42.5} & - & \ccell{ 0}{ 40}{ 0.0} \\
 & Recorder   & 16 & \ccell{16}{ 19}{84.2} &  3 & \ccell{ 3}{ 19}{15.8} & - & \ccell{ 0}{ 19}{ 0.0} & 16 & \ccell{16}{ 19}{84.2} &  3 & \ccell{ 3}{ 19}{15.8} & - & \ccell{ 0}{ 19}{ 0.0} & 12 & \ccell{12}{ 12}{100.0} & - & \ccell{ 0}{ 12}{ 0.0} & - & \ccell{ 0}{ 12}{ 0.0} & 11 & \ccell{11}{ 12}{91.7} &  1 & \ccell{ 1}{ 12}{ 8.3} & - & \ccell{ 0}{ 12}{ 0.0} & 15 & \ccell{15}{ 17}{88.2} &  2 & \ccell{ 2}{ 17}{11.8} & - & \ccell{ 0}{ 17}{ 0.0} \\
 & Camera     &  9 & \ccell{ 9}{ 15}{60.0} &  6 & \ccell{ 6}{ 15}{40.0} & - & \ccell{ 0}{ 15}{ 0.0} &  9 & \ccell{ 9}{ 17}{52.9} &  8 & \ccell{ 8}{ 17}{47.1} & - & \ccell{ 0}{ 17}{ 0.0} &  9 & \ccell{ 9}{ 18}{50.0} &  9 & \ccell{ 9}{ 18}{50.0} & - & \ccell{ 0}{ 18}{ 0.0} & 20 & \ccell{20}{ 21}{95.2} &  1 & \ccell{ 1}{ 21}{ 4.8} & - & \ccell{ 0}{ 21}{ 0.0} &  4 & \ccell{ 4}{ 11}{36.4} &  7 & \ccell{ 7}{ 11}{63.6} & - & \ccell{ 0}{ 11}{ 0.0} \\
 & Misc       & 12 & \ccell{12}{ 16}{75.0} &  3 & \ccell{ 3}{ 16}{18.8} &  1 & \ccell{ 1}{ 16}{ 6.3} & 12 & \ccell{12}{ 16}{75.0} &  3 & \ccell{ 3}{ 16}{18.8} &  1 & \ccell{ 1}{ 16}{ 6.3} & 18 & \ccell{18}{ 18}{100.0} & - & \ccell{ 0}{ 18}{ 0.0} & - & \ccell{ 0}{ 18}{ 0.0} & - & \ccell{ 0}{0.1}{  0.0} & - & \ccell{ 0}{0.1}{  0.0} & - & \ccell{ 0}{0.1}{  0.0} &  6 & \ccell{ 6}{  6}{100.0} & - & \ccell{ 0}{  6}{ 0.0} & - & \ccell{ 0}{  6}{ 0.0} \\ \hhline{*{32}{-}}

\multicolumn{2}{c}{Location} &  3 & \ccell{ 3}{  7}{42.9} &  4 & \ccell{ 4}{  7}{57.1} & - & \ccell{ 0}{  7}{ 0.0} &  3 & \ccell{ 3}{  7}{42.9} &  4 & \ccell{ 4}{  7}{57.1} & - & \ccell{ 0}{  7}{ 0.0} &  7 & \ccell{ 7}{  7}{100.0} & - & \ccell{ 0}{  7}{ 0.0} & - & \ccell{ 0}{  7}{ 0.0} &  3 & \ccell{ 3}{  3}{100.0} & - & \ccell{ 0}{  3}{ 0.0} & - & \ccell{ 0}{  3}{ 0.0} &  3 & \ccell{ 3}{  3}{100.0} & - & \ccell{ 0}{  3}{ 0.0} & - & \ccell{ 0}{  3}{ 0.0} \\ \hhline{*{32}{-}}

\multicolumn{2}{c}{Share} &  4 & \ccell{ 4}{ 12}{33.3} &  7 & \ccell{ 7}{ 12}{58.3} &  1 & \ccell{ 1}{ 12}{ 8.3} &  4 & \ccell{ 4}{ 24}{16.7} & 19 & \ccell{19}{ 24}{79.2} &  1 & \ccell{ 1}{ 24}{ 4.2} &  3 & \ccell{ 3}{  3}{100.0} & - & \ccell{ 0}{  3}{ 0.0} & - & \ccell{ 0}{  3}{ 0.0} &  5 & \ccell{ 5}{  7}{71.4} &  2 & \ccell{ 2}{  7}{28.6} & - & \ccell{ 0}{  7}{ 0.0} &  5 & \ccell{ 5}{ 14}{35.7} &  9 & \ccell{ 9}{ 14}{64.3} & - & \ccell{ 0}{ 14}{ 0.0} \\ \hhline{*{32}{-}}

\multicolumn{2}{c}{Canvas} & 60 & \ccell{60}{ 81}{74.1} & 21 & \ccell{21}{ 81}{25.9} & - & \ccell{ 0}{ 81}{ 0.0} & 60 & \ccell{60}{ 81}{74.1} & 21 & \ccell{21}{ 81}{25.9} & - & \ccell{ 0}{ 81}{ 0.0} & 46 & \ccell{46}{ 50}{92.0} &  4 & \ccell{ 4}{ 50}{ 8.0} & - & \ccell{ 0}{ 50}{ 0.0} & 49 & \ccell{49}{ 50}{98.0} &  1 & \ccell{ 1}{ 50}{ 2.0} & - & \ccell{ 0}{ 50}{ 0.0} & 48 & \ccell{48}{ 52}{92.3} &  4 & \ccell{ 4}{ 52}{ 7.7} & - & \ccell{ 0}{ 52}{ 0.0} \\ \hhline{*{32}{-}}

\multicolumn{2}{c}{File} & 39 & \ccell{39}{ 40}{97.5} &  1 & \ccell{ 1}{ 40}{ 2.5} & - & \ccell{ 0}{ 40}{ 0.0} & 39 & \ccell{39}{ 42}{92.9} &  3 & \ccell{ 3}{ 42}{ 7.1} & - & \ccell{ 0}{ 42}{ 0.0} & 35 & \ccell{35}{ 35}{100.0} & - & \ccell{ 0}{ 35}{ 0.0} & - & \ccell{ 0}{ 35}{ 0.0} & 34 & \ccell{34}{ 35}{97.1} &  1 & \ccell{ 1}{ 35}{ 2.9} & - & \ccell{ 0}{ 35}{ 0.0} & 37 & \ccell{37}{ 38}{97.4} &  1 & \ccell{ 1}{ 38}{ 2.6} & - & \ccell{ 0}{ 38}{ 0.0} \\ \hhline{*{32}{-}}

\multirow{7}{*}{Open API}
 & Login      &  2 & \ccell{ 2}{  2}{100.0} & - & \ccell{ 0}{  2}{ 0.0} & - & \ccell{ 0}{  2}{ 0.0} &  5 & \ccell{ 5}{  6}{83.3} &  1 & \ccell{ 1}{  6}{16.7} & - & \ccell{ 0}{  6}{ 0.0} &  3 & \ccell{ 3}{  7}{42.9} &  1 & \ccell{ 1}{  7}{14.3} &  3 & \ccell{ 3}{  7}{42.9} &  2 & \ccell{ 2}{  2}{100.0} & - & \ccell{ 0}{  2}{ 0.0} & - & \ccell{ 0}{  2}{ 0.0} &  2 & \ccell{ 2}{  2}{100.0} & - & \ccell{ 0}{  2}{ 0.0} & - & \ccell{ 0}{  2}{ 0.0} \\
 & Navigate   &  2 & \ccell{ 2}{  6}{33.3} &  2 & \ccell{ 2}{  6}{33.3} &  2 & \ccell{ 2}{  6}{33.3} &  2 & \ccell{ 2}{  9}{22.2} &  5 & \ccell{ 5}{  9}{55.6} &  2 & \ccell{ 2}{  9}{22.2} &  3 & \ccell{ 3}{  3}{100.0} & - & \ccell{ 0}{  3}{ 0.0} & - & \ccell{ 0}{  3}{ 0.0} &  7 & \ccell{ 7}{  7}{100.0} & - & \ccell{ 0}{  7}{ 0.0} & - & \ccell{ 0}{  7}{ 0.0} &  2 & \ccell{ 2}{  4}{50.0} &  1 & \ccell{ 1}{  4}{25.0} &  1 & \ccell{ 1}{  4}{25.0} \\
 & User Info  &  2 & \ccell{ 2}{ 12}{16.7} &  7 & \ccell{ 7}{ 12}{58.3} &  3 & \ccell{ 3}{ 12}{25.0} &  5 & \ccell{ 5}{ 21}{23.8} & 13 & \ccell{13}{ 21}{61.9} &  3 & \ccell{ 3}{ 21}{14.3} &  1 & \ccell{ 1}{ 10}{10.0} &  6 & \ccell{ 6}{ 10}{60.0} &  3 & \ccell{ 3}{ 10}{30.0} &  2 & \ccell{ 2}{ 15}{13.3} & 13 & \ccell{13}{ 15}{86.7} & - & \ccell{ 0}{ 15}{ 0.0} &  2 & \ccell{ 2}{  7}{28.6} &  4 & \ccell{ 4}{  7}{57.1} &  1 & \ccell{ 1}{  7}{14.3} \\
 & Payment    &  1 & \ccell{ 1}{ 29}{ 3.4} & 13 & \ccell{13}{ 29}{44.8} & 15 & \ccell{15}{ 29}{51.7} &  1 & \ccell{ 1}{ 31}{ 3.2} & 15 & \ccell{15}{ 31}{48.4} & 15 & \ccell{15}{ 31}{48.4} &  1 & \ccell{ 1}{  2}{50.0} & - & \ccell{ 0}{  2}{ 0.0} &  1 & \ccell{ 1}{  2}{50.0} &  1 & \ccell{ 1}{  3}{33.3} &  1 & \ccell{ 1}{  3}{33.3} &  1 & \ccell{ 1}{  3}{33.3} &  2 & \ccell{ 2}{  9}{22.2} &  7 & \ccell{ 7}{  9}{77.8} & - & \ccell{ 0}{  9}{ 0.0} \\
 & Bio-Auth   &  3 & \ccell{ 3}{ 11}{27.3} &  3 & \ccell{ 3}{ 11}{27.3} &  5 & \ccell{ 5}{ 11}{45.5} &  3 & \ccell{ 3}{ 14}{21.4} &  6 & \ccell{ 6}{ 14}{42.9} &  5 & \ccell{ 5}{ 14}{35.7} & - & \ccell{ 0}{0.1}{  0.0} & - & \ccell{ 0}{0.1}{  0.0} & - & \ccell{ 0}{0.1}{  0.0} & - & \ccell{ 0}{  1}{ 0.0} &  1 & \ccell{ 1}{  1}{100.0} & - & \ccell{ 0}{  1}{ 0.0} &  3 & \ccell{ 3}{  3}{100.0} & - & \ccell{ 0}{  3}{ 0.0} & - & \ccell{ 0}{  3}{ 0.0} \\
 & Enterprise & - & \ccell{ 0}{  1}{ 0.0} &  1 & \ccell{ 1}{  1}{100.0} & - & \ccell{ 0}{  1}{ 0.0} &  5 & \ccell{ 5}{ 28}{17.9} &  6 & \ccell{ 6}{ 28}{21.4} & 17 & \ccell{17}{ 28}{60.7} & - & \ccell{ 0}{0.1}{  0.0} & - & \ccell{ 0}{0.1}{  0.0} & - & \ccell{ 0}{0.1}{  0.0} & - & \ccell{ 0}{0.1}{  0.0} & - & \ccell{ 0}{0.1}{  0.0} & - & \ccell{ 0}{0.1}{  0.0} & - & \ccell{ 0}{0.1}{  0.0} & - & \ccell{ 0}{0.1}{  0.0} & - & \ccell{ 0}{0.1}{  0.0} \\
 & Misc       & 14 & \ccell{14}{ 72}{19.4} & 42 & \ccell{42}{ 72}{58.3} & 16 & \ccell{16}{ 72}{22.2} & 14 & \ccell{14}{ 84}{16.7} & 54 & \ccell{54}{ 84}{64.3} & 16 & \ccell{16}{ 84}{19.0} & 16 & \ccell{16}{ 28}{57.1} &  2 & \ccell{ 2}{ 28}{ 7.1} & 10 & \ccell{10}{ 28}{35.7} & 25 & \ccell{25}{ 45}{55.6} & 20 & \ccell{20}{ 45}{44.4} & - & \ccell{ 0}{ 45}{ 0.0} & 12 & \ccell{12}{ 92}{13.0} & 78 & \ccell{78}{ 92}{84.8} &  2 & \ccell{ 2}{ 92}{ 2.2} \\ \hhline{*{32}{-}}

\multirow{7}{*}{Device} 
 & Wi-Fi      &  9 & \ccell{ 9}{  9}{100.0} & - & \ccell{ 0}{  9}{ 0.0} & - & \ccell{ 0}{  9}{ 0.0} &  9 & \ccell{ 9}{  9}{100.0} & - & \ccell{ 0}{  9}{ 0.0} & - & \ccell{ 0}{  9}{ 0.0} & 10 & \ccell{10}{ 10}{100.0} & - & \ccell{ 0}{ 10}{ 0.0} & - & \ccell{ 0}{ 10}{ 0.0} &  4 & \ccell{ 4}{  4}{100.0} & - & \ccell{ 0}{  4}{ 0.0} & - & \ccell{ 0}{  4}{ 0.0} &  9 & \ccell{ 9}{  9}{100.0} & - & \ccell{ 0}{  9}{ 0.0} & - & \ccell{ 0}{  9}{ 0.0} \\
 & Bluetooth  & 18 & \ccell{18}{ 30}{60.0} & 11 & \ccell{11}{ 30}{36.7} &  1 & \ccell{ 1}{ 30}{ 3.3} & 18 & \ccell{18}{ 31}{58.1} & 12 & \ccell{12}{ 31}{38.7} &  1 & \ccell{ 1}{ 31}{ 3.2} & - & \ccell{ 0}{0.1}{  0.0} & - & \ccell{ 0}{0.1}{  0.0} & - & \ccell{ 0}{0.1}{  0.0} & - & \ccell{ 0}{0.1}{  0.0} & - & \ccell{ 0}{0.1}{  0.0} & - & \ccell{ 0}{0.1}{  0.0} & 18 & \ccell{18}{ 18}{100.0} & - & \ccell{ 0}{ 18}{ 0.0} & - & \ccell{ 0}{ 18}{ 0.0} \\
 & Contact    &  1 & \ccell{ 1}{ 10}{10.0} &  5 & \ccell{ 5}{ 10}{50.0} &  4 & \ccell{ 4}{ 10}{40.0} &  1 & \ccell{ 1}{ 11}{ 9.1} &  6 & \ccell{ 6}{ 11}{54.5} &  4 & \ccell{ 4}{ 11}{36.4} &  1 & \ccell{ 1}{  3}{33.3} &  2 & \ccell{ 2}{  3}{66.7} & - & \ccell{ 0}{  3}{ 0.0} & - & \ccell{ 0}{0.1}{  0.0} & - & \ccell{ 0}{0.1}{  0.0} & - & \ccell{ 0}{0.1}{  0.0} &  1 & \ccell{ 1}{  4}{25.0} &  2 & \ccell{ 2}{  4}{50.0} &  1 & \ccell{ 1}{  4}{25.0} \\
 & NFC        &  5 & \ccell{ 5}{ 19}{26.3} & 14 & \ccell{14}{ 19}{73.7} & - & \ccell{ 0}{ 19}{ 0.0} &  9 & \ccell{ 9}{ 23}{39.1} & 14 & \ccell{14}{ 23}{60.9} & - & \ccell{ 0}{ 23}{ 0.0} & - & \ccell{ 0}{0.1}{  0.0} & - & \ccell{ 0}{0.1}{  0.0} & - & \ccell{ 0}{0.1}{  0.0} & - & \ccell{ 0}{0.1}{  0.0} & - & \ccell{ 0}{0.1}{  0.0} & - & \ccell{ 0}{0.1}{  0.0} &  5 & \ccell{ 5}{  5}{100.0} & - & \ccell{ 0}{  5}{ 0.0} & - & \ccell{ 0}{  5}{ 0.0} \\
 & Screen     &  4 & \ccell{ 4}{ 11}{36.4} &  6 & \ccell{ 6}{ 11}{54.5} &  1 & \ccell{ 1}{ 11}{ 9.1} &  4 & \ccell{ 4}{ 11}{36.4} &  6 & \ccell{ 6}{ 11}{54.5} &  1 & \ccell{ 1}{ 11}{ 9.1} &  3 & \ccell{ 3}{  3}{100.0} & - & \ccell{ 0}{  3}{ 0.0} & - & \ccell{ 0}{  3}{ 0.0} &  9 & \ccell{ 9}{  9}{100.0} & - & \ccell{ 0}{  9}{ 0.0} & - & \ccell{ 0}{  9}{ 0.0} &  4 & \ccell{ 4}{  4}{100.0} & - & \ccell{ 0}{  4}{ 0.0} & - & \ccell{ 0}{  4}{ 0.0} \\
 & Phone      &  1 & \ccell{ 1}{ 23}{ 4.3} & 21 & \ccell{21}{ 23}{91.3} &  1 & \ccell{ 1}{ 23}{ 4.3} &  1 & \ccell{ 1}{ 23}{ 4.3} & 21 & \ccell{21}{ 23}{91.3} &  1 & \ccell{ 1}{ 23}{ 4.3} &  1 & \ccell{ 1}{  1}{100.0} & - & \ccell{ 0}{  1}{ 0.0} & - & \ccell{ 0}{  1}{ 0.0} &  1 & \ccell{ 1}{  1}{100.0} & - & \ccell{ 0}{  1}{ 0.0} & - & \ccell{ 0}{  1}{ 0.0} &  1 & \ccell{ 1}{  2}{50.0} &  1 & \ccell{ 1}{  2}{50.0} & - & \ccell{ 0}{  2}{ 0.0} \\
 & Misc       & 28 & \ccell{28}{ 44}{63.6} & 15 & \ccell{15}{ 44}{34.1} &  1 & \ccell{ 1}{ 44}{ 2.3} & 28 & \ccell{28}{ 47}{59.6} & 18 & \ccell{18}{ 47}{38.3} &  1 & \ccell{ 1}{ 47}{ 2.1} & 21 & \ccell{21}{ 26}{80.8} &  5 & \ccell{ 5}{ 26}{19.2} & - & \ccell{ 0}{ 26}{ 0.0} & 16 & \ccell{16}{ 23}{69.6} &  7 & \ccell{ 7}{ 23}{30.4} & - & \ccell{ 0}{ 23}{ 0.0} & 28 & \ccell{28}{ 34}{82.4} &  6 & \ccell{ 6}{ 34}{17.6} & - & \ccell{ 0}{ 34}{ 0.0} \\ \hhline{*{32}{-}}

\multirow{2}{*}{AI}
 & CV         & 19 & \ccell{19}{ 19}{100.0} & - & \ccell{ 0}{ 19}{ 0.0} & - & \ccell{ 0}{ 19}{ 0.0} & 19 & \ccell{19}{ 19}{100.0} & - & \ccell{ 0}{ 19}{ 0.0} & - & \ccell{ 0}{ 19}{ 0.0} & 18 & \ccell{18}{ 20}{90.0} &  2 & \ccell{ 2}{ 20}{10.0} & - & \ccell{ 0}{ 20}{ 0.0} & - & \ccell{ 0}{0.1}{  0.0} & - & \ccell{ 0}{0.1}{  0.0} & - & \ccell{ 0}{0.1}{  0.0} & - & \ccell{ 0}{0.1}{  0.0} & - & \ccell{ 0}{0.1}{  0.0} & - & \ccell{ 0}{0.1}{  0.0} \\
 & Misc       & - & \ccell{ 0}{0.1}{  0.0} & - & \ccell{ 0}{0.1}{  0.0} & - & \ccell{ 0}{0.1}{  0.0} & - & \ccell{ 0}{  1}{ 0.0} &  1 & \ccell{ 1}{  1}{100.0} & - & \ccell{ 0}{  1}{ 0.0} & 11 & \ccell{11}{ 11}{100.0} & - & \ccell{ 0}{ 11}{ 0.0} & - & \ccell{ 0}{ 11}{ 0.0} &  7 & \ccell{ 7}{  7}{100.0} & - & \ccell{ 0}{  7}{ 0.0} & - & \ccell{ 0}{  7}{ 0.0} & - & \ccell{ 0}{0.1}{  0.0} & - & \ccell{ 0}{0.1}{  0.0} & - & \ccell{ 0}{0.1}{  0.0} \\ \hhline{*{32}{-}}

\multicolumn{2}{c}{AD} & 19 & \ccell{19}{ 20}{95.0} &  1 & \ccell{ 1}{ 20}{ 5.0} & - & \ccell{ 0}{ 20}{ 0.0} & 19 & \ccell{19}{ 20}{95.0} &  1 & \ccell{ 1}{ 20}{ 5.0} & - & \ccell{ 0}{ 20}{ 0.0} &  9 & \ccell{ 9}{ 14}{64.3} &  4 & \ccell{ 4}{ 14}{28.6} &  1 & \ccell{ 1}{ 14}{ 7.1} & 13 & \ccell{13}{ 21}{61.9} &  8 & \ccell{ 8}{ 21}{38.1} & - & \ccell{ 0}{ 21}{ 0.0} &  3 & \ccell{ 3}{ 12}{25.0} &  9 & \ccell{ 9}{ 12}{75.0} & - & \ccell{ 0}{ 12}{ 0.0} \\ \hhline{*{32}{-}}

\multicolumn{2}{c}{Uncategorized} & 30 & \ccell{30}{ 78}{38.5} & 47 & \ccell{47}{ 78}{60.3} &  1 & \ccell{ 1}{ 78}{ 1.3} & 30 & \ccell{30}{ 82}{36.6} & 51 & \ccell{51}{ 82}{62.2} &  1 & \ccell{ 1}{ 82}{ 1.2} & 15 & \ccell{15}{ 28}{53.6} & 10 & \ccell{10}{ 28}{35.7} &  3 & \ccell{ 3}{ 28}{10.7} & 17 & \ccell{17}{ 25}{68.0} &  7 & \ccell{ 7}{ 25}{28.0} &  1 & \ccell{ 1}{ 25}{ 4.0} & 34 & \ccell{34}{ 50}{68.0} & 15 & \ccell{15}{ 50}{30.0} &  1 & \ccell{ 1}{ 50}{ 2.0} \\

\midrule[0.5pt]

\multicolumn{2}{c}{\textbf{All}}  & 590 & \ccell{590}{1157}{51.0} & 502 & \ccell{502}{1157}{43.4} & 65 & \ccell{65}{1157}{ 5.6} & 606 & \ccell{606}{1281}{47.3} & 593 & \ccell{593}{1281}{46.3} & 82 & \ccell{82}{1281}{ 6.4} & 464 & \ccell{464}{602}{77.1} & 113 & \ccell{113}{602}{18.8} & 25 & \ccell{25}{602}{ 4.2} & 383 & \ccell{383}{505}{75.8} & 120 & \ccell{120}{505}{23.8} &  2 & \ccell{ 2}{505}{ 0.4} & 506 & \ccell{506}{807}{62.7} & 295 & \ccell{295}{807}{36.6} &  6 & \ccell{ 6}{807}{ 0.7} \\

\bottomrule[1.5pt]
\end{tabular}
 
\caption{Categories of Documented and Undocumented APIs. ``D'' means documented APIs;  ``UU'' means undocumented unchecked APIs; ``UC'' means undocumented checked APIs.}  
 
\label{tab:compare:api}
 \vspace{-0.25in}
\end{table*}


The effectiveness evaluation aims to quantify how \sysname uncovered the hidden APIs in terms of the specific numbers for the involved analysis (which is presented in~\autoref{tab:effectiveness}), and their qualities (i.e., whether there are any false positives). It is worth noting that the manually created cases are indeed rare. For example, for Baidu, we automatically created 423 test cases, and created another 56 test cases manually, so the manual efforts are around 11\%, i.e., 56/(56+423) = 0.11. Other super apps even have a lower amount of manual efforts than Baidu (e.g., WeCom has 2.9 \% manual efforts).

Specifically, the effectiveness of our static analysis is measured by the identification of API invariants, the number of identified API candidates (i.e., the functions that are very likely to be APIs). However, whether those API candidates are really APIs are determined in dynamic API classification. For the API invariants, while we have listed four invariants in \S\ref{subsce:apifinder}, not all of them will exist in all super apps (e.g., {\sf Baidu} and {\sf QQ} do not have caller invariant), as shown in~\autoref{tab:effectiveness}. That is why \sysname aggressively identifies as many invariants as possible. With these invariants, it sufficiently recognizes the undocumented APIs even though some of them do not exist in other super apps. 
During static API recognition, \sysname recognized in total { 1,829} API candidates for these super apps. Among them, \textsf{WeCom} contains the most hidden API candidates (683), followed by \wechat (containing 575 API candidates). \textsf{Tiktok} has fewer API candidates (i.e., 124 API candidates), likely due to its smallest LoC compared to other super apps. 
 
The effectiveness of dynamic analysis is measured by the number of traced functions during API invocation identification and the number of test cases used during API classification. Among the test cases, we also quantify the number of automatically generated test cases and manually created test cases. We can see that most of the test cases are automatically generated by our test case generation algorithm, and the number of automatically generated test cases is greater than the number of API candidates due to the parameter order permutation (as discussed in \S\ref{subsce:apiexploiter}). With our dynamic classification for the identified APIs, \sysname detected a large number of hidden APIs, many of which are unchecked (as reported in \autoref{tab:effectiveness}). \wechat has more APIs (590 public APIs, 502 undocumented unchecked APIs, and 65 undocumented checked APIs) than the other super apps. However, \textsf{TikTok} has a relatively small number of APIs (383 public APIs, 120 undocumented unchecked APIs, and 2 undocumented checked APIs). With respect to the percentage of undocumented unchecked and checked APIs, \textsf{WeCom} has the most undocumented unchecked APIs (46.3\%) and undocumented checked APIs (6.4\%).


    




\paragraph{Correctness of Our Result}
We quantify whether there are any false positives or false positives for the identified hidden APIs. First, a false positive here means that the identified API is not hidden, or is not an API. By design, \sysname will not have false positives for two reasons: (1) the invariants we extracted have very strict patterns (they have to exist among all public APIs and all of them have to be present in the undocumented APIs), and (2) our dynamic probing for API classification can filter out those non-APIs, which eliminate potential false positives. Nevertheless, we still thoroughly scrutinized each API identified for \wechat by conducting a manual check to ensure that there were no false positives. In other words, we made sure that the tool did not mistakenly classify non-APIs as APIs. Thanks to our design, we did not come across any false positives during our examination.
Second, with respect to false negatives (i.e., ``true'' hidden API is missed by \sysname), we note that theoretically \sysname could have false negatives, for instance, if our invariants are too strong. However, we will not be able to quantify this, since we do not have the ground truth, unless we can manually examine each line of code. Therefore, we leave this to future work. \looseness=-1

\ignore{
\sysname has a false negative if it misses a ``true'' API. Since we do not have the source code and complete picture of the super app implementation, this is nearly impossible for us to quantify the false negatives. Nonetheless, we still inspected the decompiled code of \wechat and \textsf{Baidu}, and found no false negatives, at least for the code we analyzed. This is likely because 
%
super app developers follow the same practice (e.g., using the same superclass) when implementing both documented APIs and undocumented APIs as discussed in \S\ref{sub:observation}. However, we do not claim we find all ``true'' APIs, as it may be possible for a super app to use some other nontraditional way of invoking the layer below code (e.g., the APIs implemented by Java or native code). 
}

\paragraph{Categories of the Identified APIs} With the identified APIs, we can then obtain some insights with them, such as which category contains more hidden APIs. To this end, we manually walked through each API, and categorize them based on the categories of the documented ones, to classify the undocumented (i.e., hidden) APIs. This result is presented in ~\autoref{tab:compare:api}. Interestingly, we found that most of the categories contain undocumented unchecked APIs. In particular, for some of the super apps (e.g., \wechat), their undocumented unchecked APIs can be even more than the documented APIs in some of the categories (e.g., the API category \textsf{Payment} has 28 undocumented APIs, which is way more than their documented APIs). Finally, we found that some well-documented APIs of a specific super app may not be open to the public in other super apps. For example, \texttt{getUserInfo} is an undocumented API of \textsf{Baidu}, while \wechat has the same API with the same functionalities, which is publicly accessible. Finally, since \sysname is a systematic and mostly automated tool, it can inspect API changes based on previous versions of the super app implementations as long as we can obtain both their APKs and documentation. We have a detailed evaluation of API evaluation in Appendix-\S\ref{sec:apievolution} for interested readers. \looseness=-1

\ignore{
\sysname has a false negative if it misses a ``true'' API. Since we do not have the source code and complete picture of the super app implementation, this is nearly impossible for us to quantify the false negatives. Nonetheless, we still inspected the decompiled code of \wechat and \textsf{Baidu}, and found no false negatives, at least for the code we analyzed. This is likely because 
%
super app developers follow the same practice (e.g., using the same superclass) when implementing both documented APIs and undocumented APIs as discussed in \S\ref{sub:observation}. However, we do not claim we find all ``true'' APIs, as it may be possible for a super app to use some other nontraditional way of invoking the layer below code (e.g., the APIs implemented by Java or native code)
 }

\paragraph{Usage of Hidden APIs (Among the 1st-party Miniapps)} 
We obtained many 1st-party miniapps and classified them into categories based on their meta-data. From the data in \autoref{tab:cat_score}, we found that the use of undocumented APIs is common among 1st-party miniapps regardless of their category. \textsf{WeCom} had the highest percentage of 1st-party miniapps using undocumented APIs at 38.1\%, followed by \textsf{QQ} at 37.5\%, \textsf{WeChat} at 36.9\%, and \textsf{Baidu} at 34.7\%. We also observed that 1st-party miniapps in the Traveling, Shopping, and Finance categories were more likely to use undocumented APIs, and these APIs were often related to payment. For example, many miniapps in these categories would use the undocumented API \texttt{verifyPaymentPassword} to verify payment passwords.\looseness=-1


 


\begin{table}[]
\scriptsize
\belowrulesep=0pt
\aboverulesep=0pt
\setlength\tabcolsep{0.2pt}
\centering
\begin{tabular}{l|rrr|rrr|rrr|rrr}

\toprule[1.5pt]

    \multicolumn{1}{c|}{
        \multirow{2}{*}{
            \textbf{Category}
        }
    } & 
    \multicolumn{3}{c|}{
        \textbf{\wechat}
    } &
    \multicolumn{3}{c|}{
        \textbf{\textsf{WeCom}}
    } & 
    \multicolumn{3}{c|}{
        \textbf{\textsf{Baidu}}
    } &
    \multicolumn{3}{c}{
        \textbf{\textsf{QQ}}
    } \\ \cline{2-13} 

    \multicolumn{1}{c|}{} & 
    \multicolumn{1}{c|}{\textbf{\# U}} & 
    \multicolumn{1}{c|}{\textbf{\# App}} & 
    \multicolumn{1}{c|}{\textbf{\%}}  & 
    
    \multicolumn{1}{c|}{\textbf{\# U}} & 
    \multicolumn{1}{c|}{\textbf{\# App}} & 
    \multicolumn{1}{c|}{\textbf{\%}} & 
    
    \multicolumn{1}{c|}{\textbf{\# U}} & 
    \multicolumn{1}{c|}{\textbf{\# App}} & 
    \multicolumn{1}{c|}{\textbf{\%}} &
    
    \multicolumn{1}{c|}{\textbf{\# U}} & 
    \multicolumn{1}{c|}{\textbf{\# App}} & 
    \multicolumn{1}{c}{\textbf{\%}} \\

\midrule[0.7pt]

Business      & 14 & 49 & \ccell{14}{49}{28.6} & 16 & 49 & \ccell{16}{49}{32.7} & 21 & 38 & \ccell{21}{38}{55.3} & 1 & 3 & \ccell{ 1}{ 3}{33.3} \\
Education     &  6 & 26 & \ccell{ 6}{26}{23.1} &  7 & 26 & \ccell{ 7}{26}{26.9} &  5 & 16 & \ccell{ 5}{16}{31.3} & - & 3 &                  0.0 \\
E-learning    &  5 &  9 & \ccell{ 5}{ 9}{55.6} &  5 &  9 & \ccell{ 5}{ 9}{55.6} & 12 & 33 & \ccell{12}{33}{36.4} & - & 1 &                  0.0 \\
Entertainment &  9 & 17 & \ccell{ 9}{17}{52.9} &  9 & 17 & \ccell{ 9}{17}{52.9} & 29 & 75 & \ccell{29}{75}{38.7} & 2 & 2 &\ccell{ 2}{ 2}{100.0} \\
Finance       &  1 &  1 &\ccell{ 1}{ 1}{100.0} &  1 &  1 &\ccell{ 1}{ 1}{100.0} & 21 & 23 & \ccell{21}{23}{91.3} & - & - &                  0.0 \\
Food          &  - &  - &                  0.0 &  - &  - &                  0.0 &  - &  5 &                  0.0 & - & - &                  0.0 \\
Games         & 18 & 36 & \ccell{18}{36}{50.0} & 18 & 36 & \ccell{18}{36}{50.0} &  - &  - &                  0.0 & - & - &                  0.0 \\
Government    &  2 &  7 & \ccell{ 2}{ 7}{28.6} &  2 &  7 & \ccell{ 2}{ 7}{28.6} &  3 &  8 & \ccell{ 3}{ 8}{37.5} & 1 & 1 &\ccell{ 1}{ 1}{100.0} \\
Health        &  2 &  7 & \ccell{ 2}{ 7}{28.6} &  2 &  7 & \ccell{ 2}{ 7}{28.6} &  1 &  5 & \ccell{ 1}{ 5}{20.0} & - & 1 &                  0.0 \\
Job           &  - &  1 &                  0.0 &  - &  1 &                  0.0 &  - &  - &                  0.0 & - & - &                  0.0 \\
Lifestyle     &  2 &  5 & \ccell{ 2}{ 5}{40.0} &  2 &  5 & \ccell{ 2}{ 5}{40.0} &  3 & 15 & \ccell{ 3}{15}{20.0} & - & 1 &                  0.0 \\
Photo         &  3 &  7 & \ccell{ 3}{ 7}{42.9} &  3 &  7 & \ccell{ 3}{ 7}{42.9} &  - &  - &                  0.0 & - & - &                  0.0 \\
Shopping      &  1 &  1 &\ccell{ 1}{ 1}{100.0} &  1 &  1 &\ccell{ 1}{ 1}{100.0} &  - &  2 &                  0.0 & - & - &                  0.0 \\
Social        &  4 &  8 & \ccell{ 4}{ 8}{50.0} &  4 &  8 & \ccell{ 4}{ 8}{50.0} &  1 &  4 & \ccell{ 1}{ 4}{25.0} & - & 1 &                  0.0 \\
Sports        &  - &  - &                  0.0 &  - &  - &                  0.0 &  - &  1 &                  0.0 & - & - &                  0.0 \\
Tool          & 15 & 55 & \ccell{15}{55}{27.3} & 15 & 55 & \ccell{15}{55}{27.3} & 16 & 47 & \ccell{16}{47}{34.0} & 4 & 8 & \ccell{ 4}{ 8}{50.0} \\
Traffic       &  3 &  5 & \ccell{ 3}{ 5}{60.0} &  3 &  5 & \ccell{ 3}{ 5}{60.0} &  4 & 10 & \ccell{ 4}{10}{40.0} & - & 1 &                  0.0 \\
Travelling    &  2 &  2 &\ccell{ 2}{ 2}{100.0} &  2 &  2 &\ccell{ 2}{ 2}{100.0} &  1 & 56 & \ccell{ 1}{56}{ 1.8} & 1 & 2 & \ccell{ 1}{ 2}{50.0} \\
Uncategorized &  - &  - &                  0.0 &  - &  - &                  0.0 &  1 &  2 & \ccell{ 1}{ 2}{50.0} & - & - &                  0.0 \\

\midrule[0.7pt]

Total         & 87 &236 & \ccell{87}{236}{36.9} & 90 &236 & \ccell{90}{236}{38.1} &118 &340 & \ccell{118}{340}{34.7} & 9 & 24& \ccell{9}{24}{37.5} \\

\toprule[1.5pt]

\end{tabular}
  
\caption{The 1st party miniapps that have used the undocumented APIs.  The first column indicates the number of 1st-party mini-apps using undocumented APIs, and the second column represents the total number of 1st-party mini-apps. We calculate the percentage of mini-apps by using the first column divided by the second.
}
    \label{tab:cat_score} 
 \vspace{-0.1in}
\end{table}


\begin{table}[t]
\centering
\scriptsize
 \setlength\tabcolsep{2pt}
    \aboverulesep=0pt
    \belowrulesep=0pt
\begin{tabular}{cllrrc}
\toprule[1.5pt]

    & \textbf{API Name} & \textbf{Category} & \textbf{\# App} & \textbf{\% *App} & \textbf{w/ Check} \\

\midrule[0.7pt]




\multirow{7}{*}{\rotatebox[origin=c]{90}{\textbf{Baidu}}} &
  swan.button           & Interaction    & 104 & \ccell{104}{118}{88.14} & \xmark \\
& swan.login            & Login          & 31  & \ccell{ 31}{118}{26.27} & \cmark \\
& swan.postMessage      & Uncategorized  & 8   & \ccell{  8}{118}{ 6.78} & \xmark \\
& swan.getBDUSS         & User Info      & 4   & \ccell{  4}{118}{ 3.39} & \cmark \\
& swan.getCommonSysInfo & System         & 3   & \ccell{  3}{118}{ 2.54} & \cmark \\
& swan.getUserInfo      & User Info      & 3   & \ccell{  3}{118}{ 2.54} & \xmark \\
& swan.getChannelID     & Uncategorized  & 2   & \ccell{  2}{118}{ 1.69} & \cmark \\

\midrule[0.7pt]

%

\multirow{21}{*}{\rotatebox[origin=c]{90}{\textbf{WeChat}}} &
  wx.hideNavigationBar                          & Bar             & 28 & \ccell{28}{87}{32.18} & \xmark \\
& wx.requestSubscribeMessage                    & Subscribe       & 25 & \ccell{25}{87}{28.74} & \xmark \\
& wx.showNavigationBar                          & Bar             & 23 & \ccell{23}{87}{26.44} & \xmark \\
& wx.requestVirtualPayment                      & Payment         & 11 & \ccell{11}{87}{12.64} & \cmark \\
& wx.openUrl                                    & Misc            & 8  & \ccell{ 8}{87}{ 9.20} & \cmark \\
& wx.hideHomeButton                             & Interaction     & 8  & \ccell{ 5}{87}{ 9.20} & \xmark \\
& wx.enterContact                               & Contact         & 5  & \ccell{ 5}{87}{ 5.75} & \cmark \\
& wx.drawCanvas                                 & Canvas          & 5  & \ccell{ 5}{87}{ 5.75} & \xmark \\
& wx.setPageOrientation                         & Misc            & 4  & \ccell{ 4}{87}{ 4.60} & \xmark \\
& wx.operateWXData                              & Misc            & 4  & \ccell{ 4}{87}{ 4.60} & \xmark \\
& wx.getBackgroundFetchData                     & Misc            & 3  & \ccell{ 3}{87}{ 3.45} & \xmark \\
& wx.setBackgroundFetchToken                    & Misc            & 3  & \ccell{ 3}{87}{ 3.45} & \xmark \\
& wx.startFacialRecognitionVerify               & Bio-Auth        & 3  & \ccell{ 3}{87}{ 3.45} & \cmark \\
& wx.checkIsSupportFacialRecognition            & Bio-Auth        & 2  & \ccell{ 2}{87}{ 2.30} & \cmark \\
& wx.navigateBackApplication                    & Navigate        & 2  & \ccell{ 2}{87}{ 2.30} & \xmark \\
& wx.navigateBackNative                         & Navigate        & 2  & \ccell{ 2}{87}{ 2.30} & \cmark \\
& wx.onDeviceOrientationChange                  & Device          & 2  & \ccell{ 2}{87}{ 2.30} & \xmark \\
& wx.openBusinessView                           & View            & 2  & \ccell{ 2}{87}{ 2.30} & \xmark \\
& wx.verifyPaymentPassword                      & Payment         & 2  & \ccell{ 2}{87}{ 2.30} & \cmark \\


\midrule[0.7pt]
%
%

\multirow{23}{*}{\rotatebox[origin=c]{90}{\textbf{WeCom}}} &
  wx.hideNavigationBar                          & Bar             & 28 & \ccell{28}{90}{31.11} & \xmark \\
& wx.requestSubscribeMessage                    & Subscribe       & 25 & \ccell{25}{90}{27.78} & \xmark \\
& wx.showNavigationBar                          & Bar             & 23 & \ccell{23}{90}{25.56} & \xmark \\
& wx.requestVirtualPayment                      & Payment         & 11 & \ccell{11}{90}{12.22} & \cmark \\
& wx.openUrl                                    & Misc            & 8  & \ccell{ 8}{90}{ 8.89} & \cmark \\
& wx.hideHomeButton                             & Interaction     & 8  & \ccell{ 8}{90}{ 8.89} & \xmark \\
& wx.enterContact                               & Contact         & 5  & \ccell{ 5}{90}{ 5.56} & \cmark \\
& wx.drawCanvas                                 & Canvas          & 5  & \ccell{ 5}{90}{ 5.56} & \xmark \\
& wx.setPageOrientation                         & Misc            & 4  & \ccell{ 4}{90}{ 4.44} & \xmark \\
& wx.operateWXData                              & Misc            & 4  & \ccell{ 4}{90}{ 4.44} & \xmark \\
& wx.getBackgroundFetchData                     & Misc            & 3  & \ccell{ 3}{90}{ 3.33} & \xmark \\
& wx.setBackgroundFetchToken                    & Misc            & 3  & \ccell{ 3}{90}{ 3.33} & \xmark \\
& wx.startFacialRecognitionVerify               & Bio-Auth        & 3  & \ccell{ 3}{90}{ 3.33} & \cmark \\
& wx.checkIsSupportFacialRecognition            & Bio-Auth        & 2  & \ccell{ 2}{90}{ 2.22} & \cmark \\
& wx.navigateBackApplication                    & Navigate        & 2  & \ccell{ 2}{90}{ 2.22}  & \xmark \\
& wx.navigateBackNative                         & Navigate        & 2  & \ccell{ 2}{90}{ 2.22}  & \cmark \\
& wx.openBusinessView                           & Misc            & 2  & \ccell{ 2}{90}{ 2.22} & \xmark \\
& wx.qy.chooseAttach                            & File            & 2  & \ccell{ 2}{90}{ 2.22} & \cmark \\
& wx.qy.chooseWxworkContact                     & Enterprise      & 2  & \ccell{ 2}{90}{ 2.22} & \cmark \\
& wx.qy.chooseWxworkVisibleRange                & Enterprise      & 2  & \ccell{ 2}{90}{ 2.22} & \cmark \\
& wx.qy.openWechatWebviewUrl                    & WebView         & 2  & \ccell{ 2}{90}{ 2.22} & \xmark \\
& wx.qy.postNotification                        & System          & 2  & \ccell{ 2}{90}{ 2.22}  & \cmark \\
& wx.qy.showUserProfile                         & User Info       & 2  & \ccell{ 2}{90}{ 2.22}  & \cmark \\
& wx.qy.wwLog                                   & Uncategorized   & 2  & \ccell{ 2}{90}{ 2.22} & \xmark \\
& wx.qy.wwOpenUrlScheme                         & Uncategorized   & 2  & \ccell{ 2}{90}{ 2.22} & \cmark \\
& wx.verifyPaymentPassword                      & Payment         & 2  & \ccell{ 2}{90}{ 2.22} & \cmark \\


\midrule[0.7pt]

%

\multirow{14}{*}{\rotatebox[origin=c]{90}{\textbf{QQ}}} &
   qq.openUrl            & Misc          & 4 & \ccell{4}{9}{44.44} & \xmark \\
 & qq.addRecentColorSign & UI            & 3 & \ccell{3}{9}{33.33} & \xmark \\
 & qq.exitMiniProgram    & App           & 2 & \ccell{2}{9}{22.22} & \xmark \\
 & qq.getGroupInfo       & User Info     & 2 & \ccell{2}{9}{22.22} & \xmark \\
 & qq.getGroupInfoExtra  & User Info     & 2 & \ccell{2}{9}{22.22} & \xmark \\
 & qq.getPerformance     & System        & 1 & \ccell{1}{9}{11.11} & \xmark \\
 & qq.getQua             & Uncategorized & 1 & \ccell{1}{9}{11.11} & \xmark \\
 & qq.getUserInfoExtra   & User Info     & 1 & \ccell{1}{9}{11.11} & \xmark \\
 & qq.invokeNativePlugin & System        & 1 & \ccell{1}{9}{11.11} & \cmark \\
 & qq.notifyNative       & System        & 1 & \ccell{1}{9}{11.11} & \xmark \\
 & qq.openScheme         & Misc          & 1 & \ccell{1}{9}{11.11} & \cmark \\
 & qq.requestMidasPayment& Payment       & 1 & \ccell{1}{9}{11.11} & \xmark \\
 & qq.toggleSecureWindow & UI            & 1 & \ccell{1}{9}{11.11} & \xmark \\
 & qq.wnsRequest         & App           & 1 & \ccell{1}{9}{11.11} & \xmark \\

\bottomrule[1.5pt]

\end{tabular}
 
\caption{The popular hidden APIs invoked by the 1st-party miniapps.}
\label{tab:popular}
 \vspace{-0.25in}
\end{table}

\ignore{ 
ApiName, AppId
wx.launchApplication, wx162102d2ff543cb2
wx.makeVoIPCall, wx20e91ae3eef01590
wx.launchApplicationForNative, wx162102d2ff543cb2
wx.setBackgroundFetchToken, wxbb58374cdce267a6
wx.setBackgroundFetchToken, wx7ddcd8e668b1407c
wx.setBackgroundFetchToken, wx1183b055aeec94d1
wx.checkIsSupportFacialRecognition, wx11fb875257231897
wx.checkIsSupportFacialRecognition, wx156db85ccccb85a7
wx.openOfflinePayView, wx7ec43a6a6c80544d
wx.openUrl, wx0879aa4a0aa5392e
wx.openUrl, wx1183b055aeec94d1
wx.openUrl, wx0f92159d5cff5949
wx.openUrl, wxb036cafe2994d7d0
wx.openUrl, wx0be52f1974c8bb1c
wx.openUrl, wx95e493619d0ff929
wx.openUrl, wx5237c2a6ad938b2c
wx.openUrl, wxbfdf33f581386a7a
wx.exitVoIPChat, wx10b96a36f7e0e981
wx.bindPaymentCard, wx221a6c79f8176f0a
wx.requestVirtualPayment, wx0ad25002fecb0753
wx.requestVirtualPayment, wx1d38aba856173452
wx.requestVirtualPayment, wx2d1677e7646146bd
wx.requestVirtualPayment, wx6a4d2f40aff3d22a
wx.requestVirtualPayment, wx1f250a8d89fa3759
wx.requestVirtualPayment, wx0879aa4a0aa5392e
wx.requestVirtualPayment, wx8d4fa2e72a27aab4
wx.requestVirtualPayment, wx8a5d6f9fad07544e
wx.requestVirtualPayment, wxc88f41e36c417bff
wx.requestVirtualPayment, wxbfdf33f581386a7a
wx.requestVirtualPayment, wxe2039b83454e49ed
wx.verifyPaymentPassword, wxbb58374cdce267a6
wx.verifyPaymentPassword, wx221a6c79f8176f0a
wx.reportIDKey, wxb036cafe2994d7d0
wx.launchApplicationDirectly, wx162102d2ff543cb2
wx.startFacialRecognitionVerify, wx11fb875257231897
wx.startFacialRecognitionVerify, wx156db85ccccb85a7
wx.startFacialRecognitionVerify, wx24c18acef76f6cd3
wx.startFacialRecognitionVerifyAndUploadVideo, wx156db85ccccb85a7
wx.setPageOrientation, wx1c0d06533e7dd178
wx.setPageOrientation, wx0be52f1974c8bb1c
wx.setPageOrientation, wx95e493619d0ff929
wx.setPageOrientation, wxbfdf33f581386a7a
wx.getBackgroundFetchData, wxbb58374cdce267a6
wx.getBackgroundFetchData, wx7ddcd8e668b1407c
wx.getBackgroundFetchData, wx1183b055aeec94d1
wx.openVideoEditor, wxb036cafe2994d7d0
wx.onTouchCancel, wx2012b69e27152882
wx.onTouchCancel, wx2401c2aa0c48eff0
wx.onTouchCancel, wx200b57cf02e7ca2c
wx.onTouchCancel, wx2f7fda52d8d031ee
wx.onTouchCancel, wx10b96a36f7e0e981
wx.onTouchCancel, wx7a232d077ee64bca
wx.onTouchCancel, wx21b2f604739573d2
wx.onTouchCancel, wx04d3dc7f54781406
wx.onTouchCancel, wxcf75748f149cf808
wx.onTouchCancel, wx20afc706a711eefc
wx.onTouchCancel, wxa564723fd4246246
wx.onTouchCancel, wx22256626b215fe10
wx.onTouchCancel, wxa565c0ffda455216
wx.onTouchCancel, wx0be899404e2de1ee
wx.onTouchCancel, wx1ce18fe0bd4af481
wx.onTouchStart, wx2012b69e27152882
wx.onTouchStart, wx2401c2aa0c48eff0
wx.onTouchStart, wx200b57cf02e7ca2c
wx.onTouchStart, wx2f7fda52d8d031ee
wx.onTouchStart, wx10b96a36f7e0e981
wx.onTouchStart, wx7a232d077ee64bca
wx.onTouchStart, wx21b2f604739573d2
wx.onTouchStart, wx04d3dc7f54781406
wx.onTouchStart, wxcf75748f149cf808
wx.onTouchStart, wx20afc706a711eefc
wx.onTouchStart, wxa564723fd4246246
wx.onTouchStart, wx22256626b215fe10
wx.onTouchStart, wxa565c0ffda455216
wx.onTouchStart, wx0be899404e2de1ee
wx.onTouchStart, wx1ce18fe0bd4af481
wx.onTouchMove, wx2012b69e27152882
wx.onTouchMove, wx2401c2aa0c48eff0
wx.onTouchMove, wx200b57cf02e7ca2c
wx.onTouchMove, wx2f7fda52d8d031ee
wx.onTouchMove, wx10b96a36f7e0e981
wx.onTouchMove, wx7a232d077ee64bca
wx.onTouchMove, wx21b2f604739573d2
wx.onTouchMove, wx04d3dc7f54781406
wx.onTouchMove, wxcf75748f149cf808
wx.onTouchMove, wx20afc706a711eefc
wx.onTouchMove, wxa564723fd4246246
wx.onTouchMove, wx22256626b215fe10
wx.onTouchMove, wxa565c0ffda455216
wx.onTouchMove, wx0be899404e2de1ee
wx.onTouchMove, wx1ce18fe0bd4af481
wx.onTouchEnd, wx2012b69e27152882
wx.onTouchEnd, wx2401c2aa0c48eff0
wx.onTouchEnd, wx200b57cf02e7ca2c
wx.onTouchEnd, wx2f7fda52d8d031ee
wx.onTouchEnd, wx10b96a36f7e0e981
wx.onTouchEnd, wx7a232d077ee64bca
wx.onTouchEnd, wx21b2f604739573d2
wx.onTouchEnd, wx04d3dc7f54781406
wx.onTouchEnd, wxcf75748f149cf808
wx.onTouchEnd, wx20afc706a711eefc
wx.onTouchEnd, wxa564723fd4246246
wx.onTouchEnd, wx22256626b215fe10
wx.onTouchEnd, wxa565c0ffda455216
wx.onTouchEnd, wx0be899404e2de1ee
wx.onTouchEnd, wx1ce18fe0bd4af481
wx.onDeviceOrientationChange, wx7a232d077ee64bca
wx.onDeviceOrientationChange, wxcf75748f149cf808
wx.navigateBackApplication, wx034ecf8353c9af95
wx.navigateBackApplication, wx03efd44c4eef2ff2
wx.updateApp, wxd7bec290bace5e42
wx.notifyBLECharacteristicValueChanged, wx1614b52d93bab73c
wx.enterContact, wx1c0d06533e7dd178
wx.enterContact, wx1183b055aeec94d1
wx.enterContact, wx0be52f1974c8bb1c
wx.enterContact, wx95e493619d0ff929
wx.enterContact, wxbfdf33f581386a7a
wx.checkHandoffEnabled, wx20afc706a711eefc
wx.startHandoff, wx20afc706a711eefc
wx.hideKeyboard, wx7ddcd8e668b1407c
wx.hideKeyboard, wx1c0d06533e7dd178
wx.hideKeyboard, wx034ecf8353c9af95
wx.hideKeyboard, wx2012b69e27152882
wx.hideKeyboard, wx0879aa4a0aa5392e
wx.hideKeyboard, wx2401c2aa0c48eff0
wx.hideKeyboard, wx200b57cf02e7ca2c
wx.hideKeyboard, wx11fb875257231897
wx.hideKeyboard, wx2f7fda52d8d031ee
wx.hideKeyboard, wx10b96a36f7e0e981
wx.hideKeyboard, wx03efd44c4eef2ff2
wx.hideKeyboard, wx04d3dc7f54781406
wx.hideKeyboard, wxa564723fd4246246
wx.hideKeyboard, wx22256626b215fe10
wx.hideKeyboard, wxa565c0ffda455216
wx.hideKeyboard, wx0be899404e2de1ee
wx.hideKeyboard, wx1ce18fe0bd4af481
wx.chooseMultiMedia, wxb036cafe2994d7d0
wx.compressVideo, wxb036cafe2994d7d0
wx.getVideoInfo, wxb036cafe2994d7d0
wx.requestSubscribeMessage, wxbb58374cdce267a6
wx.requestSubscribeMessage, wx47b6a7e42066a91e
wx.requestSubscribeMessage, wx0ad25002fecb0753
wx.requestSubscribeMessage, wx0733d494160bf151
wx.requestSubscribeMessage, wxbd56b62a13d242d2
wx.requestSubscribeMessage, wx595c6409513971b0
wx.requestSubscribeMessage, wx03fef9f1ee281e70
wx.requestSubscribeMessage, wx0bde570d4b5447b7
wx.requestSubscribeMessage, wx1f718043e6351e23
wx.requestSubscribeMessage, wxa2c453d902cdd452
wx.requestSubscribeMessage, wx609a5919fded4c05
wx.requestSubscribeMessage, wx8a5d6f9fad07544e
wx.requestSubscribeMessage, wx3d5765a3af6ce55f
wx.requestSubscribeMessage, wxb036cafe2994d7d0
wx.requestSubscribeMessage, wx26da53d900421226
wx.requestSubscribeMessage, wxc88f41e36c417bff
wx.requestSubscribeMessage, wx854a7dd7103ee756
wx.requestSubscribeMessage, wx05551c5ee1fd1c12
wx.requestSubscribeMessage, wx0be52f1974c8bb1c
wx.requestSubscribeMessage, wxcf75748f149cf808
wx.requestSubscribeMessage, wx24c18acef76f6cd3
wx.requestSubscribeMessage, wx1543a5cce73d4ad7
wx.requestSubscribeMessage, wx474bd4705e1ec69e
wx.requestSubscribeMessage, wx60d176f873ca2d67
wx.requestSubscribeMessage, wx8b974282f75dcd3e
wx.navigateBackNative, wx1183b055aeec94d1
wx.navigateBackNative, wx162102d2ff543cb2
wx.onVoIPChatSpeakersChanged, wx10b96a36f7e0e981
wx.showNavigationBar, wx6f4681a24ab3b251
wx.showNavigationBar, wx0743850faf95fb1f
wx.showNavigationBar, wx474a5c3c033d43e9
wx.showNavigationBar, wx0f17839143a02bdb
wx.showNavigationBar, wx1ff509c4c7be470f
wx.showNavigationBar, wx221a6c79f8176f0a
wx.showNavigationBar, wx277c9f1d194fce2f
wx.showNavigationBar, wxbd56b62a13d242d2
wx.showNavigationBar, wx0879aa4a0aa5392e
wx.showNavigationBar, wx1d5046af9a8fc086
wx.showNavigationBar, wx1183b055aeec94d1
wx.showNavigationBar, wx0d7f9a7816347b3d
wx.showNavigationBar, wx0f92159d5cff5949
wx.showNavigationBar, wx0b1ba004e9702051
wx.showNavigationBar, wx1614b52d93bab73c
wx.showNavigationBar, wx1b138d17e3ce2ecc
wx.showNavigationBar, wxb36a705fcf2a4b5c
wx.showNavigationBar, wx0cdb410f2af4ead3
wx.showNavigationBar, wxe747e5d6f4488ca2
wx.showNavigationBar, wx5237c2a6ad938b2c
wx.showNavigationBar, wx1e47475cef10f092
wx.showNavigationBar, wx0df87e4c51a1bb56
wx.showNavigationBar, wx0ca5fcc120e80dee
wx.hideHomeButton, wx7ec43a6a6c80544d
wx.hideHomeButton, wx2d1677e7646146bd
wx.hideHomeButton, wx17995583cf175c9a
wx.hideHomeButton, wxbd56b62a13d242d2
wx.hideHomeButton, wx595c6409513971b0
wx.hideHomeButton, wx0879aa4a0aa5392e
wx.hideHomeButton, wx136e8562b6d2f23e
wx.hideHomeButton, wx24c18acef76f6cd3
wx.hideNavigationBar, wx6f4681a24ab3b251
wx.hideNavigationBar, wx0743850faf95fb1f
wx.hideNavigationBar, wx474a5c3c033d43e9
wx.hideNavigationBar, wx0f17839143a02bdb
wx.hideNavigationBar, wx1ff509c4c7be470f
wx.hideNavigationBar, wx221a6c79f8176f0a
wx.hideNavigationBar, wx0733d494160bf151
wx.hideNavigationBar, wx277c9f1d194fce2f
wx.hideNavigationBar, wxbd56b62a13d242d2
wx.hideNavigationBar, wx0c2cd8070552df7a
wx.hideNavigationBar, wx0879aa4a0aa5392e
wx.hideNavigationBar, wx1d5046af9a8fc086
wx.hideNavigationBar, wx1183b055aeec94d1
wx.hideNavigationBar, wx0d7f9a7816347b3d
wx.hideNavigationBar, wx8a5d6f9fad07544e
wx.hideNavigationBar, wx0f92159d5cff5949
wx.hideNavigationBar, wx190c364ef8e363da
wx.hideNavigationBar, wx0b1ba004e9702051
wx.hideNavigationBar, wx1614b52d93bab73c
wx.hideNavigationBar, wx1b138d17e3ce2ecc
wx.hideNavigationBar, wxb36a705fcf2a4b5c
wx.hideNavigationBar, wx0cdb410f2af4ead3
wx.hideNavigationBar, wxe747e5d6f4488ca2
wx.hideNavigationBar, wx5237c2a6ad938b2c
wx.hideNavigationBar, wx8b974282f75dcd3e
wx.hideNavigationBar, wx1e47475cef10f092
wx.hideNavigationBar, wx0df87e4c51a1bb56
wx.hideNavigationBar, wx0ca5fcc120e80dee
wx.preloadWebview, wxbfdf33f581386a7a
wx.openUserProfile, wxb036cafe2994d7d0
wx.operateWXData, wx1c0d06533e7dd178
wx.operateWXData, wx1183b055aeec94d1
wx.operateWXData, wx0be52f1974c8bb1c
wx.operateWXData, wxbfdf33f581386a7a
wx.openBusinessView, wxbb58374cdce267a6
wx.openBusinessView, wxbd56b62a13d242d2
wx.showKeyboard, wx2012b69e27152882
wx.showKeyboard, wx2401c2aa0c48eff0
wx.showKeyboard, wx200b57cf02e7ca2c
wx.showKeyboard, wx2f7fda52d8d031ee
wx.showKeyboard, wx10b96a36f7e0e981
wx.showKeyboard, wx04d3dc7f54781406
wx.showKeyboard, wxa564723fd4246246
wx.showKeyboard, wx22256626b215fe10
wx.showKeyboard, wxa565c0ffda455216
wx.showKeyboard, wx0be899404e2de1ee
wx.showKeyboard, wx1ce18fe0bd4af481
wx.joinVoIPChat, wx10b96a36f7e0e981
wx.updateVoIPChatMuteConfig, wx10b96a36f7e0e981
wx.shareAppMessage, wx790ac7a6f613a8d9
wx.shareAppMessage, wx2012b69e27152882
wx.shareAppMessage, wx2401c2aa0c48eff0
wx.shareAppMessage, wx200b57cf02e7ca2c
wx.shareAppMessage, wx2f7fda52d8d031ee
wx.shareAppMessage, wx10b96a36f7e0e981
wx.shareAppMessage, wx7a232d077ee64bca
wx.shareAppMessage, wx21b2f604739573d2
wx.shareAppMessage, wxb036cafe2994d7d0
wx.shareAppMessage, wxa564723fd4246246
wx.shareAppMessage, wx22256626b215fe10
wx.shareAppMessage, wxa565c0ffda455216
wx.shareAppMessage, wx0be899404e2de1ee
wx.shareAppMessage, wx1ce18fe0bd4af481
wx.shareAppMessageForFakeNative, wxb036cafe2994d7d0
wx.drawCanvas, wx221a6c79f8176f0a
wx.drawCanvas, wx277c9f1d194fce2f
wx.drawCanvas, wx34ceb098e2e0fef3
wx.drawCanvas, wx1468638cc72c9433
wx.drawCanvas, wx0979b92f91db4b90
}

\ignore{
ApiName, AppId
wx.onBLEConnectionStateChanged, wx1614b52d93bab73c
wx.onKeyboardComplete, wx2012b69e27152882
wx.onKeyboardComplete, wx2401c2aa0c48eff0
wx.onKeyboardComplete, wx10b96a36f7e0e981
wx.onKeyboardComplete, wx0be899404e2de1ee
wx.onKeyboardComplete, wxa565c0ffda455216
wx.onKeyboardComplete, wx1ce18fe0bd4af481
wx.onKeyboardConfirm, wx2012b69e27152882
wx.onKeyboardConfirm, wx2401c2aa0c48eff0
wx.onKeyboardConfirm, wx200b57cf02e7ca2c
wx.onKeyboardConfirm, wx2f7fda52d8d031ee
wx.onKeyboardConfirm, wx10b96a36f7e0e981
wx.onKeyboardConfirm, wx04d3dc7f54781406
wx.onKeyboardConfirm, wxa564723fd4246246
wx.onKeyboardConfirm, wx22256626b215fe10
wx.onKeyboardConfirm, wxa565c0ffda455216
wx.onKeyboardConfirm, wx0be899404e2de1ee
wx.onKeyboardConfirm, wx1ce18fe0bd4af481
wx.exitMiniProgram, wx2401c2aa0c48eff0
wx.exitMiniProgram, wx200b57cf02e7ca2c
wx.exitMiniProgram, wx2f7fda52d8d031ee
wx.exitMiniProgram, wx10b96a36f7e0e981
wx.exitMiniProgram, wx04d3dc7f54781406
wx.exitMiniProgram, wxa564723fd4246246
wx.exitMiniProgram, wx22256626b215fe10
wx.exitMiniProgram, wx1ce18fe0bd4af481
wx.qy.chooseAttach, wx162102d2ff543cb2
wx.qy.chooseAttach, wxd69b3b072c978e71
wx.qy.chooseWxworkContact, wx162102d2ff543cb2
wx.qy.chooseWxworkContact, wxd69b3b072c978e71
wx.qy.chooseWxworkVisibleRange, wx162102d2ff543cb2
wx.qy.chooseWxworkVisibleRange, wxd69b3b072c978e71
wx.qy.openWechatWebviewUrl, wx1203f207f14e4382
wx.qy.openWechatWebviewUrl, wx1c572024d87e35f6
wx.qy.postNotification, wx162102d2ff543cb2
wx.qy.postNotification, wxd69b3b072c978e71
wx.qy.wwLog, wx162102d2ff543cb2
wx.qy.wwLog, wxd69b3b072c978e71
wx.qy.wwOpenUrlScheme, wx162102d2ff543cb2
wx.qy.wwOpenUrlScheme, wxd69b3b072c978e71
wx.qy.showUserProfile, wx162102d2ff543cb2
wx.qy.showUserProfile, wxd69b3b072c978e71
}

\ignore{
ApiName, AppId
swan.getChannelID, IMqaa6Zm0hYuXGGIZ9GTxpg930Gyt3O8
swan.getChannelID, f8DgDWSy4dsfbZGIoYoOXcIHPwkPGOOo
swan.getBDUSS, DTs6YmkI2WIw7PfG2DLIBNR5vrDApGFG
swan.getBDUSS, KlFhQNb7j1B912a5YPsX4aYvtcfH8acb
swan.getBDUSS, f8DgDWSy4dsfbZGIoYoOXcIHPwkPGOOo
swan.getBDUSS, HT5DeQLqY7Tfil7GEMXGYKQd5piTdpjY
swan.button, wmZE0VI8hnvA3VhwkL8iqUrOtgjdsGpc
swan.button, XQTpby1qsABm49rec62n2s4QHSLHBYNF
swan.button, DTs6YmkI2WIw7PfG2DLIBNR5vrDApGFG
swan.button, KlFhQNb7j1B912a5YPsX4aYvtcfH8acb
swan.button, ExO62Fc08I6Ouw2Fs7f7cNcFLSNAKIpp
swan.button, Lfi6Kbvh1amhBhGpSVNSCXgoqg7eEi6F
swan.button, XCM35IGcjutvN1NzszjXHBf5wUlxqIgi
swan.button, QVgDrpPPg8G3jzwm8NafW3vxWnR4CN4d
swan.button, N1Rea9eAOg0QCfiwfvx9GKrZjQ7K7jTc
swan.button, sAz5tYyx4C1f63GrRMpxdtHKRlM4Uofr
swan.button, 8vryYKAWv4gfsUscA8bACr8qWaOAN0Zj
swan.button, VicTe0Ztrj8TDbtpRTRHr33GystrMU3Q
swan.button, iVPUHUL8VQIf7IpZGf5A0gXnu60uInwi
swan.button, iFTnOcBz8ulu65oUuajXF2uflHugy2GN
swan.button, XH7W4UxV2uasEy1Rkduyrj22bh4T66ji
swan.button, AqABikBRYO3twMgjeKxxkqWOTxdozaxa
swan.button, jQbV1PYedRB6mvn5Q02dCc6UYyi9CWbM
swan.button, GidbxGjMhcXQsHbhL7KHyLm2qWim65ct
swan.button, 7Epxy0qQR61lWeklwDYlAY06grONiSjR
swan.button, OyIvf6LYVhKkbIHS1USP7xnSKYxc36SH
swan.button, 97luawUa4at5F5r9N0FTQ2bGgoudzNRk
swan.button, WMlOeCRKLYb9Dzkl1Z3wpWtKO0uVjIdS
swan.button, IsqkqAR53S06yH9o1Mk7eMge9Rzge2ga
swan.button, bVR77sGWfvulRlCjd8DPfuqXe4ygpkg5
swan.button, 9YNeQPXCrxLexd6xfYWjQpL0wS0KKbzi
swan.button, xEGbZsljVhOaWjyrbDIlTi7BK1B50I5x
swan.button, upAGD5WYNgyzXNUXFowaFEh39RzvEviG
swan.button, vNNLUE4sqzuPIYMnxag2mrOXS7gVlZ41
swan.button, ARBRiFAe6xwQBP2Lt9OfNdY4WcUGGF2N
swan.button, Tv397hxPNa5PYBBAHds139ymOm3v2nsQ
swan.button, 3KauWziz3DoOYTekDktLoZGh7Yg9b36q
swan.button, BSpTnfTcsO2srfVEfGmC0ASUFZENwiKi
swan.button, lr0QS7IkpcnrGKKAUgXmL0NpscXW2M8O
swan.button, eCDBWuwxmfDjPGXTYL6HvSpr9SpK15S3
swan.button, jRNSdWjGrADFwaR9SYqMwYwRbhixxDty
swan.button, UE2u6CQ2gSzAvKLmnIhv1LNsbPlmpyGR
swan.button, ZoF6fsy9HSfDdLOkjs1D6ZIgOrMxeL3c
swan.button, OhOwaDaNGgUKXxBl9IOT20Imh5fWk001
swan.button, f3TuxsL47kihWbLdWMiymiYwChMr52Rd
swan.button, 2ykZUv5bFGNM7vshqk8LqCx01DAYT920
swan.button, 8qm7D3C9Hpmnzd15KPD7y2ZLU8GR9ilp
swan.button, NcrHXMWr1yeFTPtXEIBIyTlPfGhSLPFM
swan.button, N8nXgkozCmmanuV0Ha3HE8LerRHRKEs3
swan.button, c4k1rcqW5LcAv9QX8NagDuWtKb1Hu8q1
swan.button, rWU76BiGwEunlrwB1zyxYV4nYCGSKUgI
swan.button, 3ZlvT5ZUBlDjFPeIeMOvYXWLVnZtVGDS
swan.button, w5zRDkNswsGCpi5BizD8byXdo5v8jBKd
swan.button, c7ReeFQW3CAHFubvOVTBDyTM62Wo06RW
swan.button, W229ytlqVG9EQTNcafLteSymT9xWNF6C
swan.button, qAWLLcFmX2eNCZXob7gx5bvehS9S7GQY
swan.button, T43rINkXjgPfdKNXTuhQER2KdACVdB00
swan.button, 4fecoAqgCIUtzIyA4FAPgoyrc4oUc25c
swan.button, VrQjAZtjEQwFshPgVw4V7BkVowsR30cp
swan.button, 7ZGomP2s7oRxWPYiv9PKY0BIoCQp0xyM
swan.button, SuUusWn6HcegDP5SRN06mC8wF9arysFi
swan.button, f7CTdbxio81e3X7gApjBgqo4lD2VP8H9
swan.button, zoh03jWIFs6Cy373xM9yDnTRTwnKXRXH
swan.button, IlX63O5ARczW9Ga0VpubYROglIPYs4u0
swan.button, mZE47aNzWyHzj8wXOi3zvyu8rotbYkM5
swan.button, 7TKXplfglFEyLDolzeV7Uc0j3PFiPPRl
swan.button, yhGQeNFz3z9qxM78iQNaCALham4sw7zA
swan.button, GMZsjOFVkKT7iUmTSOCo86ZRMXUFlCdb
swan.button, myyhLj1HSwrQ0H595nVwN7htGAgdmUAP
swan.button, PzlGOb3V8wSY9f7itXLLAhp3aERoF1fS
swan.button, e0jv1rPiI8Mp4IpIgS1PMXywgcpNgjCN
swan.button, fepiqGSYmjf5vrVievsk0cKZwq1l9Bha
swan.button, 8GOU3BA9FvMkILn0gGzlILUvGNWxK8Gq
swan.button, LBx5ObytgpfUolp3wtxpEWv6UHqF8Hh2
swan.button, eRSS2NhafvwzibGY2oFcNN98DFSPfujf
swan.button, f8DgDWSy4dsfbZGIoYoOXcIHPwkPGOOo
swan.button, E1SQcoR9M8gecBD7lew9EkBUcfExSkdy
swan.button, 9s8DcIrc4grQcPhd8U1MmHb2FKpnUDx0
swan.button, K47hzKnfaaw86whvlDObGrFNsDevROi4
swan.button, pYiXbY1koi1PH5AIgpEAZTOyq7MI1lmA
swan.button, SZ10t2GZGeNE0FWwwqEK93FFfkDIblah
swan.button, ysnTPouxM3cxY9P4p0GZlSXfAZEraRah
swan.button, BqMNg4PRZq9U4GlZLMdOkEwBbbKLMGhu
swan.button, GI8QTdxNQrdUmX8Y45k5BxIHE8MpY0qp
swan.button, gG3rpxnTGYci41rXam5XGg0nGztS8tHr
swan.button, EbYDmUrYmMYe0O11Nt95wCFRbFUv6IGw
swan.button, ptxRLxXwa1vhFsAw4s2SqvxTHOrzaw4v
swan.button, DwY9zb4CYdRV2wZGeZ1ep1SWZHVMexwt
swan.button, pOP6hqPxQYA40Cxp5vDbVw6fooPnFO9p
swan.button, rcyY7ypuKIDQvlFW2SWxFtxL8x32rlve
swan.button, 2FbkTGgnqouBpNIdnWS1h6YqdeAWakHv
swan.button, 7u8UpxwhKkc8VsUIABh7pRyHHoSHiKWu
swan.button, fwN7fY7gG8z8tabs1yvwS7VKsviGUXqG
swan.button, UqboAdzNoLI3cvSxzbQKZ0sFCKpZDAm7
swan.button, g4X7FfGEDt7G1ksLibU22o0wB2p49W0D
swan.button, bRElFkFLhTnxBnHeiTwajvda2MQIfk6n
swan.button, 9CUuGYaoLCobPaPoVaDijzjdfGmsGIxH
swan.button, 934tI2DFoXUyBZigFM1MXR0I7nxu0cpf
swan.button, bsg4Z0GMzY0GI7ml3Ui4rkzxqgQAtmp8
swan.button, s0Ry7kRmyPlM2vCc1Or1dcHs01VKbgbF
swan.button, 6XV2DtbOrN1KUS2NioPnpzWvfcTXEGxD
swan.button, KWuetrRk75hSwoMVgf7xYSWxtqEMMUZp
swan.button, frQ5XpOCpnq00kpFyPRtIgvgMjMgQGyB
swan.button, p0ykOfx819QK8atfE5sPVTGPskXT3iyq
swan.button, eot71qyZ0ino8W34o3XG6aQ9YdAn4R1m
swan.button, Gz7Grjwr0GhpGSDIhtUk6RB1EiBCRmHK
swan.button, 7QfPuWyybXBjwgP5qybpWVlAPAb4S69G
swan.button, GeVyFwtN81ARbPF3GIbuaPlRPT3SfzYB
swan.button, DbG55dXZv0gQAbvw8XW5HWpv5M150qzA
swan.button, SNkxCYjGNiPXiT4Wo3X5ctUEfY5vWnvy
swan.postMessage, ExO62Fc08I6Ouw2Fs7f7cNcFLSNAKIpp
swan.postMessage, QVgDrpPPg8G3jzwm8NafW3vxWnR4CN4d
swan.postMessage, 7TKXplfglFEyLDolzeV7Uc0j3PFiPPRl
swan.postMessage, saQY5GEyg3QuSjDHzDyVUigL5pxqBy84
swan.postMessage, BsbLgDuz2aebjmVL8aQEGufR4iISFRxt
swan.postMessage, 2FbkTGgnqouBpNIdnWS1h6YqdeAWakHv
swan.postMessage, szSAuXUhY0KFgL7dmasAi9pZcIiotekx
swan.postMessage, GeVyFwtN81ARbPF3GIbuaPlRPT3SfzYB
swan.getCommonSysInfo, 12yo4izKNsHLyiFQ4TN31ECCtel3eyhA
swan.getCommonSysInfo, f8DgDWSy4dsfbZGIoYoOXcIHPwkPGOOo
swan.getCommonSysInfo, vzaUdvMtmlqgnqs5EuO5ZgNYPgjlXPSb
swan.getUserInfo, VicTe0Ztrj8TDbtpRTRHr33GystrMU3Q
swan.getUserInfo, GsYSPFsWSw0ldvO5joogavnqkbqRAyZh
swan.getUserInfo, BsbLgDuz2aebjmVL8aQEGufR4iISFRxt
swan.login, DTs6YmkI2WIw7PfG2DLIBNR5vrDApGFG
swan.login, KlFhQNb7j1B912a5YPsX4aYvtcfH8acb
swan.login, XCM35IGcjutvN1NzszjXHBf5wUlxqIgi
swan.login, VicTe0Ztrj8TDbtpRTRHr33GystrMU3Q
swan.login, bVR77sGWfvulRlCjd8DPfuqXe4ygpkg5
swan.login, KvFgOSHBS5QytsCMeirfsTdrv8gzBQLa
swan.login, xEGbZsljVhOaWjyrbDIlTi7BK1B50I5x
swan.login, BSpTnfTcsO2srfVEfGmC0ASUFZENwiKi
swan.login, OhOwaDaNGgUKXxBl9IOT20Imh5fWk001
swan.login, N8nXgkozCmmanuV0Ha3HE8LerRHRKEs3
swan.login, rWU76BiGwEunlrwB1zyxYV4nYCGSKUgI
swan.login, GsYSPFsWSw0ldvO5joogavnqkbqRAyZh
swan.login, c7ReeFQW3CAHFubvOVTBDyTM62Wo06RW
swan.login, 4fecoAqgCIUtzIyA4FAPgoyrc4oUc25c
swan.login, iCwO2sahQdlB1AqhlBBjBbAF2cLjwptI
swan.login, e0jv1rPiI8Mp4IpIgS1PMXywgcpNgjCN
swan.login, 8GOU3BA9FvMkILn0gGzlILUvGNWxK8Gq
swan.login, f8DgDWSy4dsfbZGIoYoOXcIHPwkPGOOo
swan.login, b0XNfxWkZqykmNXY8sB6PNZyr0QO68Ub
swan.login, vzaUdvMtmlqgnqs5EuO5ZgNYPgjlXPSb
swan.login, BqMNg4PRZq9U4GlZLMdOkEwBbbKLMGhu
swan.login, gFFTQQxVKaTIi4SRpoghkTBwFakTgT8e
swan.login, GI8QTdxNQrdUmX8Y45k5BxIHE8MpY0qp
swan.login, EbYDmUrYmMYe0O11Nt95wCFRbFUv6IGw
swan.login, pOP6hqPxQYA40Cxp5vDbVw6fooPnFO9p
swan.login, MA8pFysdzrfgZpHAW6sPQK2EbTa1DVOg
swan.login, bsg4Z0GMzY0GI7ml3Ui4rkzxqgQAtmp8
swan.login, KWuetrRk75hSwoMVgf7xYSWxtqEMMUZp
swan.login, AZQtr4jkpf90T3X9QMWVLF1bkeV4LXxD
swan.login, 7QfPuWyybXBjwgP5qybpWVlAPAb4S69G
swan.login, GeVyFwtN81ARbPF3GIbuaPlRPT3SfzYB
}

\ignore{
qq.toggleSecureWindow, E0F0A50AD1BB2DB54CCDA8E1BC06F29A_90776f8b8444e4b59d92d58eae78991a
qq.getGroupInfo, 0125203DFE2FA6671854113F6EAFC256_7193a3c20a64024c6246d5a05f1acb3b
qq.getGroupInfo, F9ACFDE1995FF6447C73E06212097A9E_311d4950096fc612df65bb719d50e958
qq.getGroupInfoExtra, 0125203DFE2FA6671854113F6EAFC256_7193a3c20a64024c6246d5a05f1acb3b
qq.getGroupInfoExtra, F9ACFDE1995FF6447C73E06212097A9E_311d4950096fc612df65bb719d50e958
qq.getPerformance, F9ACFDE1995FF6447C73E06212097A9E_311d4950096fc612df65bb719d50e958
qq.getUserInfoExtra, DCEB5DF402A461B3262F3EAAC270757D_1bc25db2b7bf0f01091bc27f8d7f43f9
qq.getQua, F9ACFDE1995FF6447C73E06212097A9E_311d4950096fc612df65bb719d50e958
qq.notifyNative, F9ACFDE1995FF6447C73E06212097A9E_311d4950096fc612df65bb719d50e958
qq.openUrl, F9ACFDE1995FF6447C73E06212097A9E_311d4950096fc612df65bb719d50e958
qq.openUrl, 05B50350D442210688521C811E86B903_dbd2183d0e0918384958c6a5a93bc6d6
qq.openUrl, 15DE31E20A7513428D0F5443717BF70C_7aa6134ce7baaa7b89f634701bdebdda
qq.openUrl, E0F0A50AD1BB2DB54CCDA8E1BC06F29A_90776f8b8444e4b59d92d58eae78991a
qq.invokeNativePlugin, 2F2D6C41AC29026A0477FDD2C1B49054_c9e681b5fbb5bd34b634e14ff6039ad4
qq.requestMidasPayment, 15DE31E20A7513428D0F5443717BF70C_7aa6134ce7baaa7b89f634701bdebdda
qq.wnsRequest, F9ACFDE1995FF6447C73E06212097A9E_311d4950096fc612df65bb719d50e958
qq.openScheme, F9ACFDE1995FF6447C73E06212097A9E_311d4950096fc612df65bb719d50e958
qq.addRecentColorSign, 67A15174E41531DADB4BE51904CF86AF_8922fb8cdf25b29a90a0c25d8e13ded1
qq.addRecentColorSign, E0F0A50AD1BB2DB54CCDA8E1BC06F29A_90776f8b8444e4b59d92d58eae78991a
qq.addRecentColorSign, 3EAAAE902E5E1B4D6E6FBA2E047CCBF1_4b63d2648fddf22bea45a221a48da899
qq.exitMiniProgram, F9ACFDE1995FF6447C73E06212097A9E_311d4950096fc612df65bb719d50e958
qq.exitMiniProgram, DCEB5DF402A461B3262F3EAAC270757D_1bc25db2b7bf0f01091bc27f8d7f43f9

E0F0A50AD1BB2DB54CCDA8E1BC06F29A_90776f8b8444e4b59d92d58eae78991a, Entertainment
0125203DFE2FA6671854113F6EAFC256_7193a3c20a64024c6246d5a05f1acb3b, Tool
F9ACFDE1995FF6447C73E06212097A9E_311d4950096fc612df65bb719d50e958, Tool
DCEB5DF402A461B3262F3EAAC270757D_1bc25db2b7bf0f01091bc27f8d7f43f9, Business
05B50350D442210688521C811E86B903_dbd2183d0e0918384958c6a5a93bc6d6, Travelling
15DE31E20A7513428D0F5443717BF70C_7aa6134ce7baaa7b89f634701bdebdda, Entertainment
2F2D6C41AC29026A0477FDD2C1B49054_c9e681b5fbb5bd34b634e14ff6039ad4, Tool
67A15174E41531DADB4BE51904CF86AF_8922fb8cdf25b29a90a0c25d8e13ded1, Tool
3EAAAE902E5E1B4D6E6FBA2E047CCBF1_4b63d2648fddf22bea45a221a48da899, Government

119E9796692342D65B9DCAC41F096D18_bb392fc8525c5952592db62ed226b35d, Travelling
1916DAFEFC5FFC918594E43B86249A9A_a176cb31ac5af128c96efb55a8138429, Traffic
23CE8D38CF8746A29620B980F0C951BE_4d6e53200be029c778a8b8166676cbb9, Education
273047CA2331E3C449D0CE8FF41FD1ED_607b7cf82b413b114bd1da7b381e2660, Education
311CADAB6C745F787276F11F8C993261_908e5c26dc07b6d3deb6405ca6587d42, Tool
35069ADD5DB0F361AE7BA6EBA5BA8471_47dd56e7add44fef8029079bf37ff0ca, Business
4FE8A1829195C4C689232ADEFC6A2AA1_936bb8b38133244231caf34d24ffbfbf, Tool
630D393DF7550BAE4DA8567E62C16182_281cfcc8748a91e9c62b34789721d935, E-learning
7A1B478E50AE49C8E8CD66597DB01105_4f1d25b69f4b4243ce11cf75012a0938, Lifestyle
8CE98FE28D8F77388FFFC197D5ABF574_72bed55e1997b70fc56c32d17ebc84d1, Business
B1DE016811178FADDE714B1324553E5A_af025063d209021a5f1d9b9488163e13, Health
C2FB8ABBE3B9D105D42738B6A40630FF_53254862a5dcf5ae31196201fcfbe751, Tool
C522D1D3BFCC5F8D984ABF7CEDE955E1_fcfe53ee7c3fe3f9318a1877807a1952, Education
E84F9DF3B2BBBF63D23163DF7528B10E_ea8b75f85167e888d5800f14a7da3b74, Social
FCB2DBE8546CC3AF93DE3A339F00ED2B_414173135949f4c652aca775b19f42db, Tool
}

 
Next, we sought to understand the most popular undocumented APIs and how often they are used by 1st-party miniapps. We grouped the APIs by name and counted the number of miniapps that used each API. This information is presented in table \autoref{tab:popular}. We found that 7 undocumented APIs provided by \textsf{Baidu} were used by their 1st-party miniapps, 34 undocumented APIs provided by \wechat were used by their 1st-party miniapps (only 19 of which are listed in \autoref{tab:popular} due to space constraints),  43 undocumented APIs provided by \textsf{WeCom} were used by their 1st-party miniapps (again, only those used by more than two miniapps are shown), and 14 undocumented APIs provided by \textsf{QQ} were used by their 1st-party miniapps.

%
Finally, we present whether there are any missing security checks for these undocumented APIs from our API classification result in the last column of \autoref{tab:popular}. We found that 3 out of 7 (42.9\%) APIs used by \textsf{Baidu}'s 1st-party miniapps do not have security checks and can be invoked and exploited by 3rd-party miniapps; {16} of {34} ({47.06}\%) APIs of \wechat; {22} of {43} ({51.16}\%) APIs of \textsf{WeCom}; and 12 of 14 (85.7\%) APIs of \textsf{QQ} can be exploited by 3rd-party miniapps. We also noticed that different vendors have different security restrictions on their undocumented APIs. For example, \wechat and \textsf{WeCom} place security checks on their undocumented APIs that are related to payment (\texttt{wx.requestVirtualPayment}), authentication (\texttt{wx.startFacialRecognitionVerify}) and access to resources (\texttt{wx.openUrl}). \looseness=-1

\paragraph{Usage of Hidden APIs (Among the 3rd-party Miniapps)} 
Based on the data presented in \autoref{tab:cat_score4},  we have discovered that the utilization of undocumented APIs is widespread among 3rd-party miniapps, regardless of their category. The percentage of 3rd-party miniapps employing undocumented APIs is 29.54\%. Our observations have further revealed that 3rd-party miniapps in the Shopping and Business categories are more inclined to use undocumented APIs, particularly those linked to sensitive operations like payment.

In addition, we conducted an analysis to comprehend the most popular undocumented APIs and the frequency of their usage by 3rd-party miniapps. We categorized the APIs by name and tallied the number of miniapps that leveraged each API. We have found that 103 undocumented APIs provided by \wechat were utilized by their 3rd-party miniapps. Among these APIs, it is notable that 79 of them lack security checks. As shown in \autoref{tab:cat_score2}, we present a summary of undocumented APIs that have been utilized by over 50 mini-apps. It is evident that a majority of these hidden APIs lack proper security measures. To further understand the details, we delved into a selection of them to uncover why 3rd-party mini-apps have knowledge of them and whether they are being exploited. 

Our investigation has yielded some intriguing findings. (i) While some APIs are not publicly documented, Tencent does share them with certain vendors who work closely with them and permit these vendors to request access. An example of such an API is \texttt{requestFacetoFacePayment}~\cite{requestF2Fpay} (which is used by 40,091 miniapps). (ii) There were some concealed APIs that were once freely available for use without any security checks. However, Tencent subsequently banned them. One such API is ``\texttt{openUrl}''~\cite{openUrl}. Interestingly, even though Tencent has banned the usage of this API, a whopping 17,140 miniapps have yet to remove the invocation of this API from their code (obviously, this will not work). This API has already been banned by Tencent prior to our report. (iii) There are still some APIs that remain usable until we notify Tencent of the issue. For example, \texttt{captureScreen} (12 miniapps used this API) can be utilized to obtain the user's sensitive information (See   \S\ref{subsec:casestudies}). \looseness=-1

\begin{table}[]
\scriptsize
\setlength\tabcolsep{15pt}
\centering
\begin{tabular}{lrrr}

\toprule[1.5pt]

    \multicolumn{1}{c}{\textbf{Category}} & 
    \multicolumn{1}{c}{\textbf{\# U}} & 
    \multicolumn{1}{c}{\textbf{\# App}} & 
    \multicolumn{1}{c}{\textbf{\%}} \\

\midrule[0.7pt]

Business      &  8,116 & 14,887 & \ccell{ 8116}{14887}{54.52} \\
E-learning    &    335 &  2,088 & \ccell{  335}{ 2088}{16.04} \\
Education     &  2,738 & 40,410 & \ccell{ 2738}{40410}{ 6.78} \\
Entertainment &  1,286 &  5,258 & \ccell{ 1286}{ 5258}{24.46} \\
Finance       &    262 &  1,408 & \ccell{  262}{ 1408}{18.61} \\
Food          &  1,107 &  6,345 & \ccell{ 1107}{ 6345}{17.45} \\
Games         &  1,777 &  4,745 & \ccell{ 1777}{ 4745}{37.45} \\
Government    &    929 &  7,808 & \ccell{  929}{ 7808}{11.90} \\
Health        &    795 &  6,422 & \ccell{  795}{ 6422}{12.38} \\
Job           &    177 &  4,399 & \ccell{  177}{ 4399}{ 4.02} \\
Lifestyle     & 11,846 & 35,371 & \ccell{11846}{35371}{33.49} \\
Photo         &    136 &  1,981 & \ccell{  136}{ 1981}{ 6.87} \\
Shopping      & 44,629 & 46,202 & \ccell{44629}{46202}{96.60} \\
Social        &    217 &  5,694 & \ccell{  217}{ 5694}{ 3.81} \\
Sports        &    312 &  3,378 & \ccell{  312}{ 3378}{ 9.24} \\
Tool          &  3,423 & 72,301 & \ccell{ 3423}{72301}{ 4.73} \\
Traffic       &    580 &  6,502 & \ccell{  580}{ 6502}{ 8.92} \\
Travelling    &    309 &  2,160 & \ccell{  309}{ 2160}{14.31} \\

\midrule[0.7pt]

Total & 78,974 & 267,359 & \ccell{78974}{267359}{29.54} \\

\toprule[1.5pt]

\end{tabular}
  
\caption{The 3rd party WeChat miniapps that have used the undocumented APIs.}
    \label{tab:cat_score4} 
 \vspace{-0.2in}
\end{table}

\newcommand{\cxcell}[3]
{
    \cellcolor{pink!\xinttheiexpr 500*#1/#2\relax}
    {#3}
}

\begin{table}[t]
\centering
\scriptsize
 \setlength\tabcolsep{3pt}
    \aboverulesep=0pt
    \belowrulesep=0pt
\begin{tabular}{llrrc}
\toprule[1.5pt]

    \textbf{API Name} & \textbf{Category} & \textbf{\# App} & \textbf{\% *App} & \textbf{w/ Check} \\

\midrule[0.7pt]

wx.requestFacetoFacePayment                   & Payment   & 40,091 & \cxcell{40091}{267545}{14.98} & \cmark \\
wx.operateWXData                              & Misc      & 21,834 & \cxcell{21834}{267545}{ 8.16} & \xmark \\
wx.setPageOrientation                         & UI        & 18,499 & \cxcell{18499}{267545}{ 6.91} & \xmark \\
wx.enterContact                               & Contact   & 17,421 & \cxcell{17421}{267545}{ 6.51} & \cmark \\
wx.openUrl                                    & Misc      & 17,140 & \cxcell{17140}{267545}{ 6.41} & \cmark \\
wx.preloadWebview                             & WebView   & 15,335 & \cxcell{15335}{267545}{ 5.73} & \cmark \\
wx.navigateBackNative                         & Navigate  & 13,407 & \cxcell{13407}{267545}{ 5.01} & \cmark \\
wx.editTextWithPopForm                        & Misc      & 13,390 & \cxcell{13390}{267545}{ 5.00} & \xmark \\
wx.openAddressWithLightMode                   & Address   & 13,390 & \cxcell{13390}{267545}{ 5.00} & \xmark \\
wx.requestPersonalPay                         & Payment   & 10,263 & \cxcell{10263}{267545}{ 3.84} & \xmark \\
wx.previewMedia                               & Media     &  6,635 & \cxcell{ 6635}{267545}{ 2.48} & \xmark \\
wx.drawCanvas                                 & Canvas    &  6,055 & \cxcell{ 6055}{267545}{ 2.26} & \xmark \\
wx.openBusinessView                           & Misc      &  3,800 & \cxcell{ 3800}{267545}{ 1.42} & \xmark \\
wx.onDeviceOrientationChange                  & Device    &  1,626 & \cxcell{ 1626}{267545}{ 0.61} & \xmark \\
wx.startFacialRecognitionVerify               & Bio-Auth  &  1,239 & \cxcell{ 1239}{267545}{ 0.46} & \cmark \\
wx.checkIsSupportFacialRecognition            & Bio-Auth  &    669 & \cxcell{  669}{267545}{ 0.25} & \cmark \\
wx.notifyBLECharacteristicValueChanged        & Bluetooth &    603 & \cxcell{  603}{267545}{ 0.23} & \xmark \\
wx.getBackgroundFetchData                     & Misc      &    498 & \cxcell{  498}{267545}{ 0.19} & \xmark \\
wx.setBackgroundFetchToken                    & Misc      &    485 & \cxcell{  485}{267545}{ 0.18} & \xmark \\
wx.startFacialRecognitionVerifyAndUploadVideo & Bio-Auth  &    464 & \cxcell{  464}{267545}{ 0.17} & \cmark \\
wx.updateApp                                  & Update    &    448 & \cxcell{  448}{267545}{ 0.17} & \xmark \\
wx.openOfflinePayView                         & UI        &    324 & \cxcell{  324}{267545}{ 0.12} & \cmark \\
wx.sendBizRedPacket                           & Payment   &    212 & \cxcell{  212}{267545}{ 0.08} & \cmark \\
wx.getVideoInfo                               & Video     &    193 & \cxcell{  193}{267545}{ 0.07} & \xmark \\
wx.compressVideo                              & Video     &    148 & \cxcell{  148}{267545}{ 0.06} & \xmark \\
wx.setBLEMTU                                  & Bluetooth &    127 & \cxcell{  127}{267545}{ 0.05} & \xmark \\
wx.getPhoneNumber                             & User Info &    122 & \cxcell{  122}{267545}{ 0.05} & \xmark \\
wx.openVideoEditor                            & Video     &    118 & \cxcell{  118}{267545}{ 0.04} & \xmark \\
wx.chooseContact                              & Contact   &    100 & \cxcell{  100}{267545}{ 0.04} & \xmark \\
wx.openChannelsLive                           & Misc      &     97 & \cxcell{   97}{267545}{ 0.04} & \xmark \\
wx.openAddress                                & Address   &     96 & \cxcell{   96}{267545}{ 0.04} & \xmark \\
wx.setMenuStyle                               & Menu      &     74 & \cxcell{   74}{267545}{ 0.03} & \xmark \\

\bottomrule[1.5pt]

\end{tabular}
 
\caption{The popular hidden APIs invoked by the 3rd-party WeChat miniapps.}
\label{tab:cat_score2}

 \vspace{-0.25in}
\end{table}

\section{Exploiting Unchecked Hidden APIs} 
\label{subsec:casestudies}

\subsection{Quantifying the Security Risks}

\paragraph{Methodology} After quantifying the number of unchecked undocumented APIs, our goal is to gain a better understanding of whether or not these APIs pose any security risks. While it is possible to manually analyze each API individually, it is not very practical or reliable, especially given the vast number of APIs we need to analyze (more than 1,500 APIs). However, our observation is that for an undocumented API to have potential security implications, it must be able to access sensitive information and resources on the Android system (e.g., location, files, and the internet). Therefore, if we find that the hidden API calls a native API, we can conclude that it has the potential to pose security risks. Otherwise, we can proceed to examine the implementation of each method within that hidden API, conducting the process recursively as needed.

However, not all invoked APIs manipulate sensitive resources within the Android system. For example, the \texttt{android.graphics} API offers graphics tools that allow developers to draw directly onto the screen. It is evident that invoking these APIs would not result in any security consequences. Therefore, we consider APIs that access resources protected by permissions (such as location, the Internet, and file system) to have security risks.  
Consequently, we opted to utilize a lightweight dynamic analysis approach to identify such APIs. Specifically, we hook all Android APIs that access sensitive resources, which are typically protected by Android permissions, and invoke unchecked undocumented APIs one by one. By monitoring whether the sensitive resource access APIs are invoked during this process, we can determine whether the undocumented APIs are implemented based on them. Furthermore, we are able to infer whether these APIs posed any security risks. While this approach may not uncover all the APIs since the execution of the hidden APIs may depend on the parameters and may not trigger the underlying security sensitive APIs, it can at least provide a lower-bound.

\paragraph{Results} 
We categorize the hidden APIs by analyzing the Android APIs that utilize the resources and grouping them accordingly. 
As shown in \autoref{tab:android_trace}, we have identified 39 APIs (7.77\%) in WeChat, 40 APIs (6.75\%) in WeCom, 8 APIs (7.08\%) in Baidu,   32 APIs (26.67\%) in Tiktok and 38 APIs (12.88\%) in QQ that invoke Android APIs that are protected by permissions. It should be noted that WeChat and WeCom have the most APIs that can access sensitive resources, while Baidu has the least number of such APIs. This is likely due to the fact that super apps require more Android permissions. To be more specific, WeChat requires 92 permissions, which is larger than that of Baidu (82). These accessed sensitive resources include camera,  location,   audio, and Internet. It is important to note that hidden APIs that access sensitive resources do not necessarily mean that they can access them without requiring permission. Specifically, in addition to the resources that are safeguarded by Android permissions, we are also including \texttt{SharedPreferences} in our checklist. This is because miniapps may utilize this Android API to store files in the space belonging to the super apps, which could potentially compromise the files of both the super apps and other apps. \looseness=-1

\begin{figure}[t]
        \centering
      \includegraphics[width=0.45\textwidth]{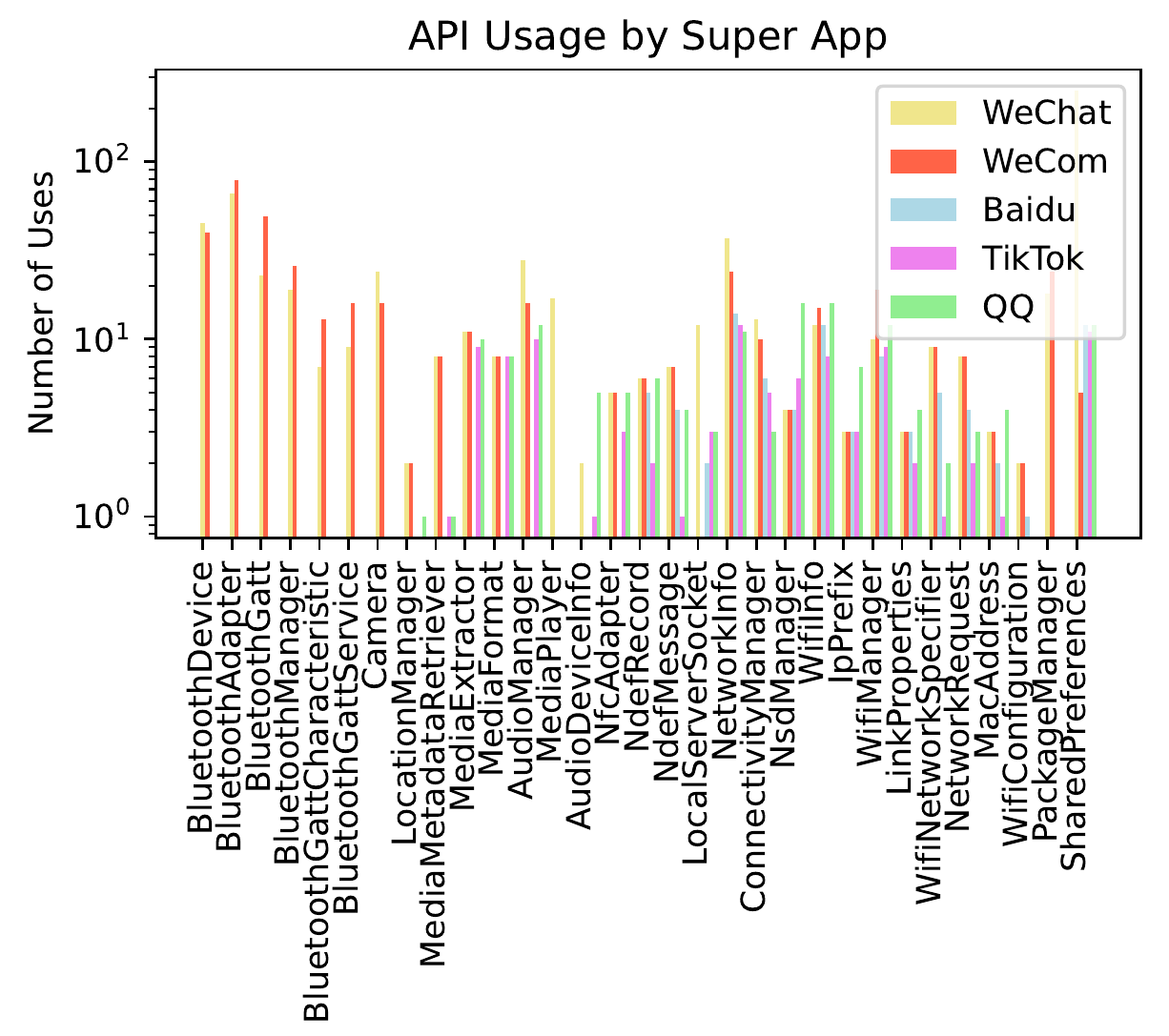}
        \caption{Android APIs used by the hidden APIs from different companies.}
        \label{fig:apiuseage}
       \vspace{-0.25in}
\end{figure}

Next, our objective is to understand the Android APIs utilized by the undocumented APIs. For this purpose, we count the number of Android APIs invoked by each hidden API of the super app, and classify them based on the names of the corresponding Android API Packages. We exclude the API packages that only be invoked once.     
It can be observed from \autoref{fig:apiuseage} that the API most commonly used is \texttt{SharedPreferences}. This is reasonable, as many of the APIs involve file operations. The available APIs consist of those  dedicated to saving screenshots onto disks, which can be utilized to launch A3.  Besides file access APIs, numerous hidden APIs make use of Internet access APIs for different purposes, including payment processing, network resource access, and more. The currently available APIs comprise those responsible for website access, which can be leveraged to trigger A1, APIs created for APK downloading and installation, which can be utilized to launch A2, and APIs for querying contact information, which can be employed to initiate A5. Please note that there are also APIs that access NFC, Camera, and Telephony Manager (which can be used to launch A4). However, since they have only been invoked once, we have excluded them from the figure. \looseness=-1


\newcommand{\cacell}[3]
{
    \cellcolor{pink!\xinttheiexpr 1000*#1/#2\relax}
    {#3}
}

\begin{table}[]
\scriptsize
\belowrulesep=0pt
\aboverulesep=0pt
\setlength\tabcolsep{2pt}

\begin{tabular}{l|rr|rr|rr|rr|rr}

\toprule[1.5pt]

    \multicolumn{1}{c|}{
        \multirow{2}{*}{
            \textbf{Resource}
        }
    } & 
    \multicolumn{2}{c|}{
        \textbf{\wechat}
    } &
    \multicolumn{2}{c|}{
        \textbf{\textsf{WeCom}}
    } & 
    \multicolumn{2}{c|}{
        \textbf{\textsf{Baidu}}
    } &
    \multicolumn{2}{c|}{
        \textbf{\textsf{Tiktok}}
    } &
    \multicolumn{2}{c}{
        \textbf{\textsf{QQ}}
    } \\ \cline{2-11} 

    \multicolumn{1}{c|}{} & 
    \multicolumn{1}{c|}{\textbf{\# UUS}} & 
    \multicolumn{1}{c|}{\textbf{\%}}  & 
    
    \multicolumn{1}{c|}{\textbf{\# UUS}} & 
    \multicolumn{1}{c|}{\textbf{\%}} & 
    
    \multicolumn{1}{c|}{\textbf{\# UUS}} & 
    \multicolumn{1}{c|}{\textbf{\%}} &
    
    \multicolumn{1}{c|}{\textbf{\# UUS}} & 
    \multicolumn{1}{c|}{\textbf{\%}} &
    
    \multicolumn{1}{c|}{\textbf{\# UUS}} & 
    \multicolumn{1}{c}{\textbf{\%}} \\

\midrule[0.7pt]

Bluetooth &  3 & \cacell{ 3}{502}{ 0.59} &  3 & \cacell{ 3}{593}{ 0.51} & - & \cacell{0}{113}{   -} &  - & \cacell{ 0}{120}{   - } &  - & \cacell{ 0}{295}{-} \\
Camera    &  1 & \cacell{ 1}{502}{ 0.20} &  1 & \cacell{ 1}{593}{ 0.17} & - & \cacell{0}{113}{   -} &  - & \cacell{ 0}{120}{   - } &  1 & \cacell{ 1}{295}{0.34} \\
Location  &  - & \cacell{ 0}{502}{   - } &  - & \cacell{ 0}{593}{   - } & - & \cacell{0}{113}{   -} &  - & \cacell{ 0}{120}{   - } &  1 & \cacell{ 1}{295}{0.34} \\
Media     &  5 & \cacell{ 5}{502}{ 0.96} &  5 & \cacell{ 5}{593}{ 0.84} & - & \cacell{0}{113}{   -} & 11 & \cacell{11}{120}{ 9.17} & 11 & \cacell{11}{295}{3.73} \\ 
NFC       &  3 & \cacell{ 3}{502}{ 0.59} &  3 & \cacell{ 3}{593}{ 0.51} & - & \cacell{0}{113}{   -} &  - & \cacell{ 0}{120}{   - } &  - & \cacell{ 0}{295}{-} \\
Network   &  16 & \cacell{ 16}{502}{ 3.19} &  16 & \cacell{ 16}{593}{ 2.70} & 7 & \cacell{7}{113}{6.19} & 20 & \cacell{20}{120}{16.67} & 24 & \cacell{24}{295}{8.14} \\
Package   &  3 & \cacell{ 3}{502}{ 0.59} &  4 & \cacell{ 4}{593}{ 0.67} & 1 & \cacell{1}{113}{0.88} &  - & \cacell{ 0}{120}{   - } &  1 & \cacell{ 1}{295}{0.34} \\
Storage   & 25 & \cacell{25}{502}{ 4.98} & 26 & \cacell{26}{593}{ 4.38} & 3 & \cacell{3}{113}{  2.65 } &  2 & \cacell{ 2}{120}{  1.67 } &  8 & \cacell{ 8}{295}{2.71} \\
Telephony &  - & \cacell{ 0}{502}{   - } &  - & \cacell{ 0}{593}{   - } & - & \cacell{0}{113}{  - } &  1 & \cacell{ 1}{120}{ 0.83} &  - & \cacell{ 0}{295}{-} \\

\midrule[0.7pt]

Total     & 39 & \cacell{39}{502}{ 7.77} & 40 & \cacell{40}{593}{ 6.75} &  8 & \cacell{8}{113}{7.08} & 32 & \cacell{32}{120}{26.67} & 38 & \cacell{38}{295}{12.88} \\

\bottomrule[1.5pt]

\end{tabular}
  
\caption{The sensitive resources that undocumented unchecked APIs accessed. UUS means undocumented unchecked sensitive APIs. Please note that a single hidden API may have access to multiple types of resources. Therefore, the total number of hidden APIs may not be equal to the sum of all the APIs that have been identified for each individual resource type.}
    \label{tab:android_trace} 
 \vspace{-0.2in}

\end{table}

\ignore{

{
"Bluetooth": [
"android.bluetooth.BluetoothDevice",
"android.bluetooth.BluetoothAdapter",
"android.bluetooth.BluetoothGatt",
"android.bluetooth.BluetoothManager",
"android.bluetooth.BluetoothGattCharacteristic",
"android.bluetooth.BluetoothGattService"
],
"Camera": [
"android.hardware.Camera"
],
"Clipboard": [
"android.text.ClipboardManager",
"android.content.ClipboardManager",
"android.content.ClipData"
],
"Intent": [
"android.content.Intent"
],
"Location": [
"android.location.LocationManager"
],
"Media": [
"android.media.MediaMetadataRetriever",
"android.media.MediaExtractor",
"android.media.MediaFormat",
"android.media.AudioManager",
"android.media.MediaPlayer",
"android.media.AudioDeviceInfo",
"android.media.AudioTrack",
"android.media.RingtoneManager",
"android.media.AudioAttributesBuilder"
],
"NFC": [
"android.nfc.NfcAdapter",
"android.nfc.NdefRecord",
"android.nfc.NdefMessage"
],
"Network": [
"android.net.LocalServerSocket",
"android.net.NetworkInfo",
"android.net.ConnectivityManager",
"android.net.nsd.NsdManager",
"android.net.RouteInfo",
"android.net.wifi.WifiInfo",
"android.net.IpPrefix",
"android.net.wifi.WifiManager",
"android.net.LinkProperties",
"android.net.wifi.WifiNetworkSpecifier",
"android.net.NetworkRequest",
"android.net.MacAddress",
"android.net.wifi.WifiConfiguration"
],
"Package": [
"android.content.pm.PackageManager"
],
"Sensor": [
"android.hardware.SensorManager"
],
"Storage": [
"android.content.res.Resources",
"android.content.SharedPreferences",
"android.os.StatFs"
],
"Telephony": [
"android.telephony.PhoneStateListener",
"android.telephony.TelephonyManager"
]
}

}

\subsection{Attack Case Studies}
\label{subsec:casestudies}






\begin{table}[]
	\centering
\scriptsize
 \setlength\tabcolsep{1pt}
\begin{tabular}{@{}llrr@{}}
\toprule[1.5pt]

    \textbf{Attacks} &
    \textbf{Targeted Resources} &
    \textbf{Exploited APIs} &
    \textbf{Vulnerable Super Apps} \\
    
\midrule[0.5pt]

    A1  & 
    Web Resources &
    \begin{tabular}[c]{@{}r@{}} \code{private\_openUrl}\\\code{openUrl}\\ \code{postMessage} \end{tabular} & 
   \begin{tabular}[c]{@{}r@{}}\textsf{WeChat},\textsf{WeCom}\\ \textsf{QQ}, \textsf{Baidu}\end{tabular} \\ \hline 
    
   A2  & 
    Web Resources &
    \begin{tabular}[c]{@{}r@{}}\code{installDownloadTask}\\  \code{addDownloadTaskStraight}\\\code{startDownloadAppTask}\\ \code{installApp}
    \end{tabular} &\begin{tabular}[c]{@{}r@{}}\textsf{WeChat},\textsf{WeCom}\\ \textsf{QQ} \end{tabular} \\ \hline

    A3 &  User information
     & \texttt{captureScreen}
     &  WeChat, WeCom
  \\ \hline  
      A4 &  User phonenumber
     & \texttt{getLocalPhoneNumber}
     &  Tiktok
      \\ \hline  
  A5 &  User contacts
     & \texttt{searchContacts}
     &  WeChat
      \\ 
 
 \ignore{   Information Stealing &
    OS Resources &
    \code{getDeviceInfo} & 
    \wechat, \textsf{WeCom}        \\ 
    }
	\bottomrule[1.5pt]
\end{tabular}
 
	\caption{Summary of the attacks we tested }
	\label{tab:vulns}
\vspace{-0.25in}
\end{table}


We present a few case studies to demonstrate how we can exploit those hidden unchecked (i.e., unprotected) APIs. For proof of concept, 
we present five case studies covering from arbitrary webpage access to information theft,   as shown in \autoref{tab:vulns}. 

\paragraph{(A1) Arbitrary Web Page Access} We made a malicious miniapp that can open any webpage using the hidden API \texttt{private\_openUrl}. Super apps usually have an allowlist of approved domains to prevent users from accessing untrusted sources (i.e., miniapps usually utilize the official API \texttt{wx.request} to access websites, and any network requests made through this API will be thoroughly vetted), but our malware can bypass these restrictions and navigate to any webpage without being vetted. This vulnerability allows our miniapp to open phishing websites and steal sensitive information, which is more powerful than previous phishing attacks~\cite{lu2020demystifying}. We were successful in this attack on several super apps but could not test it on TikTok because it does not have the necessary APIs. This vulnerability is a significant security risk for super apps because they have a unique threat model that differs from web browsers. Super apps only allow access to specific domains, unlike web browsers that can access any website. This vulnerability has been confirmed as a high-severity vulnerability by Tencent.

\paragraph{(A2) Malware Download and Installation} We developed a   malicious miniapp that can download and install malware using APIs \texttt{installDownloadTask} or \texttt{addDownloadTaskStraight}. Regular miniapps cannot download or install APK files on a mobile device because they have limited capabilities and can only download certain file types from specific servers. However, by using these APIs, a miniapp can download and install harmful APKs, which can cause significant damage to the user's mobile security and privacy. This attack works on both \wechat and WeCom.  Finally, although APKs cannot be installed without the user consent, miniApps is running inside the Super Apps, and as long as Super App has the installing permission (which most users will grant because they trust Super Apps), the malicious miniApp can install arbitrary APKs.



\paragraph{(A3) Screenshot-based  Information Theft}
We made a malicious miniapp that uses the \texttt{captureScreen}  to secretly take screenshots and store them without the user's permission. This could be used by attackers to steal sensitive information like passwords and credit card numbers from the user's screen. The consequences of this kind of attack are serious.  For example, the attacker could use them to steal the victim's identity and open fake accounts or make illegal purchases. 
 They could also use the screenshots to commit financial fraud by stealing the victim's credit card.\looseness=-1

\paragraph{(A4) Phone Number Theft} The malicious miniapps may use \texttt{getLocalPhoneNumber} to illicitly obtain the user's phone numbers. The hidden API is implemented by \texttt{getLine1Number}, which is a built-in feature of the Android SDK intended to provide the phone number associated with the SIM card currently inserted in the device. Nevertheless, access to phone number information from the SIM card may be blocked or restricted by some carriers or manufacturers, thereby rendering this attack unsuccessful in certain cases. \looseness=-1

\paragraph{(A5) Contact Information Theft} A miniapp can potentially access sensitive information, such as friend list (including the usernames and WeChat ID) using  \texttt{searchContacts}. Our experiments were conducted primarily in 2021, during which we found that this hidden API was still functional based on our raw results. Upon reporting the issue to WeChat, we were informed that another group had already reported the problem to them (CVE-2021-40180~\cite{CVE-2021}), and that the exploit no longer works on the new version of WeChat.
\section{Discussion}
\label{sec:discussion}

\ignore{
We cannot find a good reason of other than believing super app vendors had the illusion that their reserved APIs can be hidden and attackers do not know them

There could be multiple countermeasures that can prevent the undocumented API from being invoked by the third-party miniapps. Fundamentally, since the root cause of the attacks is that the host apps do not enforce the security checks for each of the undocumented API, and the most intuitive solution is to perform the security checks whenever the apps attempt to invoke those undocumented API.  To be more specific, each of the miniapp has a miniapp ID assigned by their super apps, which uniquely identifies their identities, and when any of the miniapps calls the undocumented API, the security guard will then send the miniapp ID to the backend of the super app, where they have maintained an allowlist to check whether the undocumented API is invoked by the official ones. If so, they will grant the access, and otherwise, the API invocation request will be rejected. After the API invocation request has been approved, the  JS Framework Layer then further passes the API invocation request to the underlying layers. In addition to enforcing the security checks, we have noticed that some vendors have followed other practices to prevent the unauthorized API invocation request sent from third-party miniapps. For example, although there always be interfaces in the JavaScript Framework Layer that communicate with the underlying layers, the host app can intentionally disable the availability of the interface for the third-party apps (that is, only the host app can invoke the interface). For example, we noticed that \wechat has patched their implementation used this approach after we have reported our findings to them.  
}



\paragraph{Limitations and Future Work} 
Although effective, \sysname can still be improved in various ways. It is possible for the tool to have false positives and negatives, although none have been encountered through dynamic validation and manual verification. Also, while currently tested on Android, additional work is needed to support other platforms. However, our findings are representative across different platforms, as miniapp codebases are similar. Note that \sysname is limited to super-apps that use the V8 engine and is not suitable for those that do not (e.g., Alipay).

In our study, we discovered some hidden APIs that may be vulnerable, such as the \texttt{installDownloadTask} and \texttt{addDownloadTaskStraight} APIs, which are susceptible to SQL injection attacks. Attackers can compromise super app file download tasks by replacing the download URL of the WeChat update package with a malicious one. 
We also noticed that there are two APIs called \texttt{dumpHeapSnapshot} and \texttt{HeapProfiler} that also have vulnerabilities. These APIs are designed to save data from the V8 engine to a file, but our miniapp misuses them to write to any file it wants. While Android tries to prevent this, important files like chat histories are still at risk. This could lead to serious problems because our miniapp could overwrite important files of other miniapps and their host apps, which breaks the security measures put in place by super apps. Our experiment proved that we could overwrite a file called \code{EnMicroMsg.db}, which stores chat history on \wechat. Attackers might want to make these miniapps because chat history can be used as evidence in court.
We plan to develop a tool that can identify hidden API vulnerabilities (e.g., SQL injection and buffer overflow). \looseness=-1


\ignore{
\sysname is still not perfect and can be improved in multiple ways. 
First,  while our tool does not have a false positive (confirmed with our dynamic validation) and false negative (manually verified with the code), theoretically it can have both. For instance, although \sysname has aggressively considered all the possible invariants to identify undocumened APIs, it is completely possible that there might be some corner cases that do not use these identified invariants at all. Second, we have only tested it on Android and additional work is needed to support super apps on other platforms such as Windows or iOS. However, we believe that our findings are representative and impactful as the miniapps on different platforms have similar codebase. Third, \sysname only works for super apps that utilize the V8 JavaScript engine and doesn't work for super apps (e.g., \textsf{Alipay}) that don't use it. 
\looseness=-1   
}

\paragraph{Ethics and Responsible Disclosure}
Being an attack work by nature, we must carefully address the ethical concerns. To this end, we have followed the community practice when exploiting the vulnerabilities and demonstrated our attacks. First, for proof of concept, we developed quite a number of malicious miniapps and launched attacks against our own accounts and devices. We have never uploaded our malicious miniapps onto the markets to harm other users. 
Second, we have disclosed the vulnerabilities and our attacks against \wechat to \textsf{Tencent} in September 2021, and the other four super apps in November 2021. They have all acknowledged and confirmed our findings, and so far among them \textsf{Tencent} (the biggest super app vendor with 1.2 billion monthly users) has {confirmed with 4 vulnerabilities, ranked 1 low, 2 medium, and 1 high}, and awarded us with bug bounty and fixed them. TikTok has been patched too, but not Baidu at this time of writing. 

%
 
\section{Related Work}
\label{sec:related}

\paragraph{Super Apps Security} More and more super apps have started to support the miniapp paradigm. Correspondingly, its security has received increasing attention. For instance, Lu et al.~\cite{lu2020demystifying} identified multiple flaws in \wechat, and demonstrated how an attacker would be able to launch phishing attacks against mobile users and collect sensitive data from the host apps. Zhang et al.~\cite{zhang2021measurement} developed a crawler, and understood the super apps by measuring the program practices of the provided miniapps, including how often the miniapp code will be obfuscated. Most recently, Zhang et al.~\cite{Identity:confusion:2022} studied the identity confusion in WebView-based super apps, and identified that multiple super apps contain this vulnerability. A new attack named cross-miniapp request forgery (CMRF)~\cite{yang2022cross} was also recently discovered, which exploits the missing checks of miniapp IDs for various attacks.
%
%
Differently from those works, our study  uncovers the undocumented APIs provided by the super apps and demonstrates how they can be exploited. 
In a broader scope, there is a large body of research studying the security of other super apps including web browsers and their lightweight apps, such as %
Google Instant apps~\cite{aonzo2018phishing}.  In particular, Aonzo et al.~\cite{aonzo2018phishing}, and Tang et al. ~\cite{tang2020all} point out that Google Instant Apps can be abused to mount password-stealing attacks. \looseness=-1




\paragraph{Undocumented API Detection and Exploitation} 
\sysname is the first system to detect and exploit undocumented APIs in mobile super apps like \wechat. Previous work has focused on detecting undocumented APIs in other platforms, such as Android and iOS, or on identifying missing security checks (e.g., ~\cite{livshits2013automatic,zhao2020automatic,kywe2016attacking, drakonakis2020cookie,pan2017dark,samhi2022difuzer,alhanahnah2020dina}). For example, PScout analyzed undocumented APIs in Android~\cite{au2012pscout}, and Li et al. showed that there are 17 undocumented Android APIs that are widely accessed by 3rd-party apps~\cite{li2016accessing}. Zeinab and Yousra studied access control vulnerabilities caused by residual APIs~\cite{ling2021prison}. In addition, there are ways to invoke undocumented APIs in iOS~\cite{han2013launching,wang2013jekyll} and detect their abuses~\cite{deng2015iris}. Yang et al.~\cite{yang2017precisely} proposed BridgeScope to identify sensitive JavaScript bridge APIs in hybrid apps. Undocumented APIs have also been found in the Java language and exploited by attackers~\cite{mastrangelo2015use,huang2019safecheck}. \sysname builds on this previous work to specifically focus on mobile super-apps. Finding hidden APIs in super apps using traditional techniques is difficult due to the combination of web views, host native apps, and mini app execution environments, along with code scattering and obfuscation. Our new approach monitors parameter propagation to detect API usage, using robust signatures based on super classnames and public methods. We have also created a method for automatic test case generation and API classification.
 


\section{Conclusion}
\label{sec:conclusion}

In this paper, we have revealed that super apps often contain undocumented and unchecked APIs for their 1st-party mini-apps, which can grant elevated privileges such as APK downloading, arbitrary web view accessing, and sensitive information querying. Unfortunately, these undocumented APIs can be exploited by malicious 3rd-party mini-apps, as they lack security checks. To address this issue, we have designed and implemented \sysname, a tool that can statically identify these undocumented APIs and dynamically verify their exploitability. Through our testing on five popular super apps such as \wechat and \textsf{TikTok}, we have found that all of them contain these types of APIs. Our findings suggest that super app vendors must thoroughly examine and take caution with their privileged APIs to prevent them from becoming potential exploit points.
%
\looseness=-1

\ignore{
\section*{Acknowledgments}
\label{sec:acknowledgments}
\input{paper/11_acknowledgments}

\section*{Availability}
\label{sec:availability}

Along with our research,
    we release \sysname,
    the WeChat open documentation crawler and analysis tool to reproduce our work:

\begin{itemize}
    \item Document Crawler: \url{https://github.com/MatryoshkaCracker/WeChatDocumentCrawler}
    \item Analysis Tool: \url{https://github.com/MatryoshkaCracker/WeChatJsApiMapper}
\end{itemize}

}

\bibliographystyle{IEEEtranS}

\bibliography{paper,miniapp,inputscope}


\begin{appendices}

\section{Algorithm for Invariant Extraction and Matching}
\label{appendixa}

\begin{algorithm} 
\scriptsize
\caption{\textbf{Invariant Extraction and Matching}}\label{alg:a1}
\KwIn{$ DAPI$: The Set of Documented APIs  ;  $F$: The Set of All Functions}
\KwOut{$UAPI$: The Set of Undocumented APIs }
\textbf{PROCEDURE} \textsc{InvariantExtraction}$(PAPI,F)$\\ \label{alg:i0}
$PAPI \gets \emptyset $ \;
 
    \ForEach{$~f_{j}~\in~F$}{  \label{alg:string:start}
        \ForEach{$~api_{k}~\in~DAPI$}
    {
        \If{\textsc{searchString}~($api_{k}$, $f_{j}$)}{
        
        $PAPI.add(f_{j}) $ \;
        }
  }\label{alg:string:end}
}
 
$invariant \gets \emptyset $ \; \label{alg:invariant:start}
$I \gets \emptyset $ \;
$isInvarint \gets $ \textit{\textbf{TRUE}}\;
 \ForEach{$~capi_{i}~\in~PAPI$}
  {
  $invariant \gets $\textsc{getMethodSignature}$(PAPI_{i}) $ \;
  \ForEach{$~capi_{j}~\in~PAPI$}
  {
  \If{$ invariant != $\textsc{getMethodSignature}$(capi_{j})$}{
    $isInvarint \gets $ \textit{\textbf{FALSE}}\;
     \textbf{BREAK}\; 
 
  }
  }
  \If{$isInvarint$ == \textit{\textbf{TRUE}}}{
  \If{$invariant \notin I$ }{ $I.add(invariant) $ \;}
  }
    $invariant \gets $\textsc{getSuperClass}$(capi_{i}) $ \; \label{alg:invariant:superclass:start}
  \ForEach{$~capi_{j}~\in~PAPI$}
  {
  \If{$ invariant != $\textsc{getSuperClass}$(capi_{j})$}{
    $isInvarint \gets $ \textit{\textbf{FALSE}}\; \label{alg:break}
     \textbf{BREAK}\; 
 
  }
  }
  \If{$isInvarint$ == \textit{\textbf{TRUE}}}{
  \If{$invariant \notin I$ }{ $I.add(invariant) $ \;}
  }\label{alg:invariant:superclass:end}
  $isInvarint \gets $ \textit{\textbf{TRUE}}\;
     $invariant \gets $\textsc{getSuperPackage}$(capi_{i}) $ \;
  \ForEach{$~capi_{j}~\in~PAPI$}
  {
  \If{$ invariant != $\textsc{getSuperPackage}$(capi_{j})$}{
    $isInvarint \gets $ \textit{\textbf{FALSE}}\;
     \textbf{BREAK}\; 
 
  }
  }
  \If{$isInvarint$ == \textit{\textbf{TRUE}}}{
  \If{$invariant \notin I$ }{ $I.add(invariant) $ \;}
  }
  $isInvarint \gets $ \textit{\textbf{TRUE}}\;
    $invariant \gets $\textsc{getCaller}$(capi_{i}) $ \;
  \ForEach{$~capi_{j}~\in~PAPI$}
  {
  \If{$ invariant != $\textsc{getCaller}$(capi_{j})$}{
    $isInvarint \gets $ \textit{\textbf{FALSE}}\;
     \textbf{BREAK}\; 
 
  }
  }
  \If{$isInvarint$ == \textit{\textbf{TRUE}}}{
  \If{$invariant \notin I$ }{ $I.add(invariant) $ \;}
  } 
 
 
 
  
  } \label{alg:l2}
\textbf{PROCEDURE}  \textsc{UndocumentedAPIRecognition} ($PAPI,F$) \\ \label{alg:undocumented:start}
 $UAPI \gets \emptyset $ \;
\ForEach{$~f_{j}~\in~F$}
  {

    \If{\textsc{matchinvariant}$(I, f_{j}) == $  \textbf{TRUE}}{ \label{alg:matching}
     
     \If{$f_{j} \notin IAPI$}{
     
     \If{$f_{j} \notin UAPI$}{ \label{alg:add}
     
     $UAPI.add(f_{j})$
     }\label{alg:undocumened:end}
    }
  }
 }
\end{algorithm}  

The detailed algorithm of how we extract the invariants is presented between~\autoref{alg:i0} and~\autoref{alg:l2} in~\autoref{alg:a1}.
In particular, it first identifies the implementation $f_j$ of the public APIs by searching the strings with the name of the public API (\autoref{alg:string:start}-\ref{alg:string:end}). For instance, to identify the implementation of API \texttt{wx.getLocation}, we use the string ``{\tt getLocation}'' as shown in the 5th line of \autoref{lst:jsapiclass}, if it matches, we add the implementation of the whole body into set $PAPI$ (\autoref{alg:string:end}). Next we will iterate API implementation in $PAPI$ to extract the invariants (\autoref{alg:invariant:start}-\ref{alg:l2}). For each specific invariant, e.g., the superclass (\autoref{alg:invariant:superclass:start}-\ref{alg:invariant:superclass:end}), only when this invariant exists in all APIs, we consider it is an invariant and we add it to the invariant set $I$ (\autoref{alg:invariant:superclass:end}); otherwise, we break the iteration and skip this invariant (\autoref{alg:break}). After these iterations, our invariant set will contain method signature, super class, super packages, and callers, if they exist in the corresponding public API implementations. 

With the extracted API invariants, it then becomes straightforward to identify the undocumented APIs, as shown in~\autoref{alg:undocumented:start}-\ref{alg:undocumened:end}.  Specifically, we first iterate  implementations of functions by matching the collected invariants (\autoref{alg:matching}), and if a implementation matches with all the invariants as in the public APIs (and it has not been added in the undocumented set yet),  the implementation is added as an undocumented API (\autoref{alg:add}).

\begin{figure}[t]
        \centering
      \includegraphics[width=0.35\textwidth]{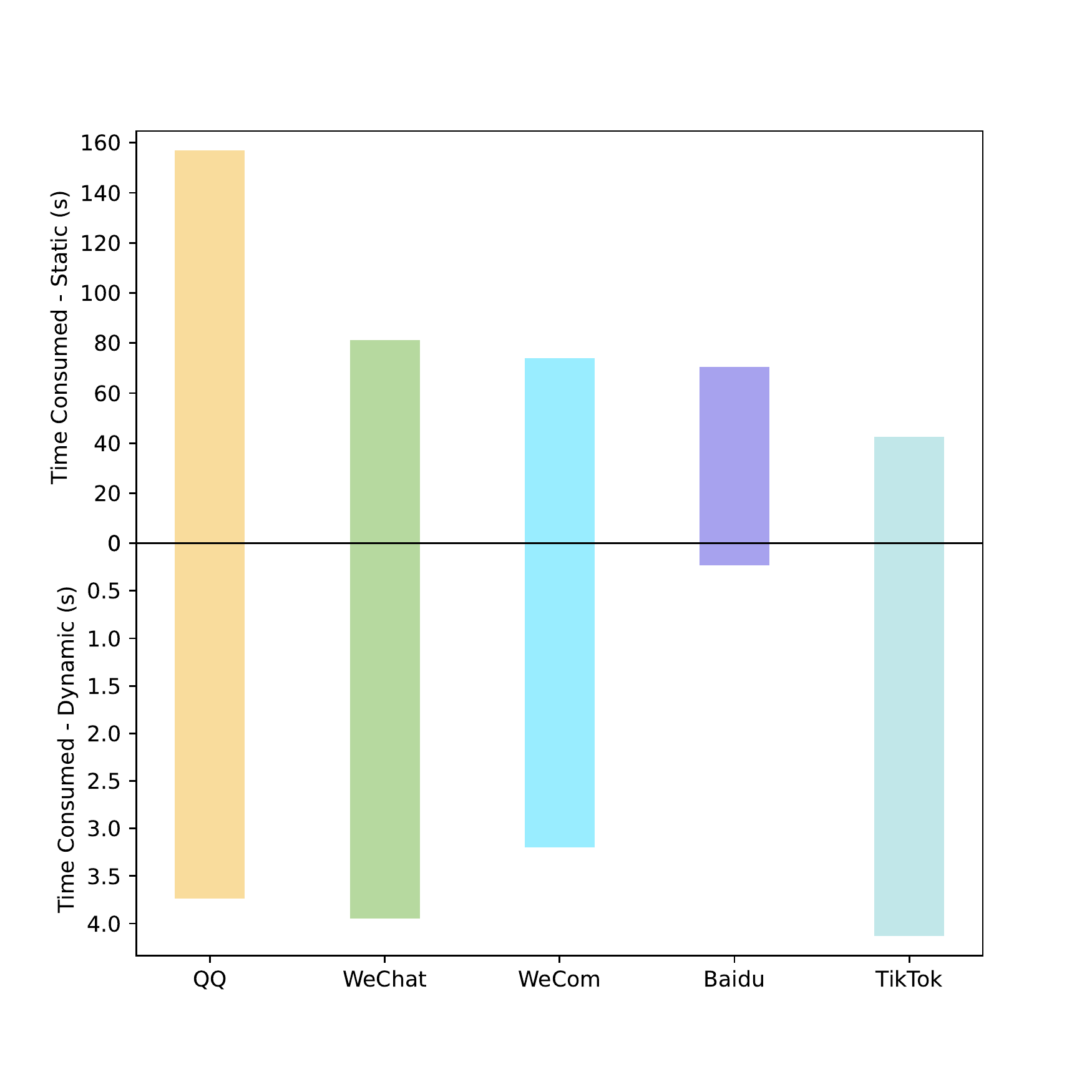}
        \caption{Time cost of \sysname in its static and dynamic analysis. The dynamic analysis only includes the time consumed for identifying API invocation points.}
        \label{fig:toolperf}
        \vspace{-0.25in}
\end{figure}
\begin{figure*}[t]
        \centering
      \includegraphics[width=0.9\textwidth]{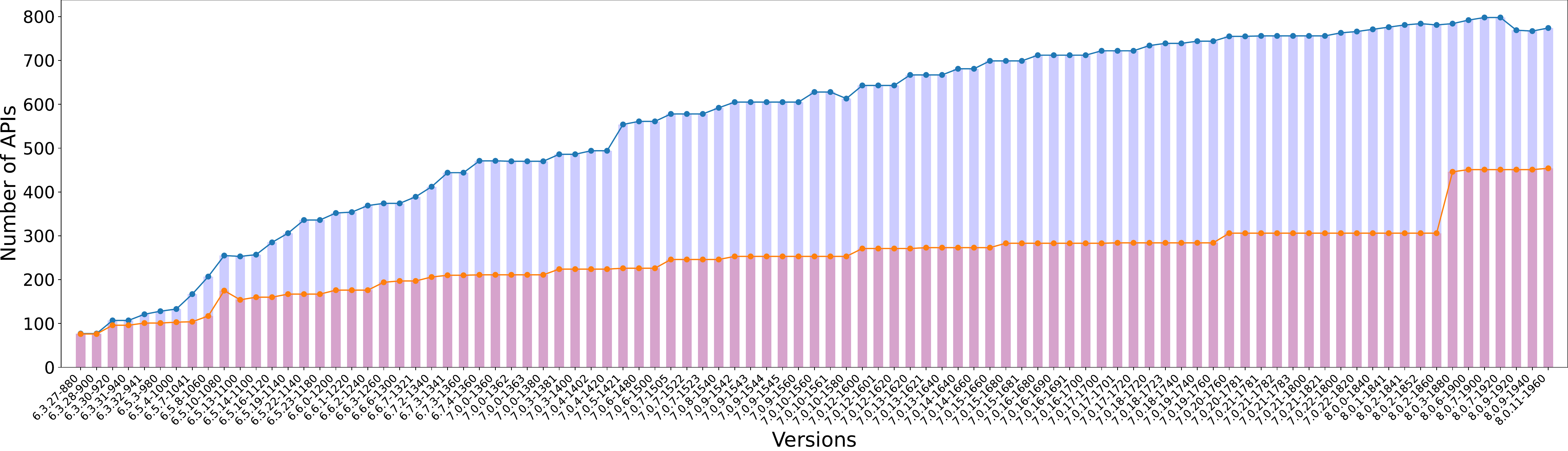}
        \caption{\# of Uncovered APIs in WeChat.  The bluebar is the \#  the APIs, and the redbar is \#  of public APIs. }
        \label{fig:wechat-old}
\end{figure*}

\begin{figure*}[t]
        \centering
      \includegraphics[width=0.7\textwidth]{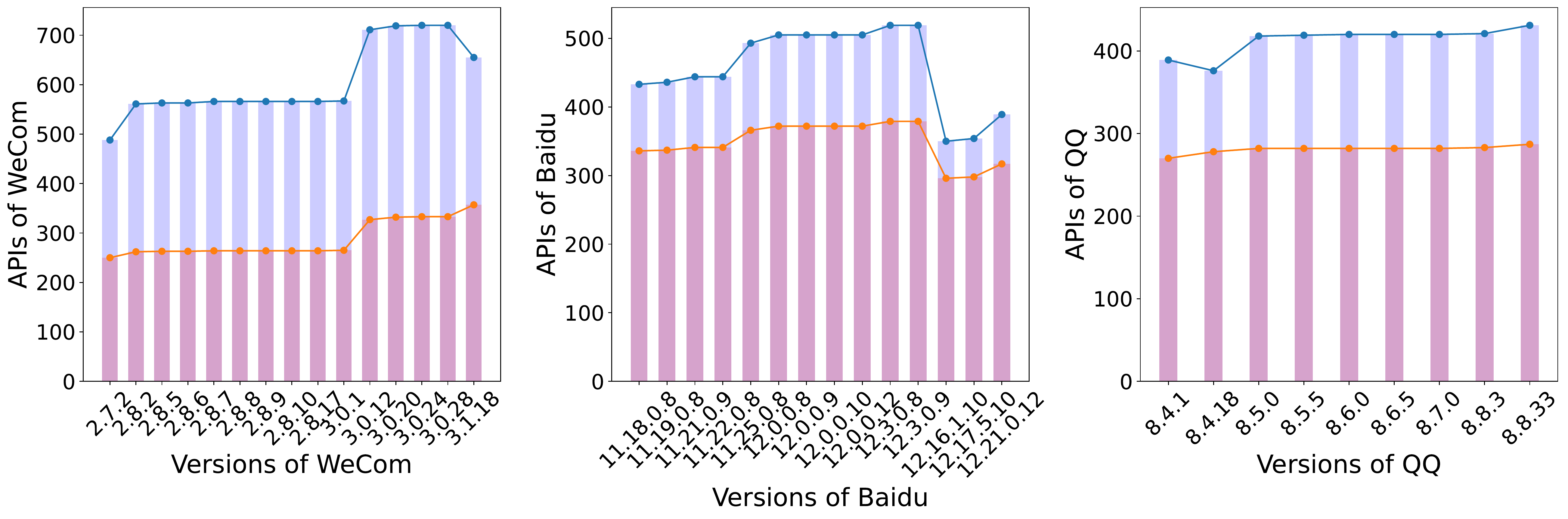}

        \caption{\# of Uncovered APIs in WeCom, Baidu and QQ.}
        \label{fig:otherold} 
\end{figure*}
\section{Efficiency} 
\label{sub:efficiency}


As our \sysname consists of two phases of analysis --- static analysis and dynamic analysis, in the following, we measure the performance of those two phases,  respectively. Specifically, the performance of the static analysis was measured by the time consumed by decompiling (via \texttt{Soot}) of the super app, and then scanning the decompiled code to identify the invariants and API candidates. 
The upper half of \autoref{fig:toolperf} shows this time cost. It can be seen that it takes 85.1  seconds to finish the static analysis of a super app on average. Among all those super apps, QQ is the most time-consuming one, which takes around 160 seconds. 

The performance of dynamic analysis was measured by the time consumed for generating test cases, identifying API invocation points, and classifying the APIs based on their executions. As shown in the bottom half of \autoref{fig:toolperf}, \sysname takes an average of 3.05 seconds to identify the API invocation point of a specific host app. It is difficult to measure the time cost for API classification, as the invocation of an API may involve user interactions that cannot be precisely measured. For example, when our tool invokes \texttt{getLocation}, the host app will pop up a dialog and ask the user to grant permission. The user's reaction time, including the time taken to press the button, will also be included in the results. As such, we can only provide approximate results, and we found that none of the dynamic API executions took hours to complete, even though there may be thousands of test cases to execute (as shown in \autoref{tab:effectiveness}). In fact, most of them just took several minutes to complete, which is acceptable since \sysname is a one-time program analysis tool for a specific super app. 

\section{The API Evolution of Super Apps} 
\label{sec:apievolution}

%
 \sysname can be used to analyze the earlier version of super apps.
 However, we have to note that its dynamic analysis component may not support the older version of the super apps (e.g., they even cannot be installed in our {\sf Google Pixel 4} phone). Also, to detect whether an API is documented or not, we need the official documentation. Unfortunately, among all five tested super apps, we cannot obtain all the historical documentations. Therefore, eventually, we collected 93 historic \wechat apps, 15 historic {\sf Wecom}, 14 historic {\sf Baidu}, and 9 historic {\sf QQ}, together with their corresponding documentation. 



The detailed changes of APIs (including documented and undocumented) with these super apps over the previous versions have been reported in \autoref{fig:wechat-old} and \autoref{fig:otherold}.  We can clearly see that
most of the super apps when started contain undocumented APIs except the first two version of \wechat, and also all of them contain significantly number of undocumented APIs. 
Meanwhile, through our manual investigations on the historical versions, we also obtained two interesting findings: (i) the documented APIs in earlier version may later become undocumented available. 
For example, API {\tt captureScreen}, which is used to capture a screenshot, has been removed from their documentation and become an undocumented one; (ii) the undocumented APIs can be released to the public. For example, an API named ``chooseContact'' was an undocumented API, and since 7.0.12, it has become a documented API. \looseness=-1

\end{appendices}

\end{document}